\documentclass[12pt]{article}
\usepackage[english]{babel}

\usepackage{amsmath}
\usepackage{amssymb}
\usepackage{amsfonts}
\usepackage{amsthm}
\usepackage{graphics,graphicx}
\usepackage{color}
\usepackage{epstopdf}
\usepackage{cite}
\usepackage{hyperref}

\date{\today}

\title{A temporal-averaging based approach to toughness homogenisation in heterogeneous material}
\author{G. Da Fies$^{(1)}$, D. Peck$^{(1)}$, M. Dutko$^{(2,1)}$, G. Mishuris$^{(1*)}$
\\
{\it $^{(1)}$Department of Mathematics, Aberystwyth University, }\\
{\it Aberystwyth, Wales, United Kingdom}
\\
{\it $^{(2)}$Rockfield Ltd, Swansea, UK}
}

\begin{document}

\maketitle

\begin{abstract}
A new approach to defining the effective fracture toughness for heterogeneous materials is proposed. This temporal averaging approach is process-dependent, incorporating the crack velocity and material toughness. The effectiveness of the new technique is investigated in the context of hydraulic fracture through heterogeneous rock with a periodic material toughness. The plane strain model is considered without fluid leak-off, to more easily investigate different regimes (toughness/viscosity). Numerical simulations are used to examine the effectiveness of the new homogenisation strategy, with comparison against the recently-proposed maximum toughness strategy. Simulations are conducted using an extremely effective (in house-built) time-space adaptive solver. The regimes in which each strategy is effective are determined. 
\end{abstract}

\section{Introduction}\label{Introduction}

Most materials, whether natural or man-made, are heterogeneous on the micro level (and many on the macro level). When creating mathematical or numerical models of complex processes however, incorporating this behaviour explicitly is often not possible, or is extremely costly. Instead, it is typically beneficial to implement some `averaging' concept, whereby the heterogeneous material is replaced in the model with an equivalent homogeneous one which preserves the properties considered most important for the analysis. This broad branch of mathematical and material sciences, comprising a wide array of homogenisation techniques and strategies (see e.g.\ textbooks and references therein \cite{KachanovSevostianov,Kanaun2008a,Kanaun2008b,Kanaun2020,Kushch2020,Movchan2002}), has developed for over a century. Starting with Maxwell \cite{maxwell_2010}, it now offers homogenised equivalents for elastic, thermal-electro-magnetic and other physical fields, having even established clear linkages between them (cross-property connections) depending upon the type of heterogeneity (see \cite{KachanovSevostianov}). Moreover, various types of heterogeneity (periodic, quasi-periodic, random, etc.) have their own array of analyses and methods, relating to the nonlocal nature of the material characteristics.

Unfortunately, the approach to successful homogenisation becomes far more complicated when dealing with local material properties. One of the most crucial examples of this is the toughness, which defines the resistance of the material to fracture propagation. It has long been known that a general homogenisation theory for the toughness was not possible, with \emph{Kachanov (and Sevostianov)} \cite{Kachanov1994,KachanovSevostianov} providing several illuminating examples to explain this unfortunate fact, while further overviews can be found in the theses \cite{HsuehThesis,LebihainThesis}, as well as the conclusion of \cite{HOSSAIN201415}. While we can't provide a full summary here, a few relevant aspects will be discussed below.




Firstly, for problems of fracture, the heterogeneity of the elastic constants (Young's modulus, Poisson's ratio) can significantly effect the crack behaviour in ways that the homogenisation needs to account for. In \cite{HOSSAIN201415}, it was demonstrated that, even if the material toughness is constant, elastic heterogeneity can result in a toughening mechanism even if the crack path remains unchanged, while an asymmetric elastic distribution (in the direction of propagation) leads to a corresponding asymmetric effective toughness.

Heterogeneity of the Young's modulus or material toughness may also cause the fracture to redirect to the more malleable material, resulting in a tortuous path and thus increasing the effective toughness. For example, the changing crack paths around material inclusions greatly complicates the prediction of the (effective) material toughness of designer materials (see e.g. \cite{10.1115/1.4045682,LEBIHAIN2020103876,LI2018128}). For the case of hydraulic fracture it was demonstrated \cite{10.2118/184873-MS} that, on the pore-scale for granular materials, while the crack follows the least tortuous path over long time periods, this may not be true over short time. Instead, the crack may branch, with some paths closing at a later time as the least tortuous path becomes the primary path. In addition, the weakening/damaging of material by the propagating crack(s) may also influence which path is least tortuous. The solid-fluid coupling inherent to hydraulic fracture complicates this further, as the nature of the path may effect the distribution of particles within the fluid (proppant) \cite{XU2020103285}, altering the fluid properties and thus the resulting fracture development \cite{10.2118/184873-MS}. The increased path length may effect the leak-off of fluid to the surrounding domain, or the transfer of heat to the fluid in geothermal applications \cite{MA202281}. The heterogeneity can also simply redirect the crack to a new primary direction of propagation \cite{OUCHI2017384}, affect the fracture profile \cite{ZOU2020103652}, alter the crack initiation and propagation conditions \cite{JU2016614,LIU2016541}, impact the fracture breakdown pressure \cite{Li2013}, alter the rock mass stability \cite{Kong2019} and influence the formation/extent of complex fracture networks \cite{Li2018}. 

The dynamic nature of fracture must also be considered. For instance, the potential rate-dependence of the energy release rate may need to be accounted for \cite{10.1007/978-94-011-4736-1_18,PhysRevLett.127.035501}. In particular, the relation may vary between different regimes of crack propagation (e.g. the subcritical and critical regimes in rock, see e.g. \cite{PhysRevLett.Ponson}). For heterogeneous materials this is complicated by crack re-nucleation, or the stopping and restarting of the crack as it encounters an area of higher toughness (see Sect.~\ref{DeltaSect}). This is of fundamental importance in hydraulic fracture applications, where predicting fracture arrest is one of the key aims of applied models, but is complicated further by the already piece-wise nature of the fracture growth \cite{CAO201724,Schrefler2019a,Schrefler2019b}. Predicting this phenomena requires a linking of the toughness and fracture criterion, necessitating an understanding of the energy driving the crack. As such, an effective toughness homogenisation strategy requires some knowledge of the process governing the fracture evolution to be incorporated. 




In spite of these difficulties, among others, a number of different approaches to homogenising the fracture toughness have been developed, only a few of which can be mentioned here. Most generalised approaches are based on the Griffith relation, and operate by estimating the fracture energy, for example \cite{10.1007/978-3-319-21611-9_19,10.1007/978-3-319-42195-7_3,HOSSAIN201415,PhysRevLett.127.035501,10.1115/1.4045682}. 
More specialised approaches have been developed for specific applications, including the prediction of the material toughness of metamaterials with inclusions \cite{LEBIHAIN2020103876,LEBIHAIN2021104463}, materials containing fibres \cite{Abrahams2017}, or those with anisotropic crack resistance \cite{Ernesti2022}. Some authors have even suggested replacing the concept of material toughness with a more specialised energy-based analysis \cite{REIFSNIDER201620}. In simple cases, it may also be possible to approximate the effective toughness using the maximum material toughness found in the body\footnote{See e.g. \cite{HOSSAIN201415}, which notes: ``After an initial transient stage, this [macroscoptic energy release rate] $J(t)$ falls into a periodic pattern as long as the crack set is away from the boundary and define the maximum macroscopic energy release rate to be the effective toughness.''}, with this approach recently being investigated for the case of hydraulic fracture in \cite{DONTSOV2021108144}.

While these methods have demonstrated appreciable successes in defining the effective toughness for heterogeneous materials, they are not applicable in all cases. Most notably, the majority of these approaches make assumptions, either explicitly or implicitly, about the nature of the crack tip velocity. This may either take the form of presuming crack propagation remains within a certain regime (typically close to the wave-speed of the medium), or that the velocity is in some way smooth, such that the tip can be examined using a roaming boundary condition $x-vt$. While in some applications these requirements are satisfied that is not the case in general, for example, when crack re-nucleation plays a role (see e.g.\ the velocity profile in Fig.~\ref{fig:12b}). 
As such, there is room for a complementary approach, particularly in applications which necessitate process-dependence of the homogenisation strategy, or where fracture arrest is of primary importance.




\begin{figure}[t]
 \centering
 \includegraphics[width=0.8\textwidth]{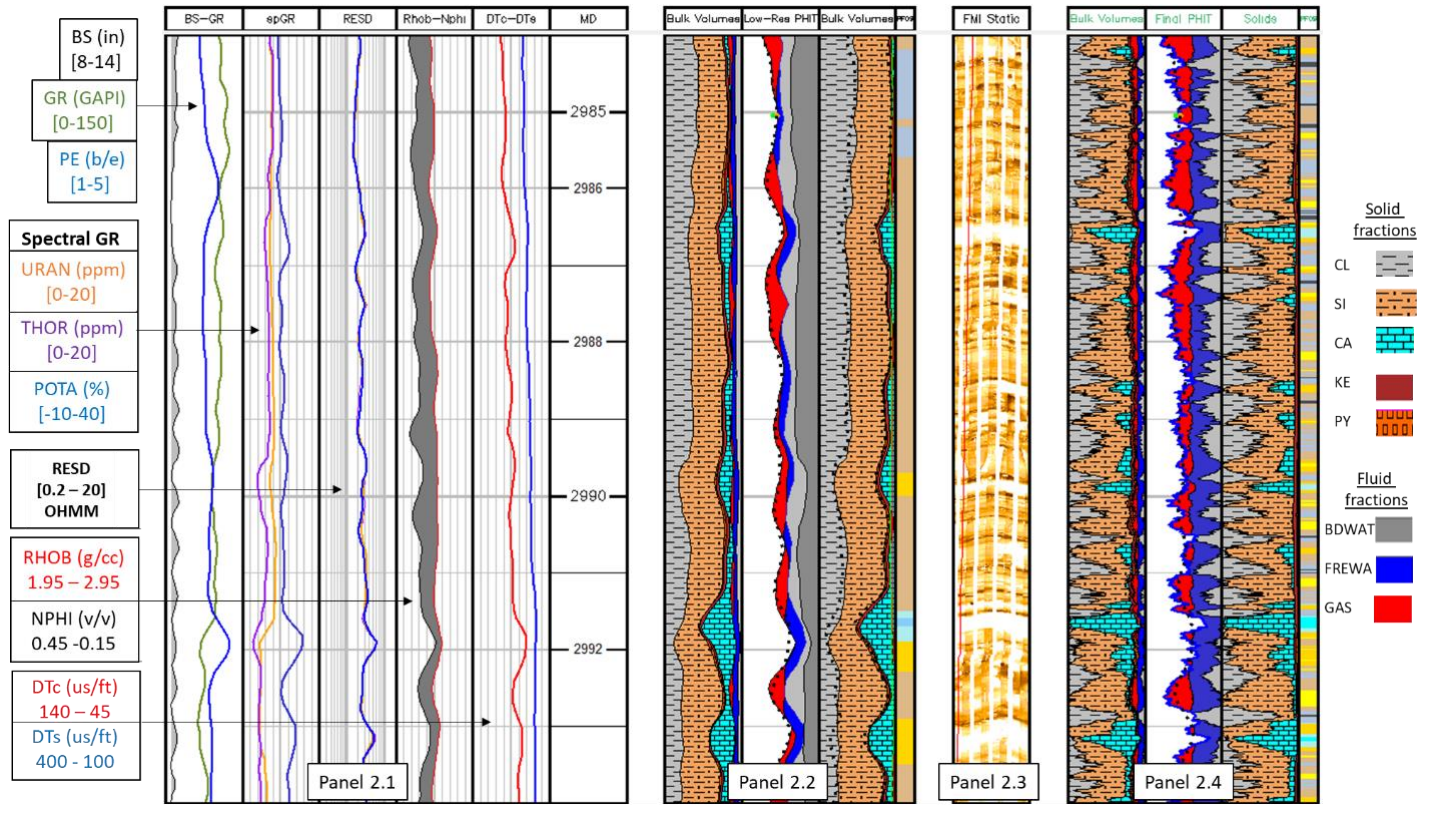}
 \caption{Different levels of resolution (from Galliot, 2020): Input logs (Panel p2.1), standard log vertical resolution petrophysical outputs, (p2.2), resistivity borehole image (p2.3), and high-resolution outputs (p2.4).}
 \label{Fig:Intro}
\end{figure}

The case of hydraulic fracture (HF), where the crack propagation in a solid body is driven by the action of a fluid, is one such area. In this instance, there is both a pressing practical need for effective homogenisation strategies, while the applications also allow for potential experimental verification. Of particular importance is its application within the energy sector, including the extraction of hydrocarbons from unconventional reservoirs, carbon sequestration, geothermal energy, among others. With these processes taking place deep underground, the depositional history and subsequent deformation means that the rock strata in which HF is employed are typically characterised by highly complex structures.  It has long been understood that heterogeneity is a key component for understanding the influence of rock mechanical behaviour on HF outcomes (see e.g.\ \cite{10.2523/IPTC-17018-MS,ARMA2021}). Attempts have been made to incorporate this directly, utilizing core samples taken along the expected fracture path, from which details of the fracture toughness can be obtained (typically determined by experiment, following e.g. the standard \ ASTM E1820/E399 \cite{ASTME1820}). In \cite{10.15530/urtec-2020-2180,GJRichards2020}, \emph{Gaillot} examined the significance of local, discrete features in determining macro-scale geological and fracture characteristics. They utilized an FEM-based approach (see e.g. \cite{Profit2016} and reference therein) to model fracture development for domain geometries obtained using the standard vertical resolution ($\sim 1$m) petrophysical measurements against those complemented by high resolution ($\sim 1$cm) borehole images (see Fig.~\ref{Fig:Intro} for an example of resolution levels). It clearly showed that using low-res parameter approximation ahead of the tip can lead to erroneous prediction of fracture evolution, including predicting fracture arrest. Given the difficulty of obtaining such high resolution data, and difficulty computing with such high resolutions, a homogenisation-based strategy may therefore be necessary or preferable in many cases. However, a more rigorous approach to defining the most appropriate - efficient and accurate - approximations of the material parameters is required.

In this paper, we propose a new, process dependent approach to homogenising the fracture toughness, based on the concept of temporal averaging. We investigate its effectiveness in the context of hydraulic fracture (HF), which involves fluid driven cracks propagating through a solid medium and is of keen interest due to its broad applicability in both industrial and natural contexts. We will also restrict the analysis to a domain with periodic toughness for simplicity. We will focus in this work on the plane-strain model, for a number of reasons. Firstly, it presents a case where the fracture propagation is (relatively) well-behaved, in that it varies significantly on the local level but is never comparable to the wave-speed of the medium, while still potentially featuring crack re-nucleation (see Fig.~\ref{Fig:Delta1}c,d). The inherently nonlinear nature of the relations governing the process allow the process-dependence to be investigated. In the context of HF, approaches to homogenising the other material parameters, such as the Young's modulus, are already well known, while developing effective homogenisation procedures for the plane strain formulation is an essential first step to incorporating the issues related to elastic heterogeneity mentioned above. Working with the plane strain formulation, with only heterogeneity of the material toughness, therefore greatly simplifies the analysis. It allows this work to concentrate solely on the effect of varying toughness on the homogenisation strategy, without needing to consider issues such as crack redirection, domain geometry, inclusions, and others mentioned previously. 


The paper is structured as follows. In Section 2 we first introduce the mathematical formulation of analysed problem and then provide short description of our motivations to undertake the work. Section 3 describes our proposal for improved homogenization technique including its formulation, efficiency and consistency.  The final conclusions are presented in Section 4.

\section{Motivations and problem formulation}

\subsection{The KGD model with inhomogeneous toughness}

\begin{figure}[h]
 \centering
 \includegraphics[trim = 2.5cm 12cm 2.5cm 2.5cm, clip,width=0.45\textwidth]{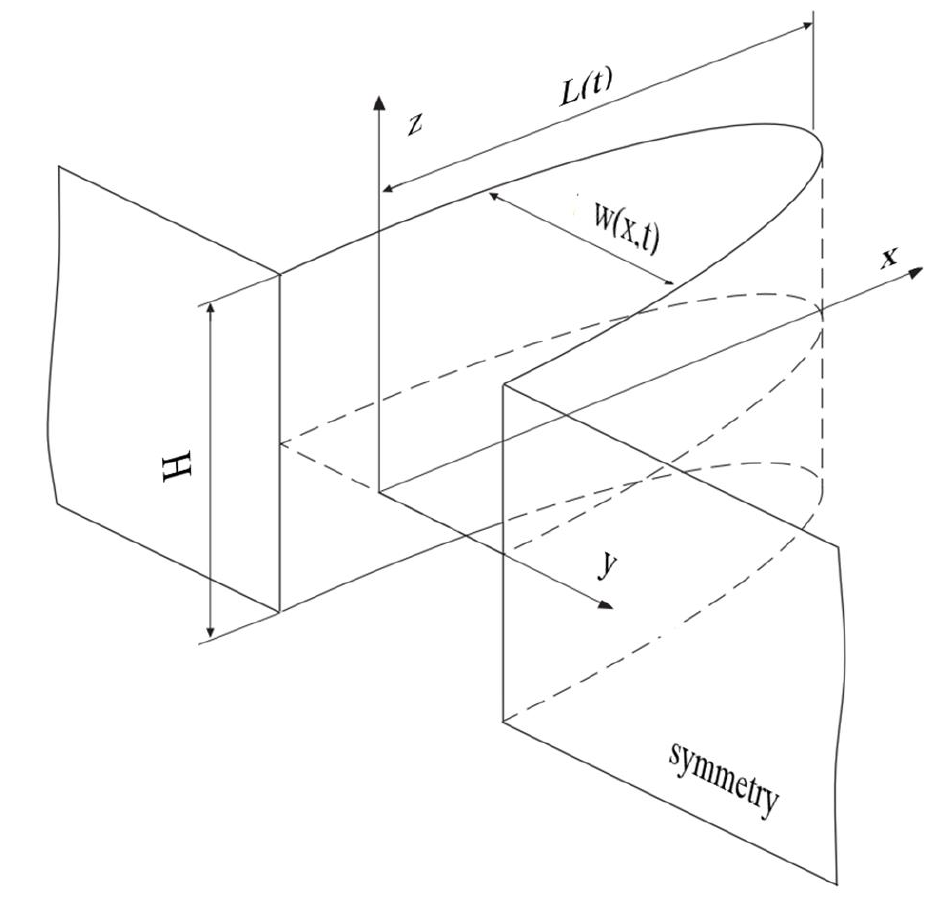} 
 \caption{The KGD fracture geometry.}
 \label{Fig:KGD}
\end{figure}

We consider a hydraulic fracture (HF) approximated using the KGD (plane strain) formulation first developed by \emph{Kristianovich}, \emph{Geertsma \& De Klerk} \cite{Geertsma1969}, modified to account for the inhomogeneity of the material toughness. The crack is of fixed height $H$ and length $2L(t)$, with the fractured domain given in terms of spatial coordinate $x\in [-L , L]$ (see Fig.~\ref{Fig:KGD}). Newtonian fluid is injected at the center $x=0$, with a known rate $q_0 (t)$. The fracture propagates as a result of the normal pressure induced by the fluid $p(t,x)$, with the resulting width of the crack along the length being expressed as $w(t,x)$. Due to the symmetry of the problem, we can consider only the positive domain $x\in [0,L(t) ]$. %
As we are primarily concerned with the initial fracture development, we consider here only the storage dominated regime. As such, we assume that the rock is impermeable, i.e.\ the fluid leak-off into the surrounding domain is negligible.

The governing equations for this system follow mostly the standard form (see e.g.\ \cite{Wrobel2015}), with some modification to account for the variation in the material toughness $K_{Ic}$.

The continuity equation, which follows from the conservation of mass (assuming no fluid leak-off), is given by:
\begin{equation}\label{eq:1}
\frac{\partial w}{\partial t} + \frac{\partial q}{\partial x} = 0, \quad t>0 , \quad 0<x<l(t),
\end{equation}
while the global fluid balance equation takes the form:
\begin{equation} \label{eq:1b}
\int_0^{L(t)} w(t,x) - w(0,x) \, dx =  \int_0^t \frac{ q_0 (\tau )}{2H} \, d\tau .
\end{equation}
The fluid flow inside the fracture is described by the Pouseuille equation (laminar flow), which for a Newtonian fluid takes the form:
\begin{equation}\label{eq:2}
q = -\frac{1}{M} w^3 \frac{\partial p}{\partial x} , \quad t>0 , \quad 0<x<l(t),
\end{equation}
where $q$ is the fluid flow rate, while the constant $M=12\mu$ where $\mu$ is the fluid viscosity. Note that this equation degenerates at the crack tip, as the crack opening $w$ tends to zero while the derivative of the fluid pressure $p$ tends to negative infinity. To avoid handling the degeneration of this equation, we instead work with the fluid velocity $v$:
\begin{equation}\label{eq:3}
v = \frac{q}{w} = -\frac{1}{M}  w^2 \frac{\partial p}{\partial x} , \quad t>0 , \quad 0<x<l(t).
\end{equation}
While the right-hand side of this equation also degenerates, in the case without fluid lag the fluid front and fracture tip coincide, yielding the speed equation:
\begin{equation}\label{eq:4}
\frac{dL}{dt} = v\left(t,L(t) \right) , \quad t>0,
\end{equation}
allowing the value of $v$ to be determined at the crack tip without the use of \eqref{eq:3}. The advantages of this formulation, and corresponding method of solution for HF problems, have been outlined in e.g.\ \cite{Perkowska2015,Wrobel2015}.

Considering the crack evolution within the framework of Linear Elastic Fracture Mechanics, the fracture extension is determined using the standard Irwin criterion:
\begin{equation}
\label{toughness_cond}
K_I (t) = K_{Ic} \left( L(t) \right),
\end{equation}
where $K_{Ic}$ is the material toughness at the crack tip, while $K_I$ is the Mode-I stress intensity factor, given as:
\begin{equation}
K_I (t) = 2\sqrt{\frac{L(t)}{\pi}} \int_0^{L(t)} \frac{p(t,s)}{\sqrt{L^2 (t) - s^2}} \, ds.
\end{equation}
Note that the material toughness $K_{Ic}$ varies throughout the domain, with the particular distributions considered in this paper outline in the next subsection.\footnote{\label{Foot1}{Since we consider in this paper the case when all elastic properties remain constant, the condition \eqref{toughness_cond} is equivalent to the energy condition ${\cal E}={\cal E}_c$ - see for example discussion in \cite{DONTSOV2021108144,HOSSAIN201415}.}}

One crucial difference between the standard KGD formulation and that utilized in this work is the elasticity equation, relating the fracture width $w$ and the normal pressure induced by the fluid on the crack walls $p$, which takes the form \cite{Wrobel2015}:
\begin{equation}
\label{elasticity}
w(t,x)=\underbrace{\frac{4L(t)}{\pi E'}\int_0^{L(t)} K\left(\frac{s}{L(t)},\frac{x}{L(t)}\right)\frac{\partial p}{\partial s}(t,s)\,ds}_{w_1 (t,x)}+\underbrace{\frac{4}{E'}K_{IC}(L(t))\sqrt{\frac{L^2(t)-x^2}{\pi L(t)}}}_{w_2 (t,x)},
\end{equation}
where:
$$
K(\eta , \xi)= - (\eta - \xi ) \ln \left| \frac{\sqrt{1-\xi^2} + \sqrt{1-\eta^2}}{\sqrt{1-\xi^2}-\sqrt{1-\eta^2}}\right| - \xi \ln \left( \frac{1+\eta \xi + \sqrt{1-\xi^2}\sqrt{1-\eta^2}}{1+\eta\xi - \sqrt{1-\xi^2}\sqrt{1-\eta^2}}\right).
$$
Here, the only modification to the elasticity equation from the standard form is to account for $K_{Ic}(L)$ being the toughness of the material at the crack tip, and thus dependent on the position of the tip (or equivalently here on the fracture length). Note that the first term, which we label $w_1 (t,x)$, describes the effect of the (viscous) fluid pressure on the fracture walls, while the second term $w_2 (t,x)$ describes an impact of the material toughness.

\subsection{Computational algorithm}

The results provided in this paper are produced using a previously-developed code, based on the ``universal algorithm'' approach to the KGD model outlined in \cite{Wrobel2015}. This includes utilizing the fluid velocity and associated Stefan-type condition \eqref{eq:3}-\eqref{eq:4}, alongside employing the asymptotics at all stages of the algorithm, to properly treat the singular points of the domain and trace the fracture front. The inverse elasticity operator \eqref{elasticity} (the BEM formulation) is utilized to relate the solid and fluid phases in the form of a compact operator acting on the pressure gradient, while also allowing direct implementation of the variable toughness. The algorithm is adaptive in both the spatial and temporal dimensions, allowing for a high level of solution accuracy over the whole domain. The specifics of the algorithm are outlined in detail in \cite{Gaspare2020}.

Throughout this paper, all parameters except for the toughness are taken as constant values for simplicity (see footnote~\ref{Foot1}, page \pageref{Foot1}). Thus, the Poisson's ratio and Young's modulus are the same in all rock layers and, within the fracture, the fluid is attributed to a constant effective viscosity $\mu$. To ensure that only the impact of variable toughness is being considered, we take the fluid injection rate $q_0$ to be constant as well. The problem parameters (excluding toughness) taken in all simulations in this paper are provided in Table~\ref{Table:parameters}.

\begin{table}[h]
 \centering
 \begin{tabular}{c|c|c|c|c}
$E$ & $\nu$ & $\mu$ & $H$ & $q_0 $ \\
\hline \hline
&&&&\\
$2.81 \times 10^{10}$ \, [Pa] & $0.25$ & $1 \times 10^{-3}$ \, [Pa s] & $15$ \, [m] & $6.62 \times 10^{-2}$ \, [m$^3$ / s] \\
&&&&\\
\hline \hline
 \end{tabular}
\caption{Problem parameters used in simulations, with $H$ denoting the fracture height. Note that the pumping rate $q_0 $ is taken to be constant.}
\label{Table:parameters}
\end{table}

\begin{figure}[h!]
\centering
 \includegraphics[width=0.45\textwidth]{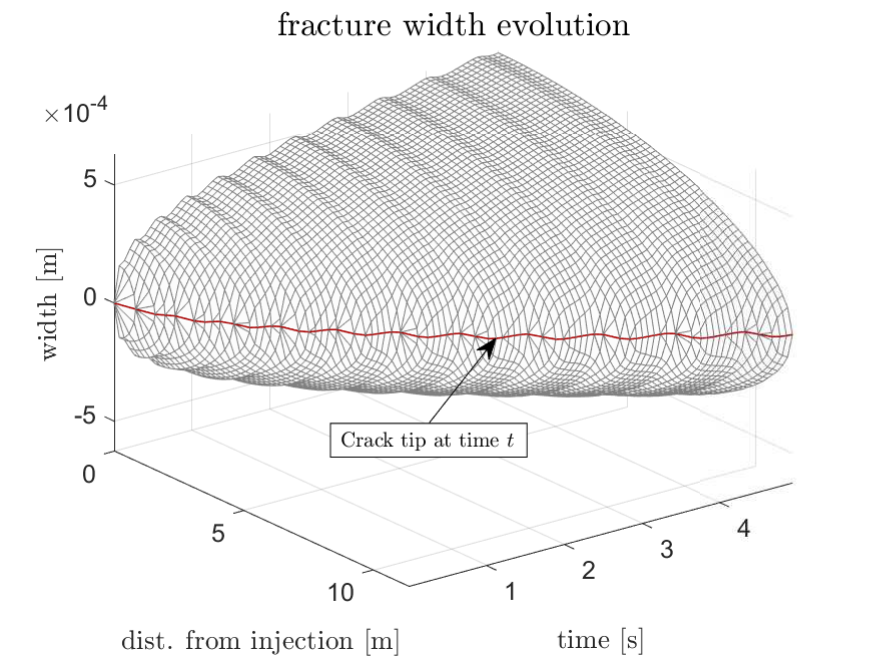}
 \put(-215,155) {{\bf (a)}}
 \hspace{12mm}
 \includegraphics[width=0.45\textwidth]{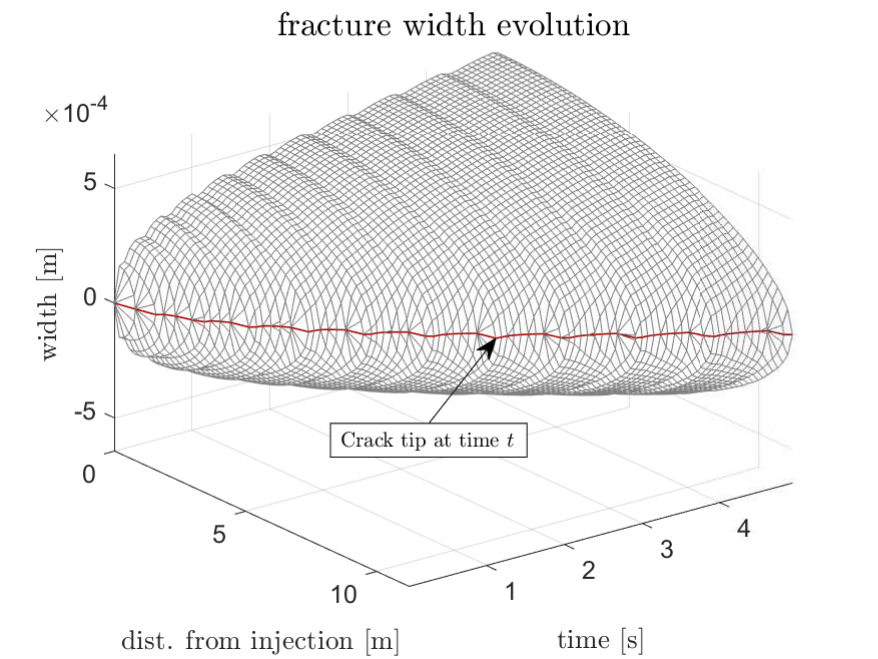}
  \put(-215,155) {{\bf (b)}}
\\%
        \includegraphics[width=0.45\textwidth]{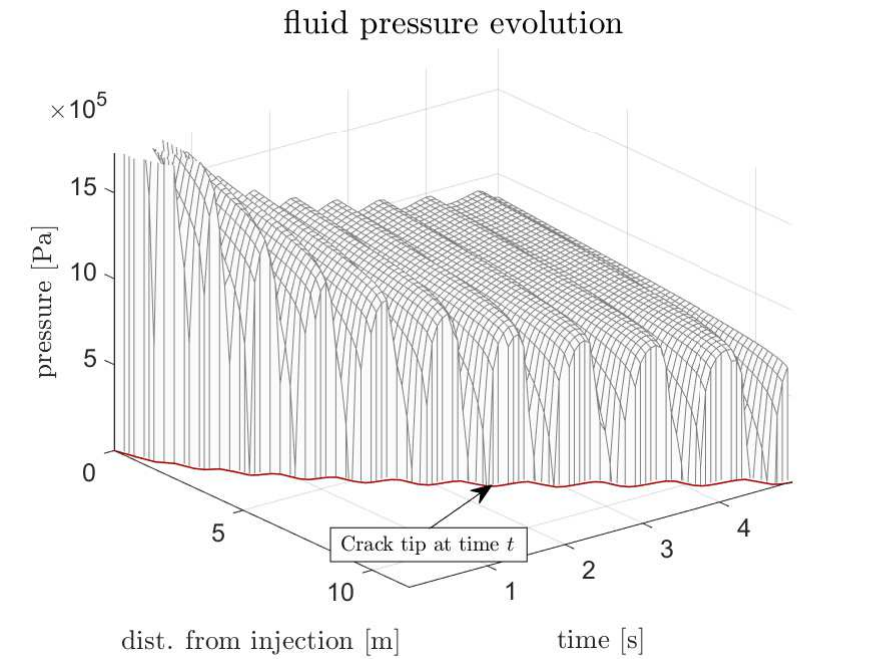}
 \put(-215,155) {{\bf (c)}}
 \hspace{12mm}
        \includegraphics[width=0.45\textwidth]{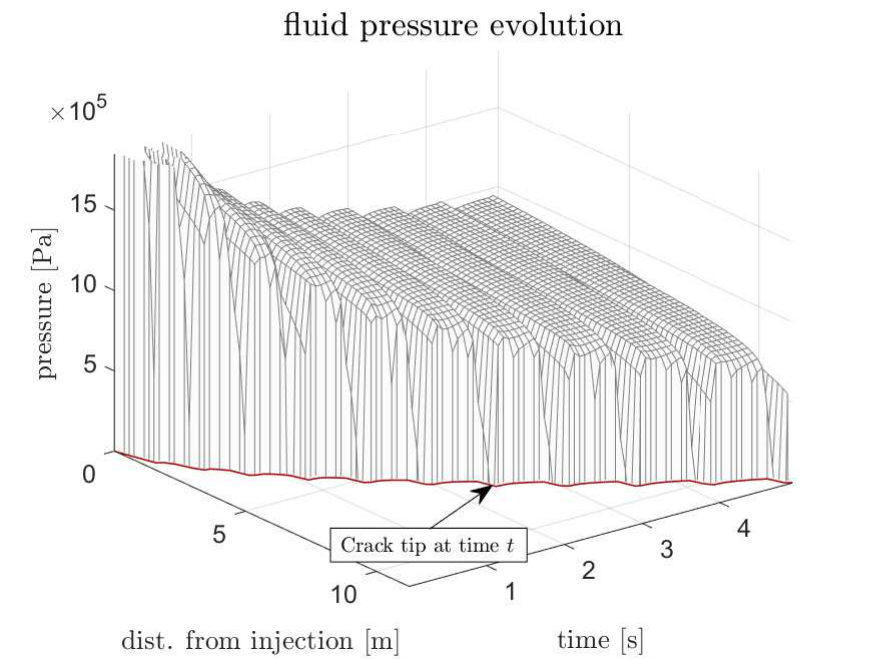}
  \put(-215,155) {{\bf (d)}}
 \caption{The {\bf (a)}, {\bf (b)} aperture $w$ and {\bf (c)}, {\bf (d)} fluid pressure $p$ in the first five seconds for a sinusoidal ((a) and (c)) and a step-wise periodic toughness with $\delta_{max}=10$ and $\delta_{min} = 1$ (see Sect.~\ref{ToughnessDistExp}-\ref{DeltaSect}). }
 \label{Fig:ExampleSols1}
 \end{figure}

Example solutions for the aperture and fluid pressure are provided in Fig.~\ref{Fig:ExampleSols1} for the case of a fracture propagating through distinct layers of rock with periodically distributed toughness distribution (defined such that $\delta_{max}= 10$ and $\delta_{min} = 1$ in alternating layers, see Sect.~\ref{DeltaSect}). The differing propagation behaviour between the two layers can clearly be seen, and will be discussed in detail in Sect.~\ref{Motivational}. Meanwhile, the high gradient of the net pressure on the crack front is a consequence of its asymptotic (logarithmic) behaviour (note that in Fig.~\ref{Fig:ExampleSols1}c-d we cut the negative value in the immediate neighbourhood of the crack tip). 

\subsection{Form of the material toughness}\label{ToughnessDistExp}

For simplicity we consider a rock structure with a toughness distribution that is periodic in space. The period is taken as $X$, while it is assumed that $0<K_{min}<K_{max}<\infty$ where
$$
K_{min} = \min_{0\leq x \leq X} K_{Ic} (x) , \quad K_{max} = \max_{0\leq x \leq X} K_{Ic} (x) .
$$

\begin{figure}[h]
\centering
 \includegraphics[width=0.40\textwidth]{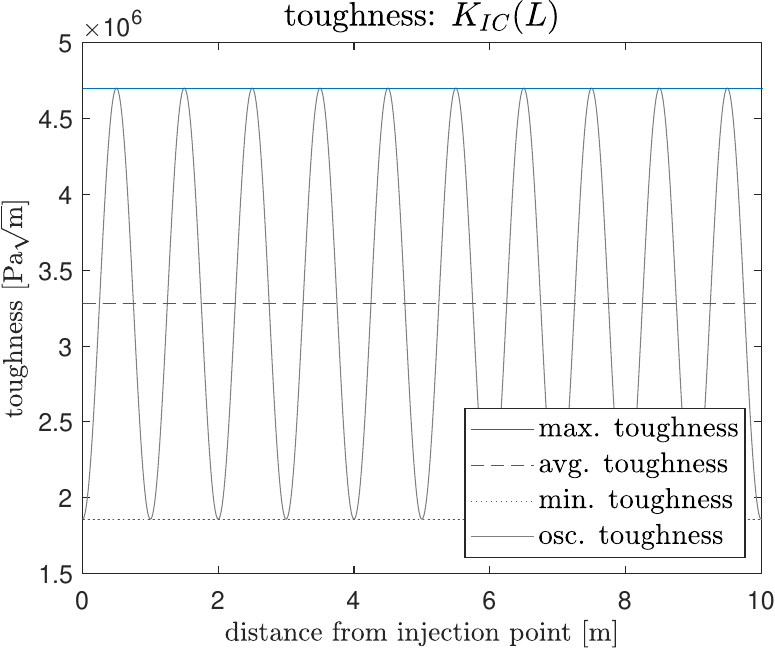}
 \put(-205,155) {{\bf (a)}}
 \put(-115,170) {{\bf Sinusoidal}}
 \hspace{12mm}
 \includegraphics[width=0.40\textwidth]{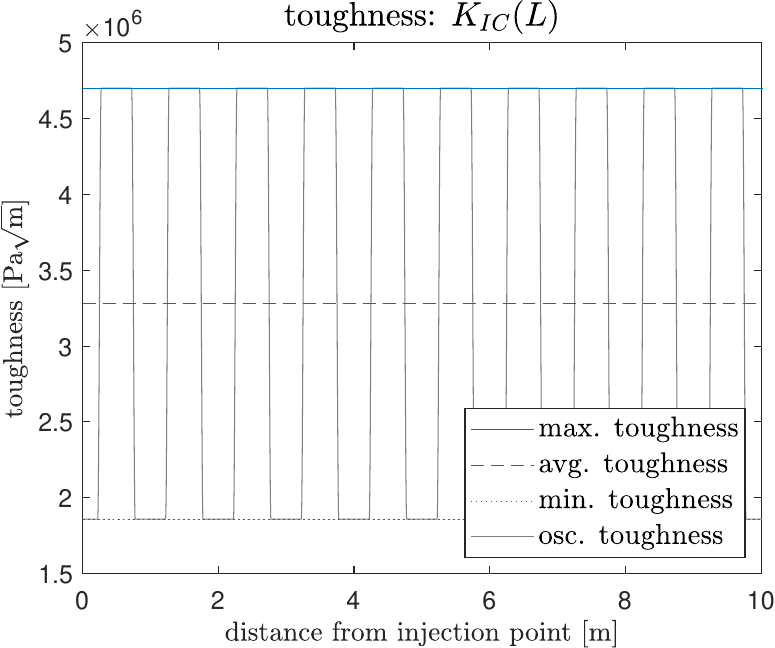}
  \put(-205,155) {{\bf (b)}}
   \put(-115,170) {{\bf Step-wise}}
\\%
 \includegraphics[width=0.40\textwidth]{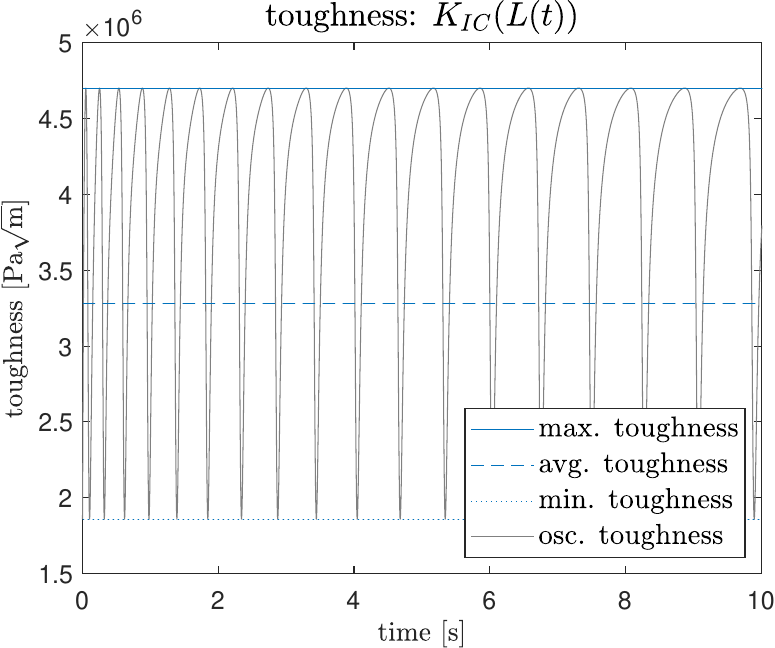}
 \put(-205,155) {{\bf (c)}}
 \hspace{12mm}
 \includegraphics[width=0.40\textwidth]{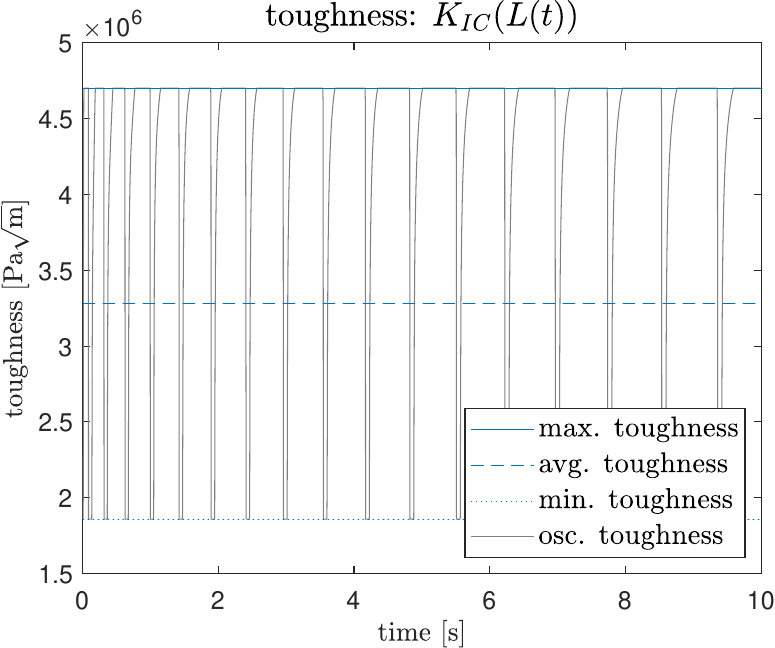}
  \put(-205,155) {{\bf (d)}}
 \caption{The toughness distribution  {\bf (a)}, {\bf (b)} over the spatial length $x$ of the material, {\bf (c)}, {\bf (d)} experienced at the crack tip $L$ over time $t$ (for a material with $\delta_{max}=10$, $\delta_{min}=1$, see Sect.~\ref{DeltaSect}). The two distributions considered are: {\bf (a)}, {\bf (c)} sinusoidal and {\bf (b)}, {\bf (d)} step-wise.}
 \label{Fig:ToughDis1}
 \end{figure}

Two separate distributions are considered, both of which are considered symmetrical about the injection point, allowing only the region $x\geq 0$ to be considered. The first represents a transition between two distinct rock layers, such that the spatial toughness distribution is a step-wise function alternating between $K_{max}$ and $K_{min}$. The second distribution is a sinusoidal oscillation between these values. Examples of these two distributions for some arbitrary $K_{max}$ and $K_{min}$ are represented in Fig.~\ref{Fig:ToughDis1}. Throughout this paper, figures placed on the left will correspond to the sinusoidal distribution, while those on the right will provide results for the step-wise distribution.

We can consider the average of the toughness $K_{Ic}(x)$, defined as:
\begin{equation} \label{K_AvSpace}
 K_{aver} = \langle K_{IC}\rangle_X = \frac{1}{X}\int_0^{X}K_{IC}(x)dx.
\end{equation}
An alternative measure which places a greater focus on the local behaviour of the toughness is the roaming average, defined over some length $Y$, with $0<Y\leq X$, as:
\begin{equation} \label{K_MovSpace}
K_{mov}(x,Y) = \frac{1}{Y} \int_x^{x+Y} K_{Ic} (\xi ) \, d\xi .
\end{equation}
Note that this function $K_{mov}$ is also periodic, but will typically be smoother than the original function $K_{Ic}$. The choice of the spatial frame plays a crucial role on the final results, 
as is evident from the limiting behaviour:
$$
\lim_{Y\to 0+} K_{mov}(x,Y) = K_{Ic}(x) , \quad K_{mov}(x,nX)= K_{mov}(x,X) = K_{aver},\quad n=2,3,\ldots .
$$

\subsection{Motivation}\label{Motivational}

To consider the handling of the the fracture toughness heterogeneity, the difficulties associated with this and the rational behind proposed techniques, we must begin by examining the effect that a variable toughness distribution has upon the fracture behaviour. As the effect will be regime dependent, we start by introducing an approach to approximately determine which regime a fracture will experience when propagating through each rock layer. Then, we can conduct an examination of existing strategies and their relative effectiveness in differing propagation regimes.

\subsubsection{Parameterising the fracture regime}\label{DeltaSect}

We can consider the fluid volume stored in the fracture:
\begin{equation}
V(t) = 2H \int_0^{L(t)} w(t,x) \, dx .
\end{equation}
Noting the form of the elasticity equation given in \eqref{elasticity}, where the term $w_1 (t,x)$ largely describes the effect of the (viscous) fluid pressure on the fracture walls, while the term $w_2 (t,x)$ describes the toughness effects, we can decompose the volume into the sum of those resulting from the viscosity and toughness dominated terms:
$$
V(t) = V_v (t) + V_T (t),
$$
with
$$
V_v (t) = 2H \int_0^{L(t)} w_1 (t,x) \, dx , \quad V_T (t) = 2H \int_0^{L(t)} w_2 (t,x) \, dx  .
$$
As a result, we can define the ratio between these two:
\begin{equation} \label{delta1}
\delta (t) = \frac{V_T (t)}{V_v (t)} .
\end{equation}
which will clearly provide an approximation of the extent to which the fracture evolution is governed by viscosity dominated or toughness dominated effects, providing a basic measure for the regime a particular fracture is in at a certain moment. This can be demonstrated by considering the self-similar solution, where the material is homogeneous and pumping is constant in time. It is easy to see that in this case $\delta (t)$ is constant in time as well. Furthermore, it can be verified that if two solutions, having different parameters, have the same value of $\delta$ then they differ only by a rescaling. In fact, $\delta$ is directly related to the known natural scaling ${\cal M}$ and ${\cal K}$, for the viscosity and toughness dominated regimes of a storage dominated fracture respectively, given in \cite{garagash_large_toughenss} (for more information on these scalings, see also e.g.\  \cite{Garagash2000,GaragashSummary,Peirce2008}) and we have:
\begin{eqnarray}
\text{Viscosity dominated:}\quad & \delta \sim 0.9195 \left(\frac{K_{Ic}^4 (1-\nu^2)^3 H}{q_0 \mu E^3}\right)^{\frac{1}{4}},\quad \delta \ll 1,  
\label{eq:st}\\[2mm]
\text{Toughness dominated:}\quad  & \delta \sim 1.9744  \frac{K_{Ic}^4 (1-\nu^2)^3 H}{q_0 \mu E^3},\quad \delta\gg 1.
\label{eq:em}
\end{eqnarray}
Consequently, the general form of the self-similar solution may be considered as only depending on $\delta$, that represents the regime under which the fracture is propagating. As a result, it is natural to choose in our study values of $K_{IC}$ such that $\delta$ corresponds to specific regimes within the ${\cal K} - {\cal M}$ parametric space, which are provided in Table.~\ref{Table:delta}. 

\begin{figure}[t!]
 \centering
 \includegraphics[width=0.45\textwidth]{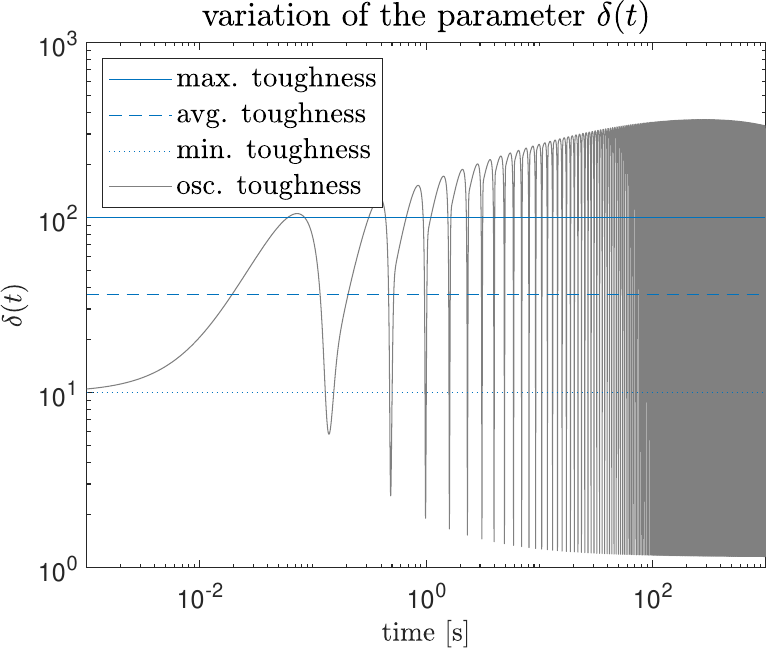}
	 \put(-220,175) {{\bf (a)}}
 \hspace{12mm}
 \includegraphics[width=0.45\textwidth]{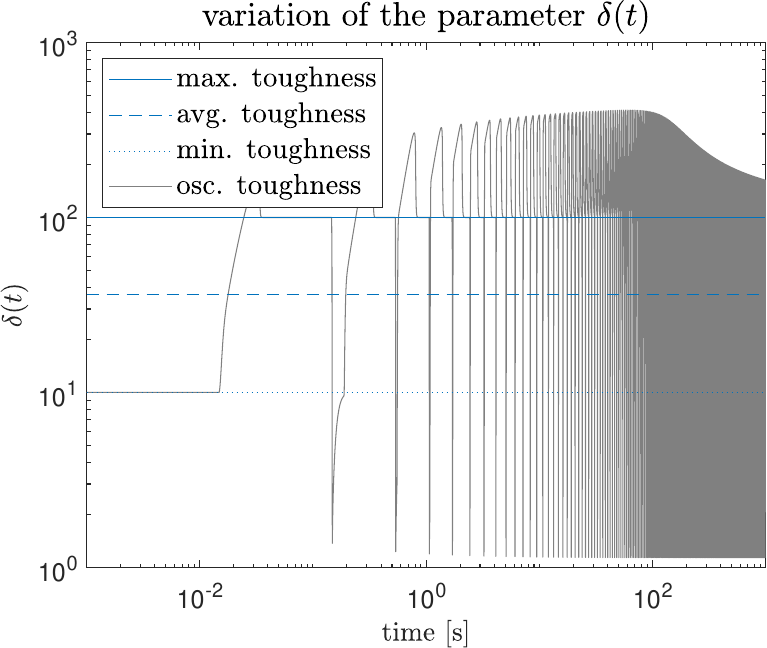}
	 \put(-220,175) {{\bf (b)}}
\\
 \includegraphics[width=0.45\textwidth]{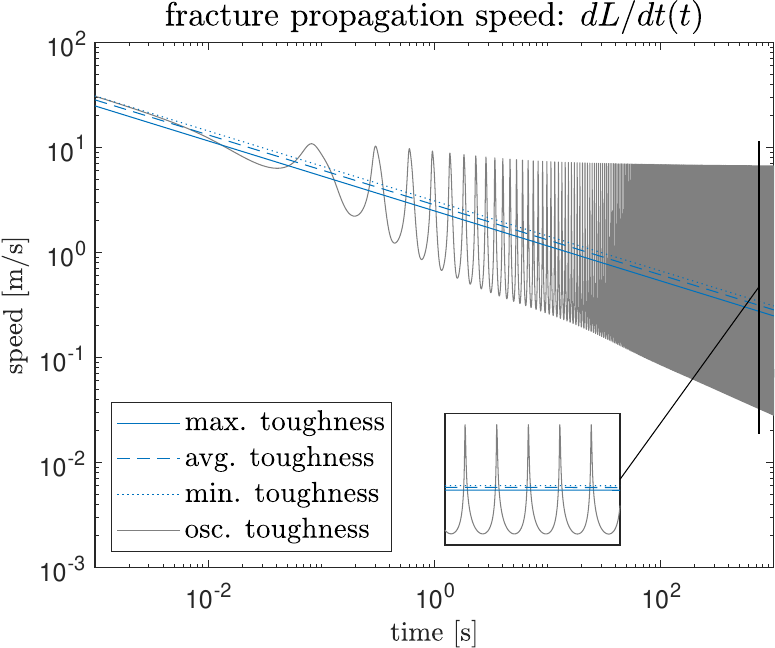}
	 \put(-220,175) {{\bf (c)}}
 \hspace{12mm}
 \includegraphics[width=0.45\textwidth]{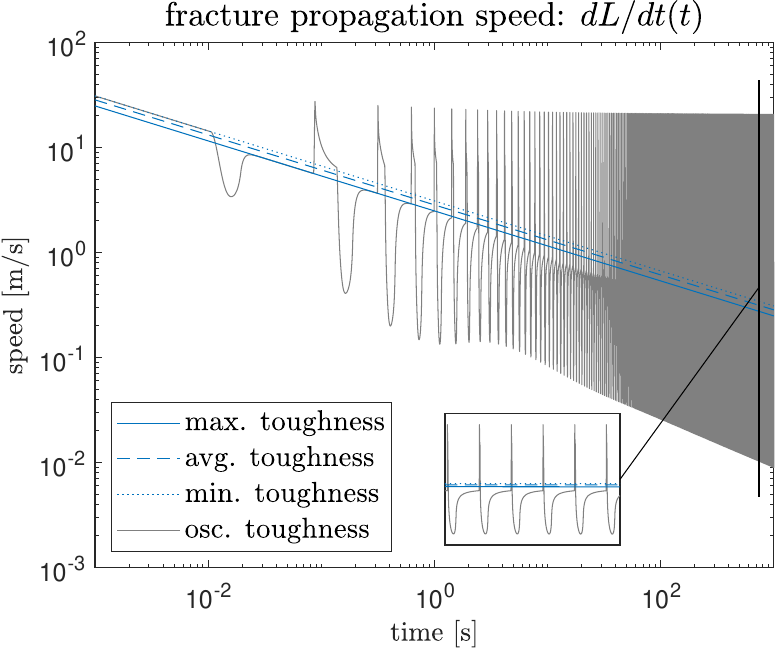}
	 \put(-220,175) {{\bf (d)}}
\caption{For the oscillatory toughness such that $\delta_{max} = 100$ and $\delta_{min} = 10$, we show: {\bf (a)}, {\bf (b)} The value $\delta (t)$ \eqref{delta1}, the ratio between the total fracture fluid volume resulting from the viscosity and toughness dominated terms of the elasticity equation, over time. {\bf (c)}, {\bf (d)} the fracture velocity over time. The toughness distributions in each case are: {\bf (a)}, {\bf (c)} sinusoidal toughness, {\bf (b)}, {\bf (d)} step-wise toughness.}
\label{Fig:Delta1}
\end{figure}

Recall that, in the case with oscillating toughness $K_{Ic}(L)$, the toughness distribution is defined by the maximum and minimum values of the toughness over the period ($K_{max}$ and $K_{min}$ respectively, see Sect.~\ref{ToughnessDistExp}). From the above, it is clear that the self-similar solutions for constant toughness $K_{max}$ and $K_{min}$ will have corresponding values of $\delta$, which we denote $\delta_{max}$ and $\delta_{min}$ respectively\footnote{Note that here $\delta_{max}$ and $\delta_{min}$ correspond to the values of $K_{max}$ and $K_{min}$, not the maximum and minimum of the $\delta(t)$. In general, $\delta_{max} \neq \max \delta (t)$, and $\delta_{min} \neq \min \delta(t)$, as can be seen in Fig.~\ref{Fig:Delta1}a,b.}. It can be seen that a similar property holds to the self-similar case, whereby if the problem parameters are changed but the values of $\delta_{max}$, $\delta_{min}$ and the shape of $K_{Ic}$ remains the same (with possibly a different period length), then the solution of the new problem will differ from the old one only by a rescaling in space and time. 
Therefore, if we define the problem in terms of the parameters $\delta_{max}$ and $\delta_{min}$, and the shape of $K_{Ic}(L)$ inside the period, we can perform a fully general investigation irrespective of the values of the remaining parameters (injection rate, toughness period length and elastic/fluid constants). The specific values of $\delta_{max}$ and $\delta_{min}$ are taken to investigate the differing fracture regimes (viscosity/toughness dominated).

It should be noted that, while one might expect that the value of $\delta (t)$ in a material with variable toughness would simply vary between the values associated with the maximum and minimum toughness, this is not the case as shown in Figs.~\ref{Fig:Delta1}a,b. Here, the values of $K_{max}$ and $K_{min}$ are taken such that, in the case of a homogeneous material, that with toughness $K_{max}$ would have constant $\delta (t) = \delta_{max} = 100$, while that for $K_{min}$ would have $\delta(t) = \delta_{min} = 10$. The final simulation results, for a crack traveling through a medium whose toughness varies periodically between $K_{max}$ and $K_{min}$, has a $\delta(t)$ that is clearly not bounded by either of these values.

\begin{table}[h!]
 \centering
 \begin{tabular}{|l||c|c|c|c|}
 \hline \hline
${\delta} $ & $0.1$ & $1$ & $10$ & $100$ \\
\hline
$K_{Ic}$ [Pa $\sqrt{\text{m}}$] & $3.26 \times 10^{5}$ & $1.86 \times 10^{6}$ & $4.70 \times 10^{6}$ & $8.76\times 10^{6}$  \\
\hline \hline
 \end{tabular}
\caption{Values of the material toughness $K_{Ic}$ (to $3$ s.f.) corresponding to the given value of the ratio $\delta$ \eqref{delta1}, for a rock whose other material parameters are as stated in Table.~\ref{Table:parameters}.}
\label{Table:delta}
\end{table}

The physical explanation follows from considering the perspective of the crack tip, as partially indicated by the propagation velocity in Figs.~\ref{Fig:Delta1}c,d. Consider the case of a fracture traveling between two distinct materials of differing toughness. If the fracture is traveling from the material with higher toughness to that with lower toughness, then when the crack tip arrives at the boundary the fracture is overpressurised for extending the crack in the new material. As such, the fracture moves rapidly, with the process governed almost solely by the fluid behaviour not the toughness, to an even greater extent than in a homogenous material. Conversely, a fracture moving from a low toughness layer to one with high toughness will encounter a material for which the fluid pressure is far lower than that required to fracture the rock. Thus, the fracture growth slows significantly, with the toughness playing an exaggerated role in the crack evolution\footnote{This physical explanation for the behaviour when transitioning between rock layers of differing toughness has previously been noted by others, (see presentation \href{https://congress.cimne.com/eccm_ecfd2018/admin/files/fileabstract/A2222.pdf}{Benedetti \& Lecampion at the ECCM-ECFD 2018 conference}).}. Note that this also holds true even when the material toughness changes continuously, as in the case of a sinusoidal distribution provided in Fig.~\ref{Fig:Delta1}a.

This is indicative of the problem with handling the toughness, which does not hold for other material constants. Having multiple layers of differing toughness alters the physical process. As such, an effective strategy must account for this change.

\subsubsection{Strategies for handling the material toughness}

\begin{figure}[t!]
\centering
 \includegraphics[width=0.4\textwidth]{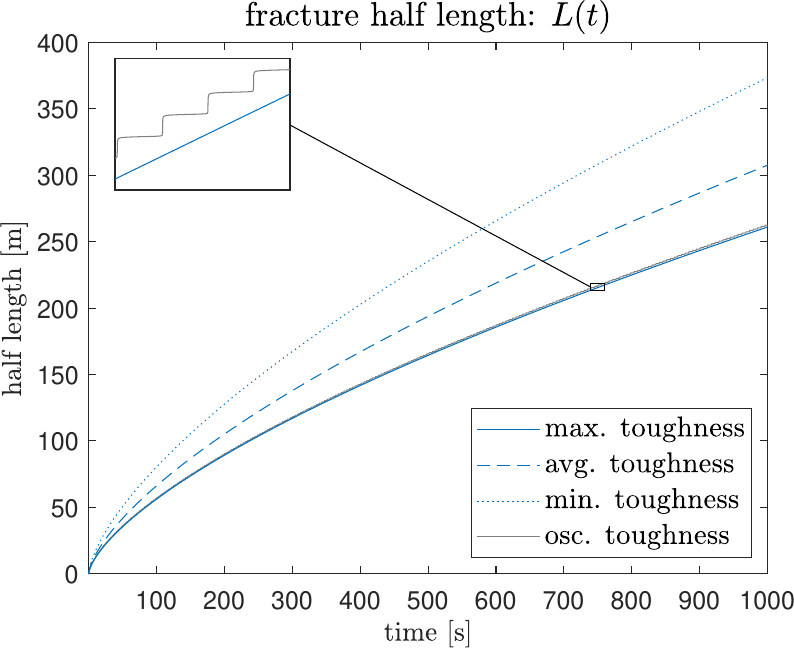}
	 \put(-205,155) {{\bf (a)}}
 \hspace{12mm}
 \includegraphics[width=0.4\textwidth]{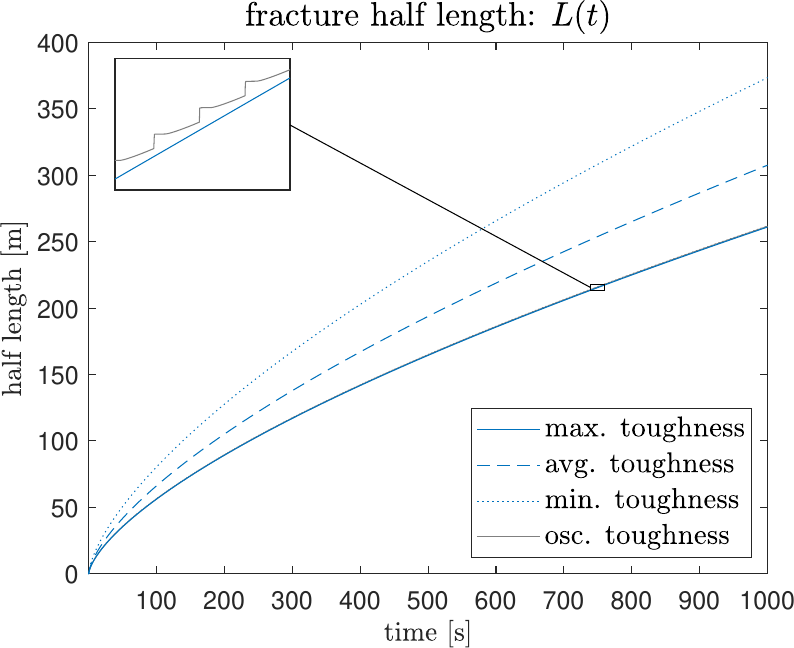}
	 \put(-205,155) {{\bf (b)}}
\\
 \includegraphics[width=0.4\textwidth]{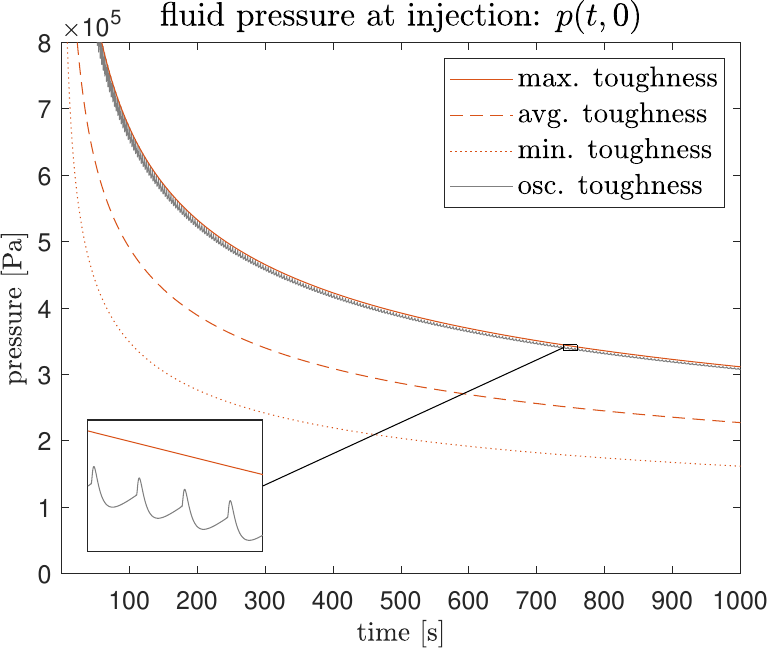}
	 \put(-205,155) {{\bf (c)}}
 \hspace{12mm}
 \includegraphics[width=0.4\textwidth]{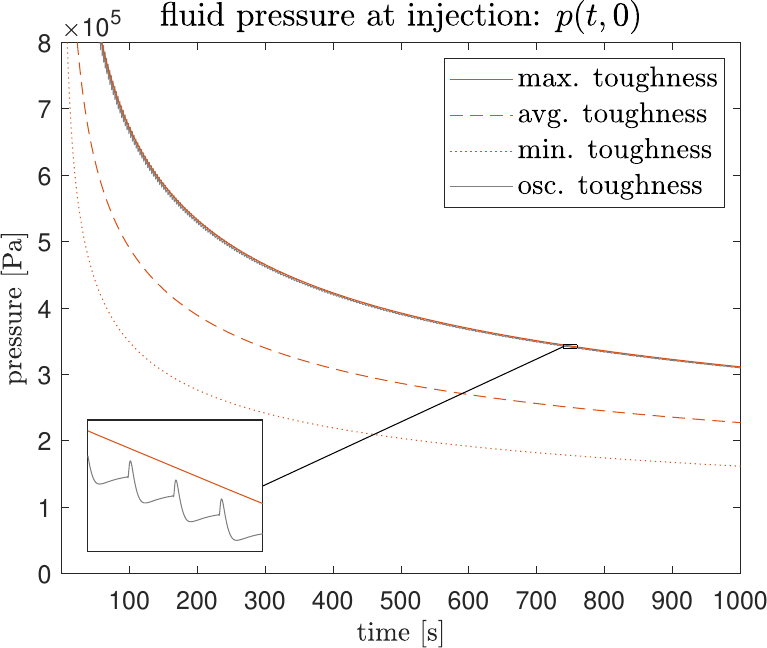}
	 \put(-205,155) {{\bf (d)}}
\\
 \includegraphics[width=0.4\textwidth]{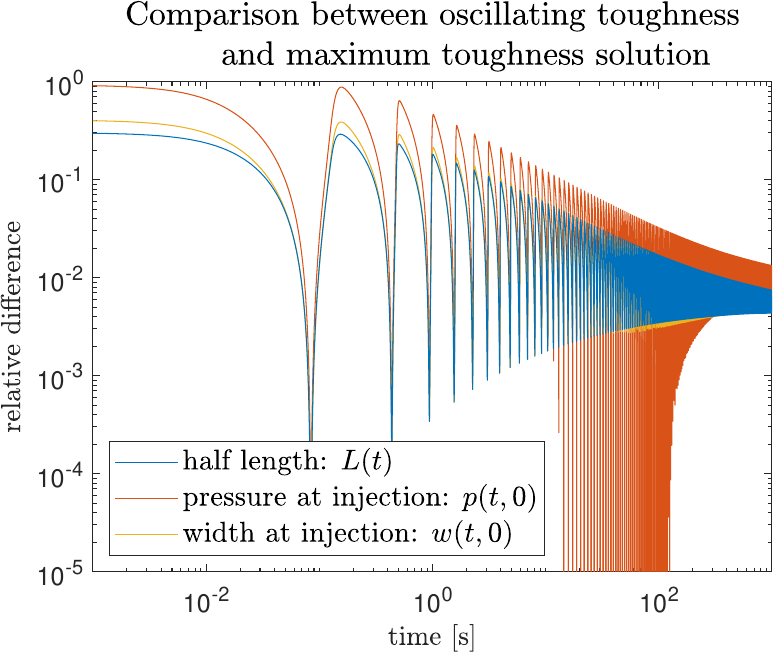}
	 \put(-205,155) {{\bf (e)}}
 \hspace{12mm}
 \includegraphics[width=0.4\textwidth]{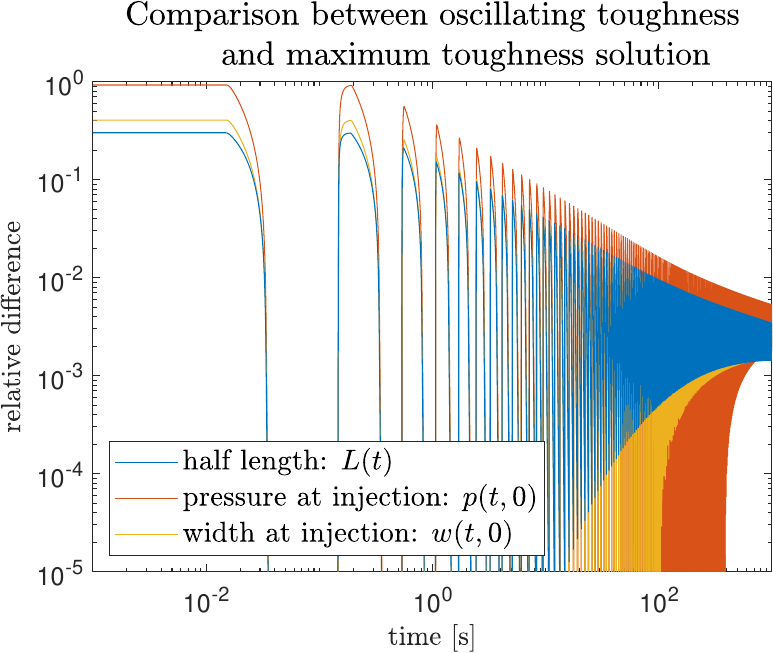}
	 \put(-205,155) {{\bf (f)}}
\caption{Various process parameters for the case of oscillating toughness when $\delta_{max} = 100$ and $\delta_{min}=10$, alongside those for a homogeneous material with the minimum, average and maximum toughness. Here we show {\bf (a)}, {\bf (b)} the fracture length, {\bf (c)}, {\bf (d)} the pressure at the injection point $x=0$, and {\bf (e)}, {\bf (f)} the relative difference between the parameters in the oscillating toughness and maximum toughness case. Here {\bf (a)}, {\bf (c)}, {\bf (e)} show the sinusoidal toughness and {\bf (b)}, {\bf (d)}, {\bf (f)} the step-wise toughness.}
	\label{fig:1}
\end{figure}

\begin{figure}[t!]
\centering
 \includegraphics[width=0.4\textwidth]{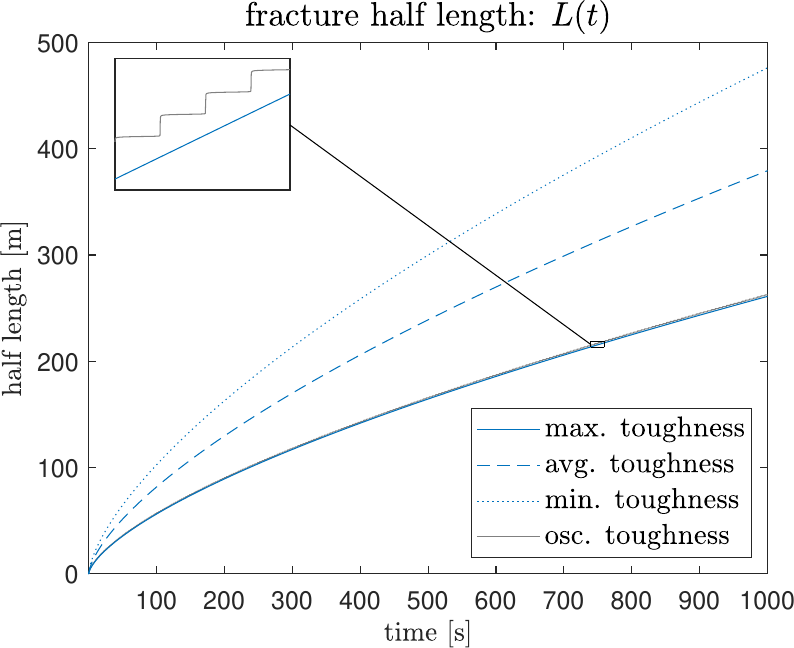}
	 \put(-205,155) {{\bf (a)}}
 \hspace{12mm}
 \includegraphics[width=0.4\textwidth]{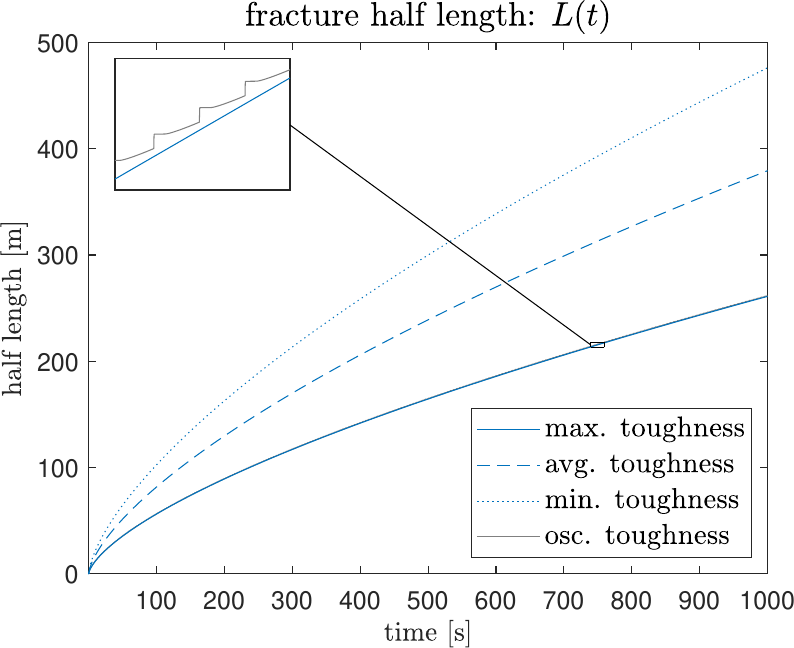}
	 \put(-205,155) {{\bf (b)}}
\\
 \includegraphics[width=0.4\textwidth]{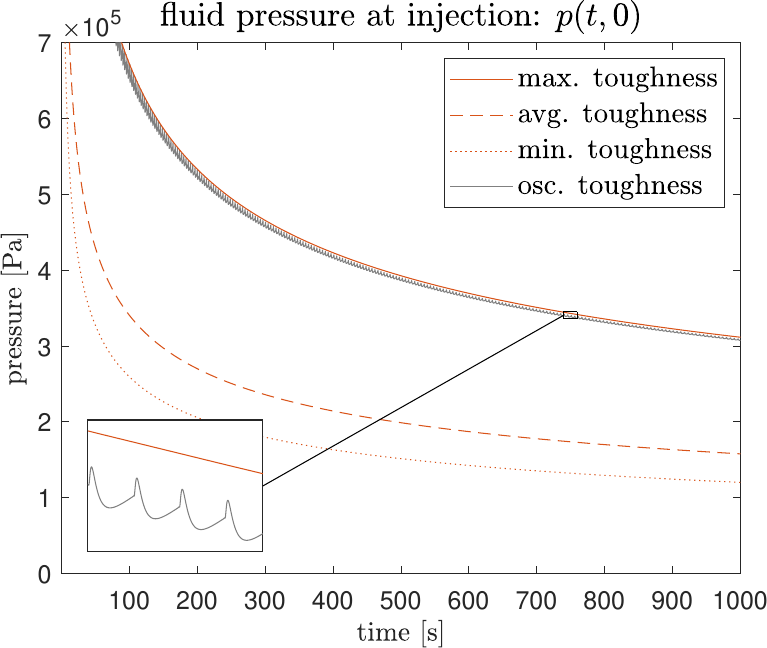}
	 \put(-205,155) {{\bf (c)}}
 \hspace{12mm}
 \includegraphics[width=0.4\textwidth]{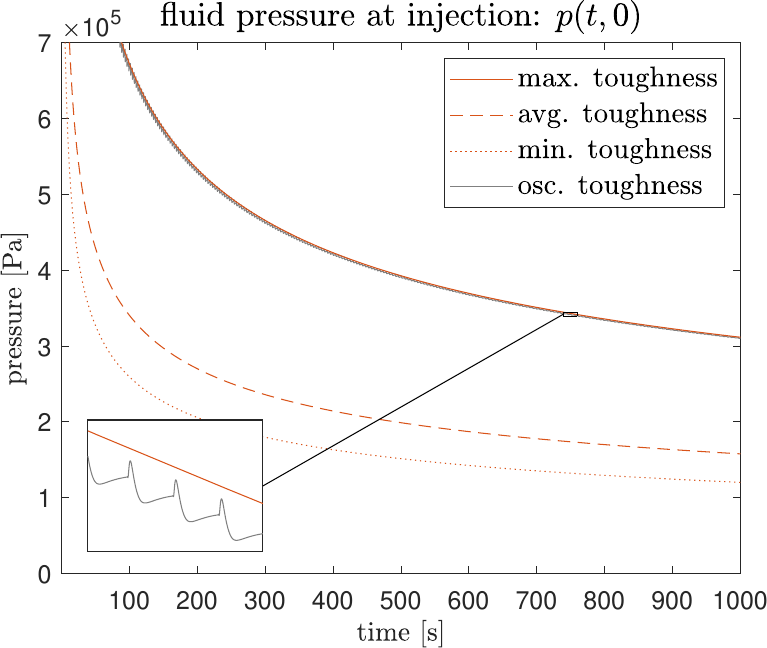}
	 \put(-205,155) {{\bf (d)}}
\\
 \includegraphics[width=0.4\textwidth]{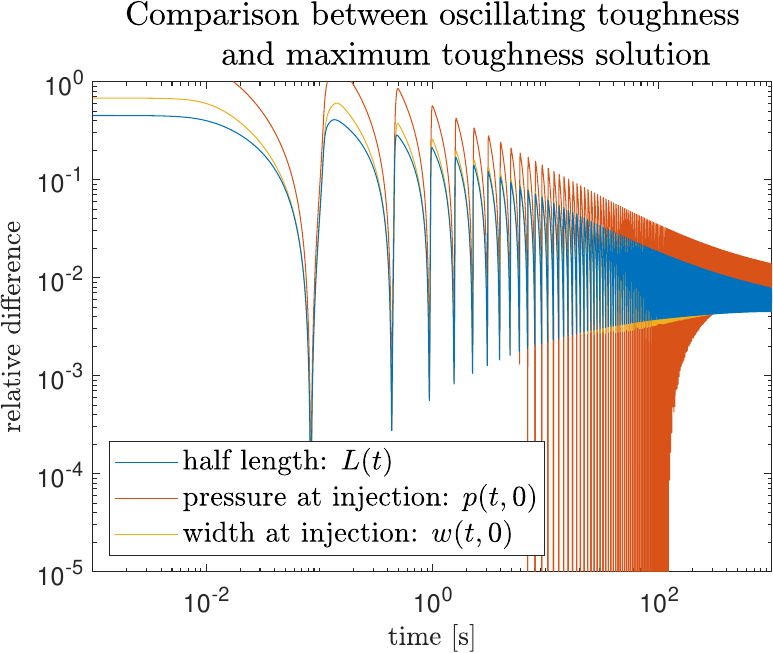}
	 \put(-205,155) {{\bf (e)}}
 \hspace{12mm}
 \includegraphics[width=0.4\textwidth]{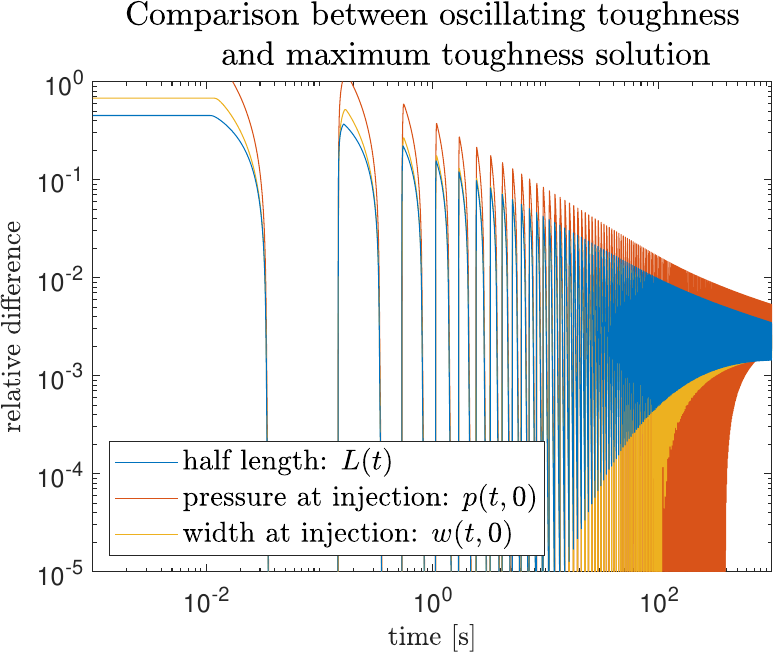}
	 \put(-205,155) {{\bf (f)}}
\caption{Various process parameters for the case of oscillating toughness when $\delta_{max} = 100$ and $\delta_{min}=0.1$, alongside those for a homogeneous material with the minimum, average and maximum toughness. Here we show {\bf (a)}, {\bf (b)} the fracture length, {\bf (c)}, {\bf (d)} the pressure at the injection point $x=0$, and {\bf (e)}, {\bf (f)} the relative difference between the parameters in the oscillating toughness and maximum toughness case. Here {\bf (a)}, {\bf (c)}, {\bf (e)} show the sinusoidal toughness and {\bf (b)}, {\bf (d)}, {\bf (f)} the step-wise toughness.}
	\label{fig:3}
\end{figure}

\begin{figure}[t!]
\centering
 \includegraphics[width=0.4\textwidth]{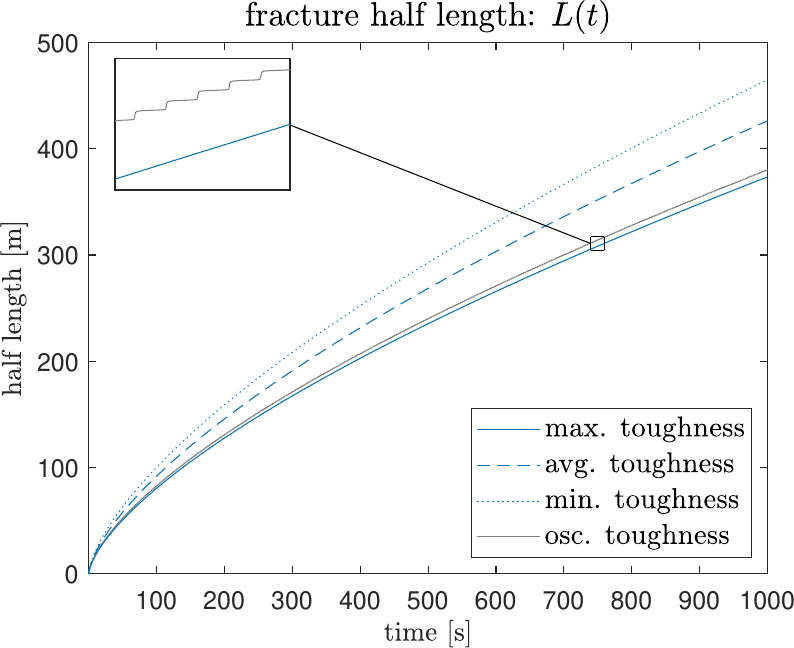}
	 \put(-205,155) {{\bf (a)}}
 \hspace{12mm}
 \includegraphics[width=0.4\textwidth]{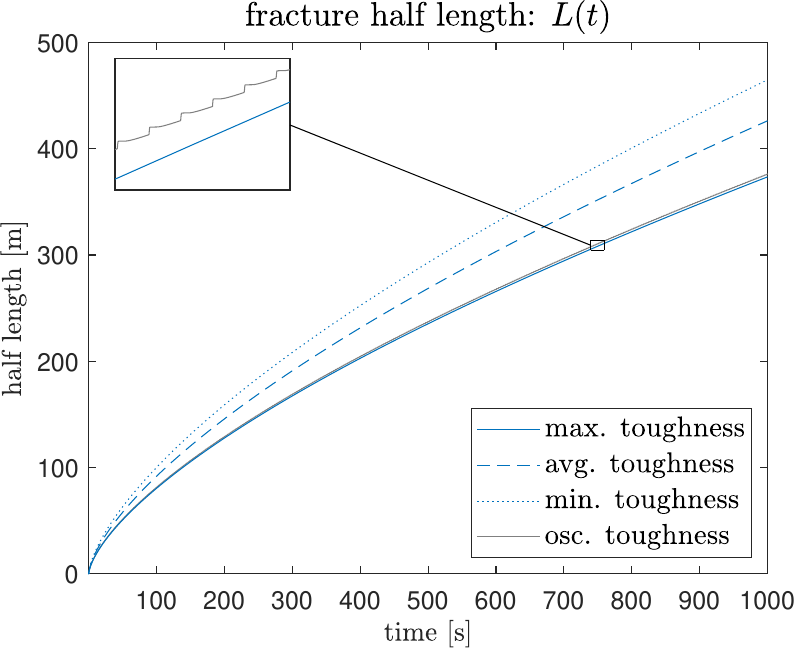}
	 \put(-205,155) {{\bf (b)}}
\\
 \includegraphics[width=0.4\textwidth]{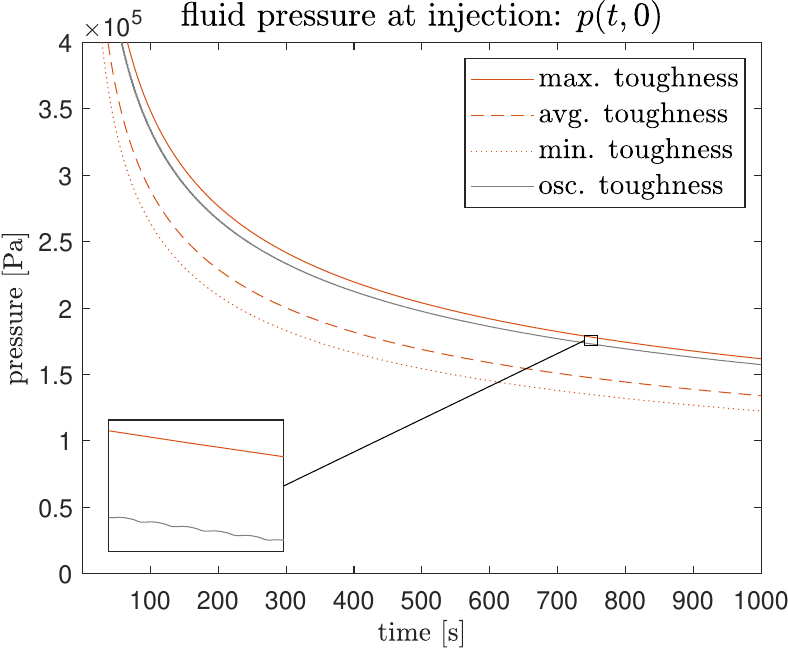}
	 \put(-205,155) {{\bf (c)}}
 \hspace{12mm}
 \includegraphics[width=0.4\textwidth]{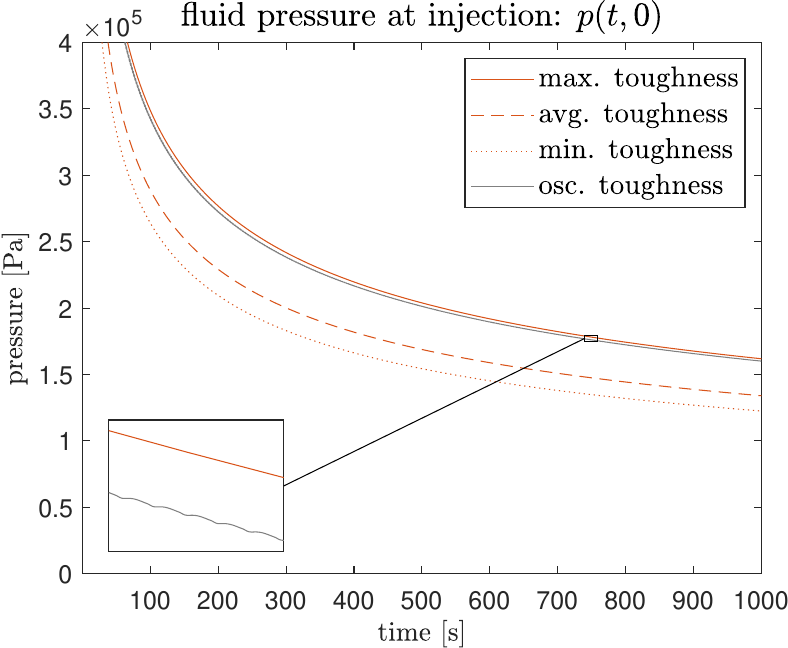}
	 \put(-205,155) {{\bf (d)}}
\\
 \includegraphics[width=0.4\textwidth]{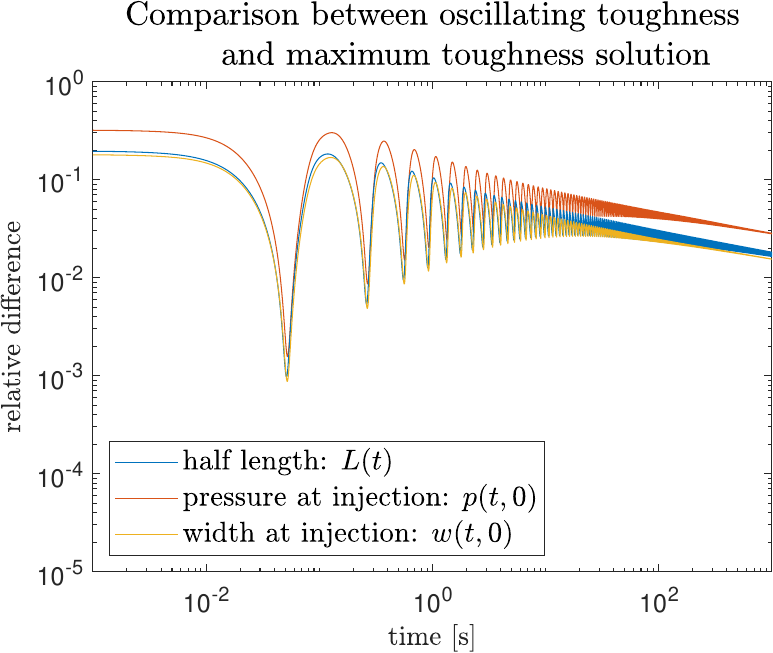}
	 \put(-205,155) {{\bf (e)}}
 \hspace{12mm}
 \includegraphics[width=0.4\textwidth]{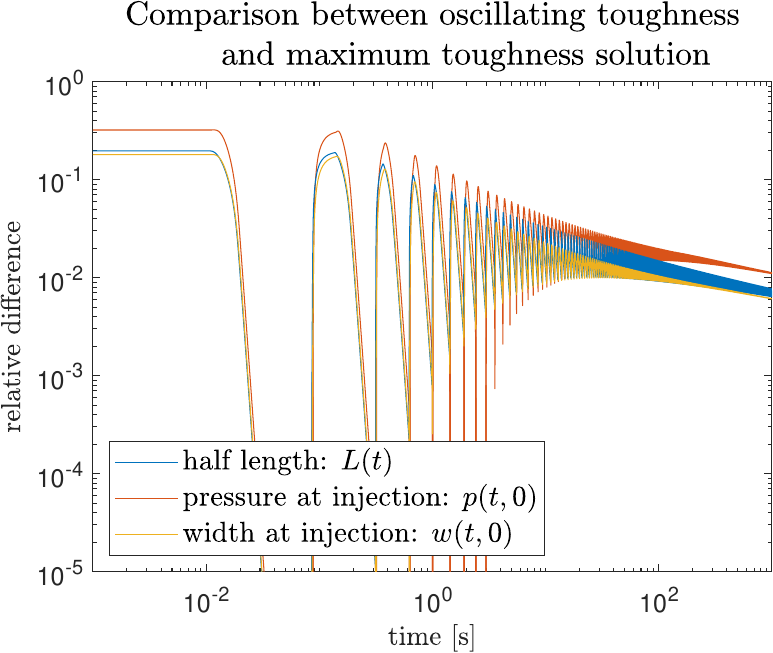}
	 \put(-205,155) {{\bf (f)}}
\caption{Various process parameters for the case of oscillating toughness when $\delta_{max} = 10$ and $\delta_{min}=1$, alongside those for a homogeneous material with the minimum, average and maximum toughness. Here we show {\bf (a)}, {\bf (b)} the fracture length, {\bf (c)}, {\bf (d)} the pressure at the injection point $x=0$, and {\bf (e)}, {\bf (f)} the relative difference between the parameters in the oscillating toughness and maximum toughness case. Here {\bf (a)}, {\bf (c)}, {\bf (e)} show the sinusoidal toughness and {\bf (b)}, {\bf (d)}, {\bf (f)} the step-wise toughness}
	\label{fig:4}
\end{figure}

\begin{figure}[t!]
\centering
 \includegraphics[width=0.4\textwidth]{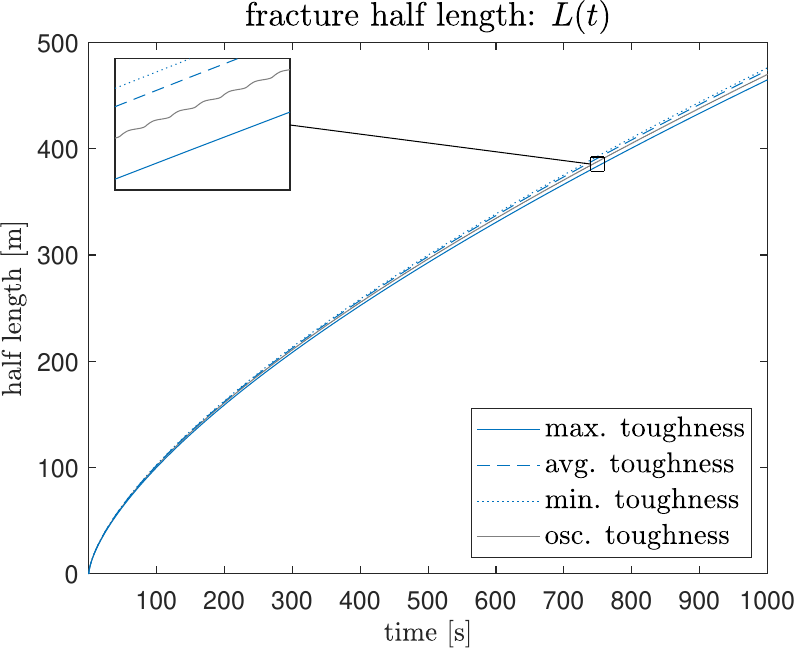}
	 \put(-205,155) {{\bf (a)}}
 \hspace{12mm}
 \includegraphics[width=0.4\textwidth]{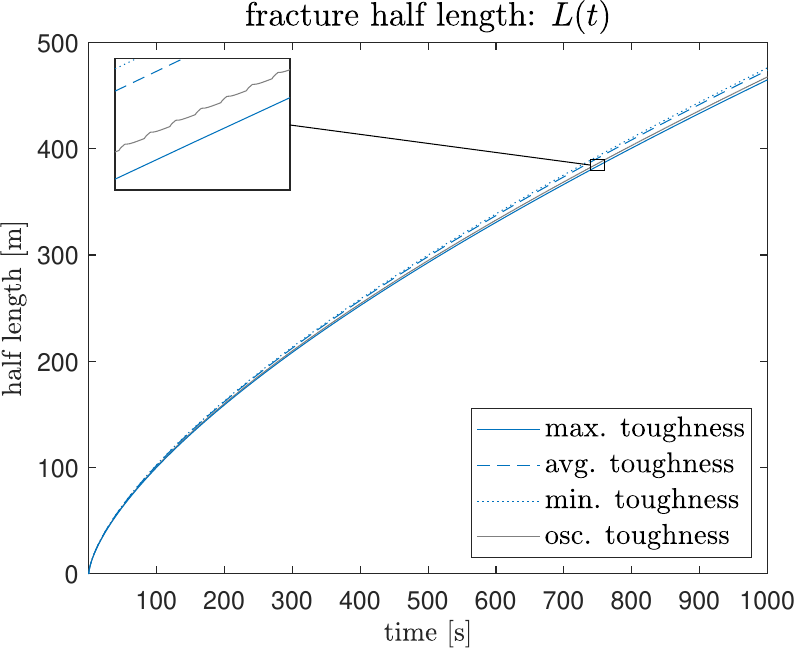}
	 \put(-205,155) {{\bf (b)}}
\\
 \includegraphics[width=0.4\textwidth]{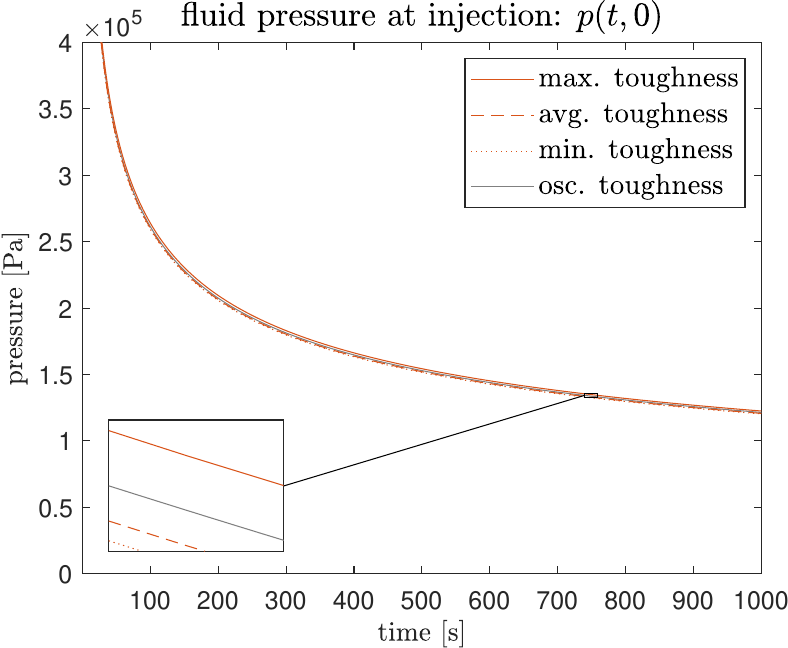}
	 \put(-205,155) {{\bf (c)}}
 \hspace{12mm}
 \includegraphics[width=0.4\textwidth]{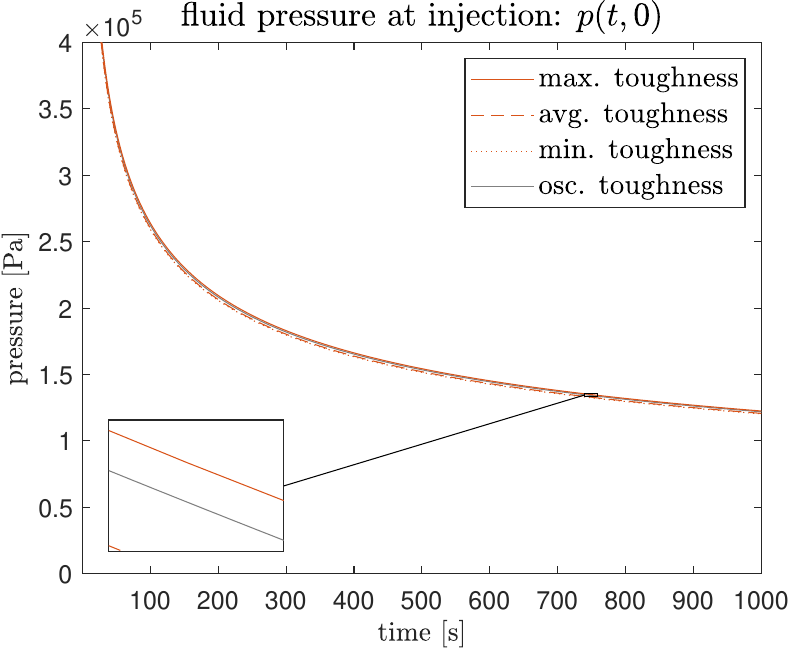}
	 \put(-205,155) {{\bf (d)}}
\\
 \includegraphics[width=0.4\textwidth]{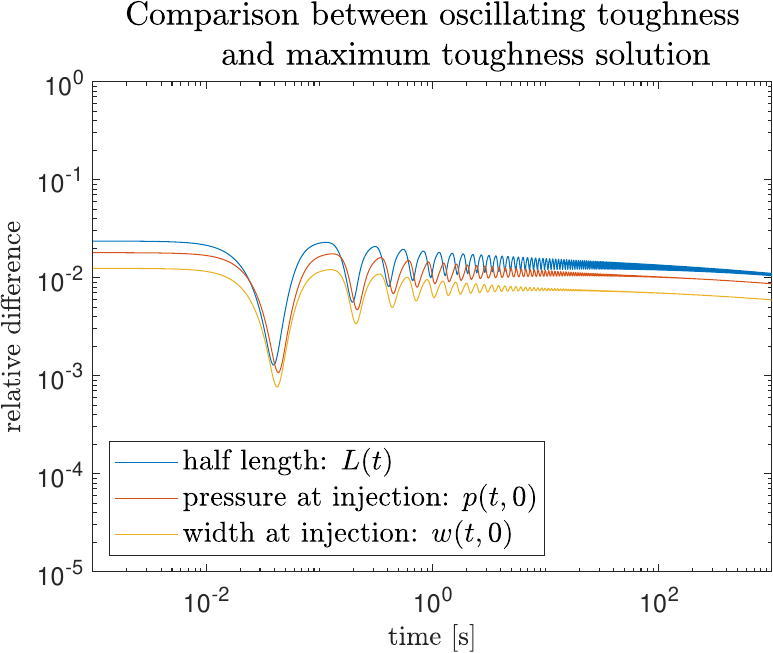}
	 \put(-205,155) {{\bf (e)}}
 \hspace{12mm}
 \includegraphics[width=0.4\textwidth]{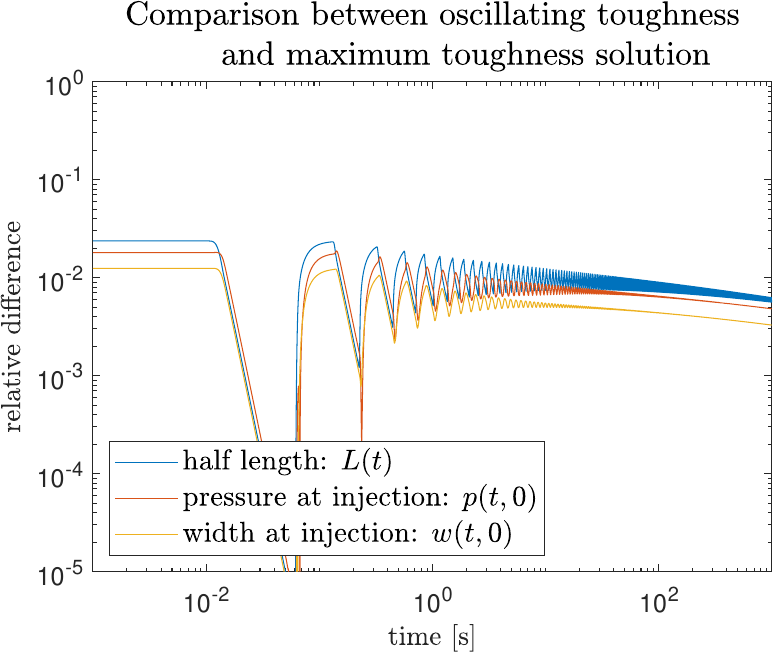}
	 \put(-205,155) {{\bf (f)}}
\caption{Various process parameters for the case of oscillating toughness when $\delta_{max} = 1$ and $\delta_{min}=0.1$, alongside those for a homogeneous material with the minimum, average and maximum toughness. Here we show {\bf (a)}, {\bf (b)} the fracture length, {\bf (c)}, {\bf (d)} the pressure at the injection point $x=0$, and {\bf (e)}, {\bf (f)} the relative difference between the parameters in the oscillating toughness and maximum toughness case. Here {\bf (a)}, {\bf (c)}, {\bf (e)} show the sinusoidal toughness and {\bf (b)}, {\bf (d)}, {\bf (f)} the step-wise toughness}
	\label{fig:5}
\end{figure}

It is well known that classical averaging of the toughness does not provide a satisfactory prediction for the crack length or other process parameters. It has previously been suggested that, away from domain boundaries, the toughness could be approximated using the maximum toughness after some period of time (see e.g.\ discussion of the energy-release rate in \cite{HOSSAIN201415}). This `maximum toughness strategy' was later proposed to be effective for the case for HF in \cite{DONTSOV2021108144}. The investigation provided there, as well as by the authors in \cite{ARMA2021}, confirmed that taking the classical averaging of the toughness does not provide a satisfactory prediction for the crack length and other process parameters, with the maximum toughness providing a significantly better approximation. This is further demonstrated in Figs.~\ref{fig:1} -- \ref{fig:5}, where results for the fracture length $L(t)$ and fluid pressure at the injection point $p(t,0)$ are provided for a variety of toughness inhomogeneities. It is worth noting that in all cases the fracture length $L(t)$ is similar to that observed for layered materials in \cite{HOSSAIN201415}.

As can be seen in Fig.~\ref{fig:1}, where both the maximum and minimum values of $K_{Ic}$ are in the toughness dominated regime ($\delta_{max}=100$ and $\delta_{min}=10$), the strategy of taking the maximum toughness provides a significantly better approximation than taking the average. Further, in Figs.~\ref{fig:1}e,f it can be seen that over long time ($>500$ seconds) the maximum toughness approximation achieves a very low level of error (of order $1$\% or below) for the three key process parameters shown. However, over short time the error is significantly higher, reaching and exceeding $90$\% for the pressure at the start of the process and remaining of order $10$\% even after $10$ seconds, and order $5$\% after 100 seconds. Note that it is the high value of $\delta_{max}$ which is the primary factor in this trend, while the value of $\delta_{min}$ only has a partial relevance. This is because reducing the value of $\delta_{min}$ after a certain threshold does not produce a meaningful difference\footnote{Note however, that if the difference between $\delta_{max}$ and $\delta_{min}$ becomes sufficiently large then the fracture velocity in the weaker region may approach the wave-speed of the medium and should be accounted for. A preliminary investigation by the authors found that this does not impact the results presented here.}. This can be seen from the results in Fig.~\ref{fig:3}, where the value of $\delta_{min}$ is significantly decreased, but almost exactly the same result is obtained for the effectiveness of the maximum toughness approximation.

This trend of the maximum toughness approximation being significantly less effective over short time periods reduces as the maximum toughness decreases. As can be seen in Fig.~\ref{fig:4} (where $\delta_{max}=10$, $\delta_{min}=1$, the latter corresponding to the transient regime), the maximum toughness approximation still has an error of order $10$\% for the fluid pressure at the inlet during the first second of the process, which decreases to below $5$\% after $100$ seconds. While the effect is less significant than the previous case, it is clear that the high level of error at the start of the process would cause issues when attempting to approximate small fractures, or the initial moments after fracture initiation, when in the toughness dominated regime.

Finally, in Fig.~\ref{fig:5} we consider a fracture passing through rock layers in the transient and viscosity dominated regimes (with $\delta_{max}=1$, $\delta_{min}=0.1$). Here, the error of the maximum toughness approximation never exceeds order $3$\% throughout the entire process. This is not surprising, as in the viscosity dominated regime the toughness plays a significantly reduced role in the process. The effectiveness of the maximum toughness approximation for fractures propagating in the viscosity dominated regime has been confirmed in other research \cite{ARMA2021} carried out by the authors.

Combining these results, we draw the following preliminary conclusions:
\begin{enumerate}
 \item[(i)] if the maximum toughness is low (viscosity dominated regime) then approximating using the maximum toughness (as well as the average one) is always effective,
  \item[(ii)] if considering a material with high toughness over long time-periods (or a large fracture length), then the maximum toughness can be consider as an effective homogenisation technique, with the error of order $1$\%,
  \item[(iii)] when considering short fractures, or the initial stages of fracture, then using the maximum toughness strategy may lead to a large error.
\end{enumerate}
Note that the distances over which this final statement is valid depends upon all of the process parameters (not only the toughness distribution and period length), because it is related to the propagation regime corresponding to the maximum possible toughness. However, point (iii) is particularly important when considering cases such as mini-frac testing, as it implies that the maximum toughness strategy would not provide an accurate approximation. This raises the question, is there a better approach for short fractures?

\newpage

$\quad$

\newpage

$\quad$

\newpage

$\quad$
\newpage

$\quad$
\newpage

\section{Improved strategies for short and intermediate fractures}

\subsection{A temporal toughness averaging approach}

\begin{figure}[h!]
\centering
		\includegraphics[width=0.45\textwidth]{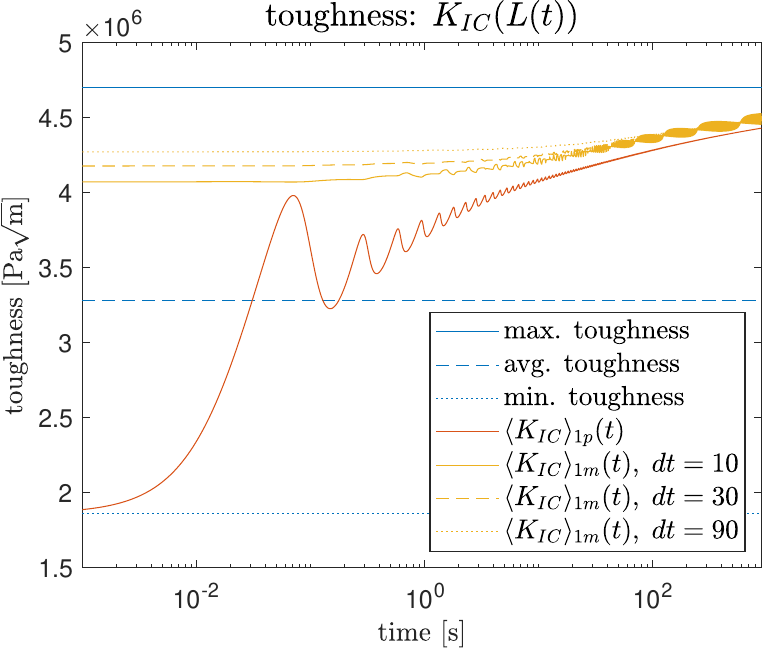}
 \put(-215,160) {{\bf (a)}}
\hspace{5mm}
		\includegraphics[width=0.45\textwidth]{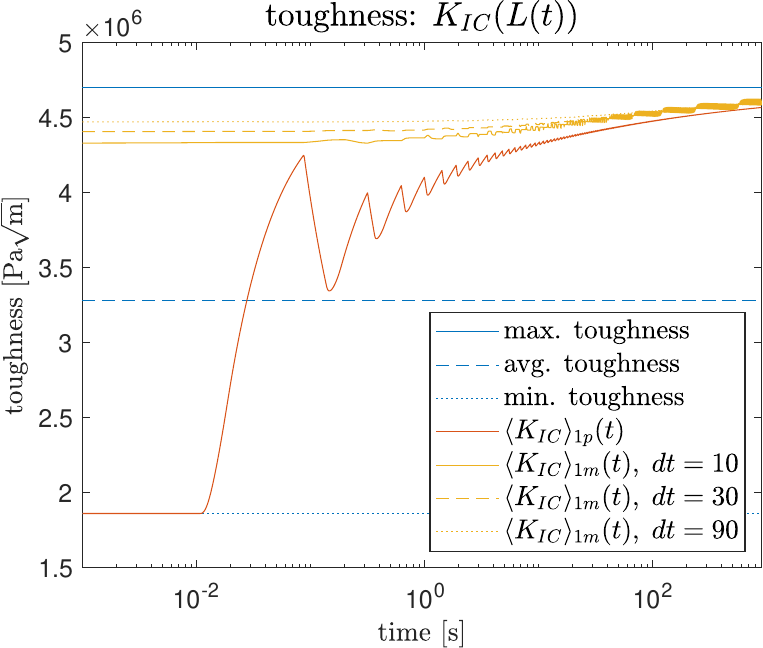}
 \put(-215,160) {{\bf (b)}}
\caption{Temporal averaging of the toughness utilizing definitions \eqref{approx_K_t1}
and \eqref{approx_K_t2} for various integration ranges $d t=10,30,90$ [s].}
	\label{fig:8}
\end{figure}

Towards developing an improved technique for handling the fracture toughness, let us consider the case in which parameters are averaged over time rather than space. As such, we can consider two measures (equivalent to \eqref{K_AvSpace}-\eqref{K_MovSpace}, except averaged over time):
\begin{equation}
\label{approx_K_t1}
\langle K_{IC}\rangle_{1p} (t) = \frac{1}{t}\int_{0}^{t}K_{IC}\big(L(\xi)\big)d\xi.
\end{equation}
and
\begin{equation}
\label{approx_K_t2}
\langle K_{IC}\rangle_{1m} (t) = \frac{1}{d t}\int_{t}^{t+d t}K_{IC}\big(L(\xi)\big)d\xi .
\end{equation}
It is clear from looking at the toughness distribution over time, see Figs.~\ref{Fig:ToughDis1}c,d, that this will give a different approximation than the spatial average, while tending to a larger value of the average toughness over time (although not necessarily $K_{max}$). An example of the values of $\langle K_{Ic} \rangle_{1p}$ and $\langle K_{Ic} \rangle_{1m}$ obtained for the case with $\delta_{max}=10$, $\delta_{min}=1$ are provided in Fig.~\ref{fig:8} for different time frames $d t$.

A drawback of the definition of $\langle K_{Ic} \rangle_{1m}$ is, however, the fact that a choice of the time frame $d t$ used in computation cannot be reasonably justified. For example, if we take $d t\to0$, then $\langle K_{Ic} \rangle_{1m}\to K_{IC}(L(t))$ (for comparison see Fig.~\ref{Fig:Delta1}) which, as could be expected, corresponds to the original problem (taking into account the initial toughness). Alternatively, if $dt$ is taken as constant, then the measure $\langle K_{Ic} \rangle_{1m}$ may experience increasingly large oscillations as $t\to \infty$ as the slowing crack growth causes the (spatial) distance covered by the integral to decrease (unless the velocity profile is such as to counter this effect). 

\begin{figure}[t!]
\begin{center}
		\includegraphics[width=0.45\textwidth]{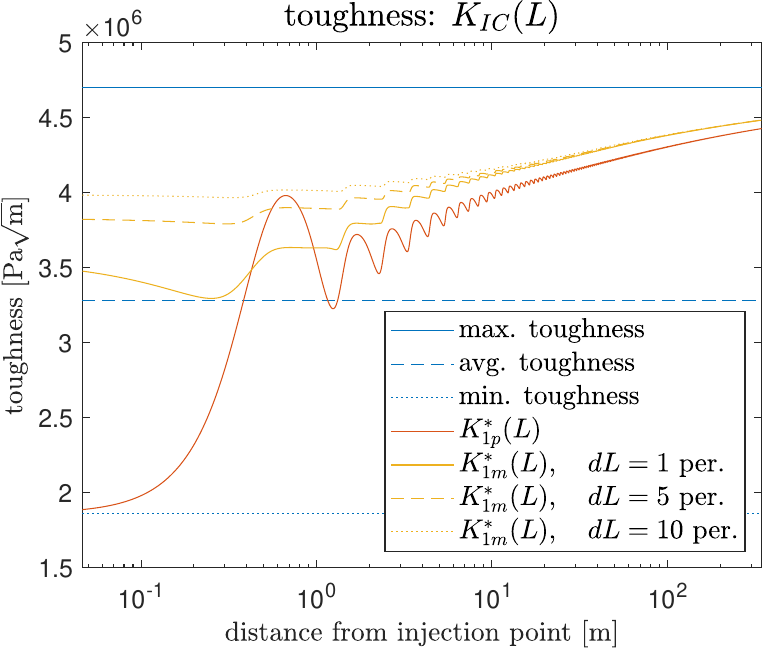}
 \put(-215,160) {{\bf (a)}}
\hspace{5mm}
		\includegraphics[width=0.45\textwidth]{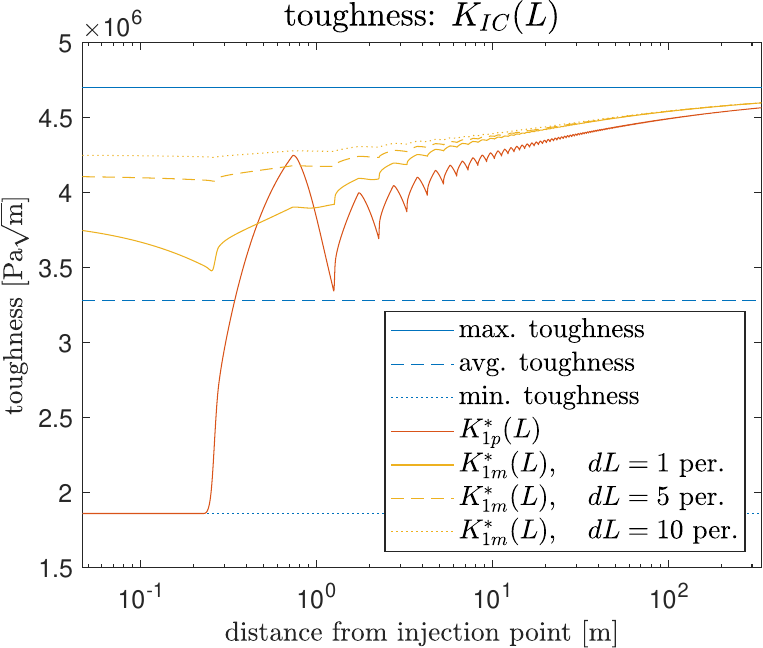}
 \put(-215,160) {{\bf (b)}}
\end{center}
\caption{Temporal averaging of the toughness using the spatial definitions $K^*_{1p} (L)$ \eqref{approx_K_L1} and $K^*_{1m} (L, dL)$ \eqref{approx_K_L2} for various integration ranges $dL= X,5X,10X$.}
	\label{fig:9}
\end{figure}

On the other hand, we can transform these into expressions in terms of the spatial coordinate, by defining inverses in the form:
\begin{equation}
\label{approx_K_L_new}
K_{1p(m)}^* (L) = \langle K_{Ic} \rangle_{1p(m)} \left( t^{-1} (L) \right) .
\end{equation}
While the definition of $K_{1p}^*$ is straightforward, in the case of $K_{1m}^*$ the additional parameter $d t$ must be dealt with. As such, we can utilize a change in \eqref{approx_K_t1}-\eqref{approx_K_t2} to better represent these:
\begin{equation}
\label{approx_K_L1}
K_{1p}^*(L)=\frac{1}{t^{-1}(L)}\int_{0}^{L}K_{IC}(x)\, d\!\left(t^{-1}(x)\right),
\end{equation}
and
\begin{equation}
\label{approx_K_L2}
K_{1m}^*(L)\equiv K_{1m}^*(L,dL)=\frac{1}{t^{-1}(L+dL)-t^{-1}(L)}\int_{L}^{L+dL}K_{IC}(x)\, d\!\left(t^{-1}(x)\right).
\end{equation}
Here the time interval $(t,t+d t)$ corresponds to its spatial image $(L,L+d L)$ when considering a fixed point in time. It is clear that if $d t$ is a constant then $dL=dL(t,d t)$ and visa versa. In practice, this also allows a more physically meaningful interpretation of this roaming length to be taken, as while $d t$ was largely arbitrary, the value of $d L$ can be taken corresponding to the expected `period' of the inhomogeneity (i.e.\ rock layer length). In addition, while there is some dependence of the second measure $K_{1m}^*$ on the choice of $d L$, it does approach a `representative' value over time, as can be seen in Fig.~\ref{fig:9}. Here, the values of $K_{1p}^*$ and $K_{1m}^*$ computed using \eqref{approx_K_L1}--\eqref{approx_K_L2} when $\delta_{max}=10$ and $\delta_{min}=1$ are given for a variety of $d L=X,5X,10X$. One can observe that, at the distance larger than several periods ($L>nX$), the measures
$K_{1m}^*(L,X)$ and $K_{1m}^*(L,nX)$ differ rather insignificantly.

In the remainder of the paper, we will take the value of $d L = X$, that is the minimal 'Representative Volume Element' (RVE), to capture as much of the process behaviour as possible\footnote{\label{Footy5} The main goal of developing this averaging procedure is to implement it into commercial solvers that, for evident reasons, cannot deliver computations with a highly detailed toughness distribution. Here however, apart from the size of the RVE there is a minimal size of the Finite Element, $e_{min}>0$, allowing for effective computations. In turn, in practical application the size of the Finite Element, $e=dL$, should be taken as $dL=nX>e_{min}$ (with a minimal natural number $n$.)}

\begin{figure}[t!]
\begin{center}
		\includegraphics[width=0.45\textwidth]{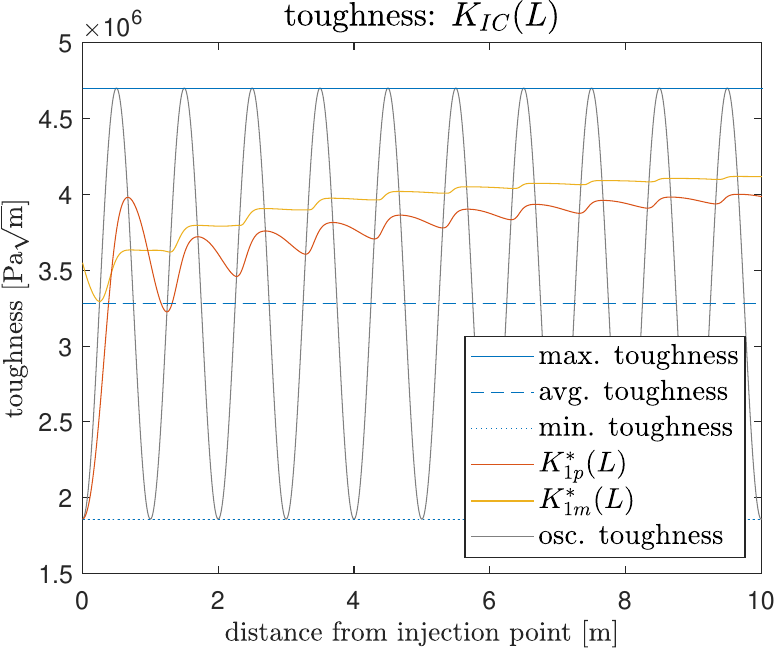}
 \put(-215,160) {{\bf (a)}}
\hspace{5mm}
		\includegraphics[width=0.45\textwidth]{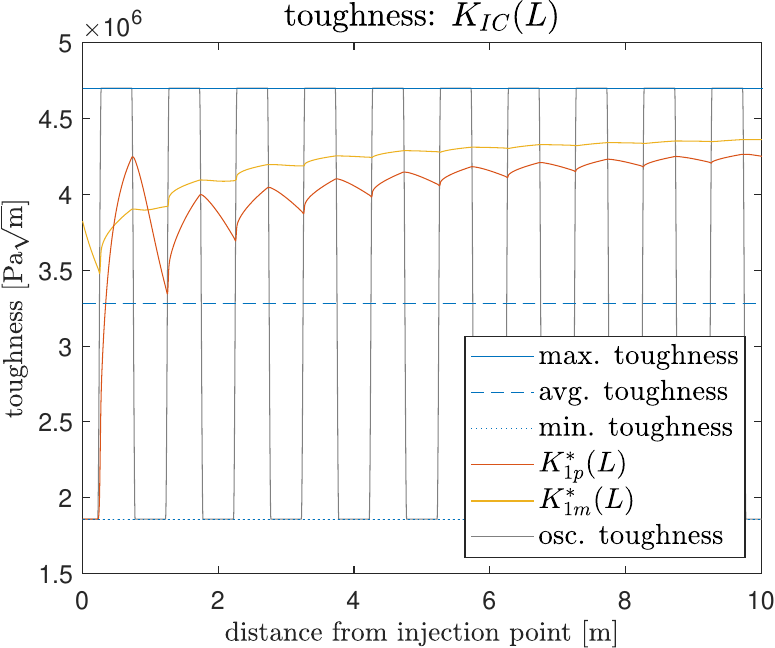}
 \put(-215,160) {{\bf (b)}}
\end{center}
\caption{Temporal averaging of the toughness using the spatial definitions \eqref{approx_K_L1} and \eqref{approx_K_L2} for $dL= X$, compared to the initial oscillating distribution.}
	\label{fig:10}
\end{figure}

In Fig.~\ref{fig:10}, values of $K_{1p}^*$ and $K_{1m}^*$ obtained when $d L=X$, the period of the toughness, are compared with the spatial distribution of the toughness $K_{Ic}$.

Finally, noting the crucial role of the fracture velocity on both the process and the influence of the toughness inhomogeneity (see Figs.~\ref{Fig:Delta1}c,d, Sect.~\ref{DeltaSect}), we can also modify the expressions \eqref{approx_K_L1}-\eqref{approx_K_L2} to be in the equivalent form:
\begin{equation}
\label{approx_K_L1v}
K_{1p}^*(L)=\left(\int_{0}^{L}\frac{dx}{v(x)}\right)^{-1}\int_{0}^{L}K_{IC}(x)\,\frac{dx}{v(x)},
\end{equation}
and
\begin{equation}
\label{approx_K_L2v}
K_{1m}^*(L,dL)=\left(\int_{L}^{L+dL}\frac{dx}{v(x)}\right)^{-1}\int_{L}^{L+dL}K_{IC}(x)\,\frac{dx}{v(x)},
\end{equation}
where $v(L)$ is the velocity the fracture tip takes for a given fracture length. This final form of the measure helps to demonstrate that, while $K_{1p}^*$ and $K_{1m}^*$ are still computing an averaging over the fracture, they are incorporating some specifics of the fracture process.

%
%

\subsection{Effectiveness of the temporal averaging approach}\label{Sect:Effective}

An immediate question arises concerning these new measures. Firstly, do they provide a more effective approximation of the material toughness for short fractures than the maximum toughness? If so, then is one of these measures \eqref{approx_K_L1v}-\eqref{approx_K_L2v} more effective than the other?

\begin{figure}[h!]
\centering
 \includegraphics[width=0.4\textwidth]{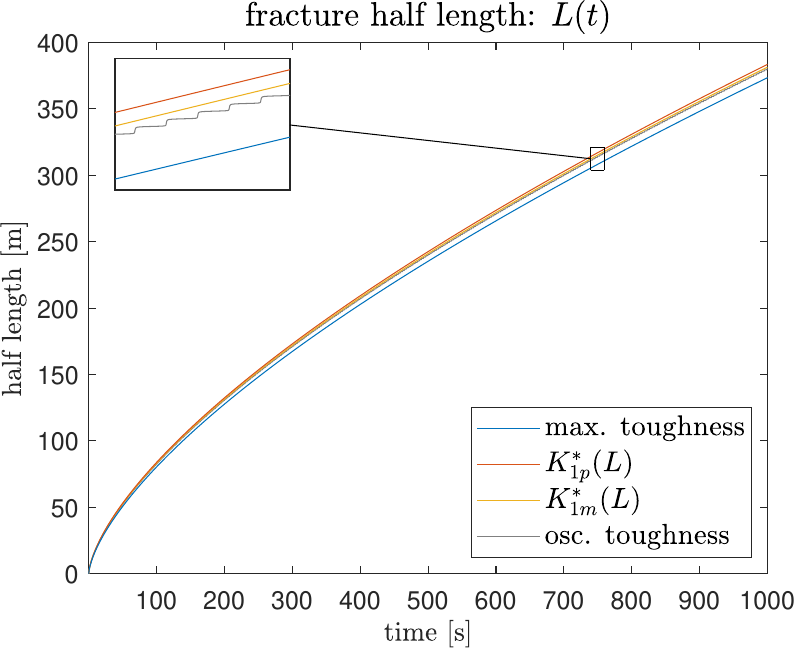}
	 \put(-205,155) {{\bf (a)}}
 \hspace{12mm}
 \includegraphics[width=0.4\textwidth]{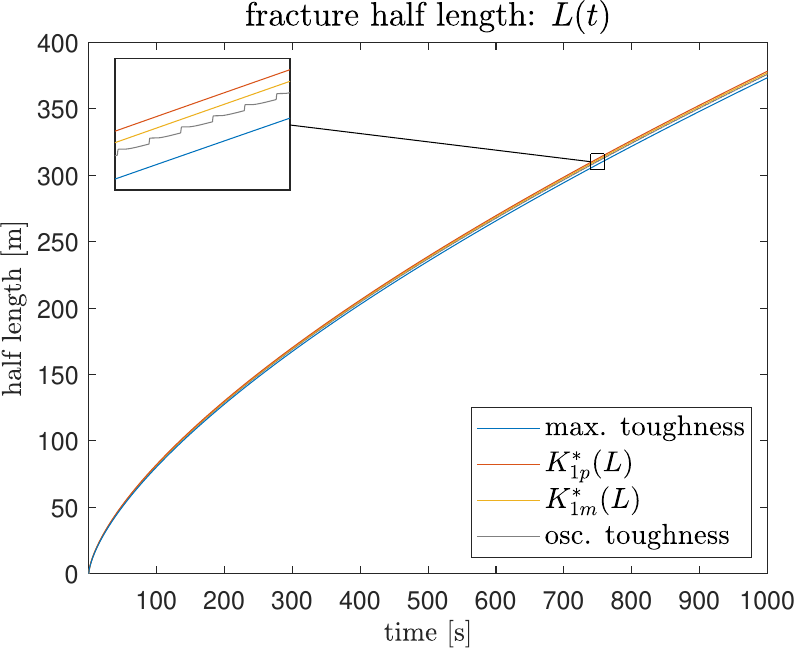}
	 \put(-205,155) {{\bf (b)}}
\\
 \includegraphics[width=0.4\textwidth]{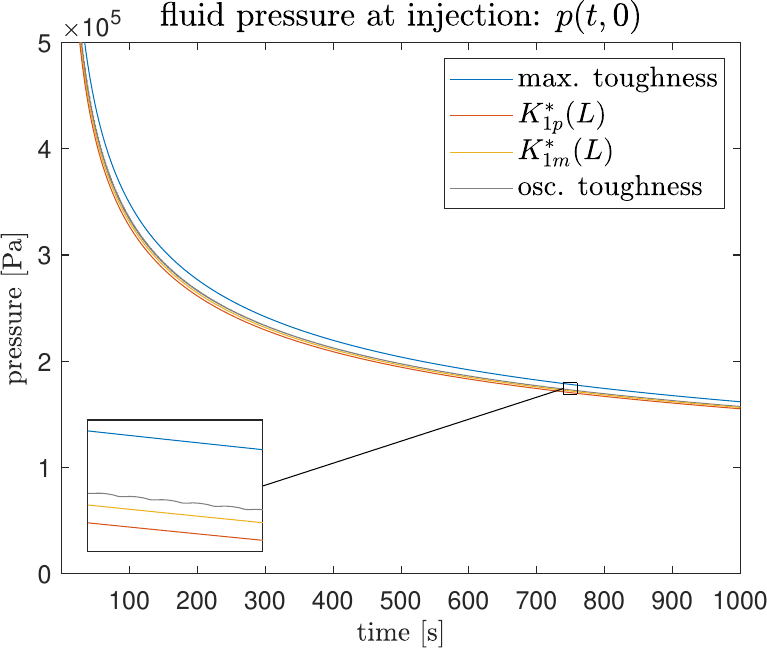}
	 \put(-205,155) {{\bf (c)}}
 \hspace{12mm}
 \includegraphics[width=0.4\textwidth]{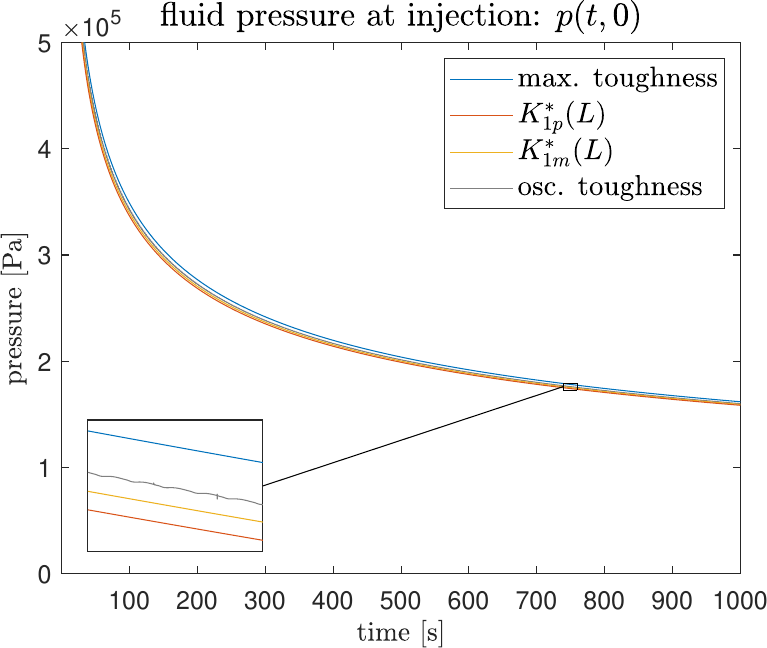}
	 \put(-205,155) {{\bf (d)}}
\caption{The {\bf (a)}, {\bf (b)} fracture half-length $L(t)$ and {\bf (c)}, {\bf (d)} fluid-induced pressure at the point of injection $p(t,0)$, estimated via homogenization using the maximum toughness $K_{max}$ and the temporal averages $K_{1p(m)}^*$ from \eqref{approx_K_L1v}-\eqref{approx_K_L2v}, together with the real oscillatory behaviour. Here we consider the oscillating toughness given by $\delta_{max}=10$ and $\delta_{min}=1$ with distribution: {\bf (a)}, {\bf (c)} sinusoidal, {\bf (b)}, {\bf (d)} step-wise.}
	\label{fig:12a}
\end{figure}

\begin{figure}[t!]
\centering
 \includegraphics[width=0.4\textwidth]{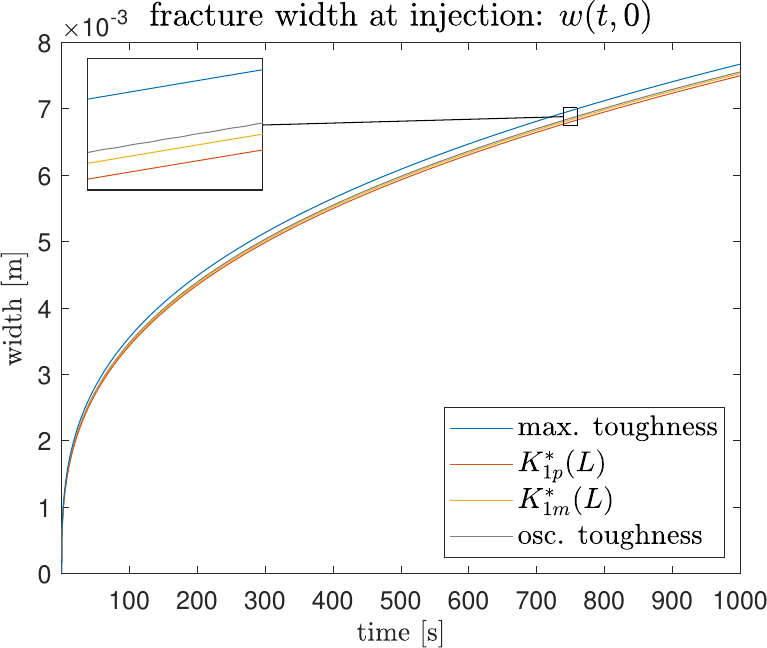}
	 \put(-205,155) {{\bf (a)}}
 \hspace{12mm}
 \includegraphics[width=0.4\textwidth]{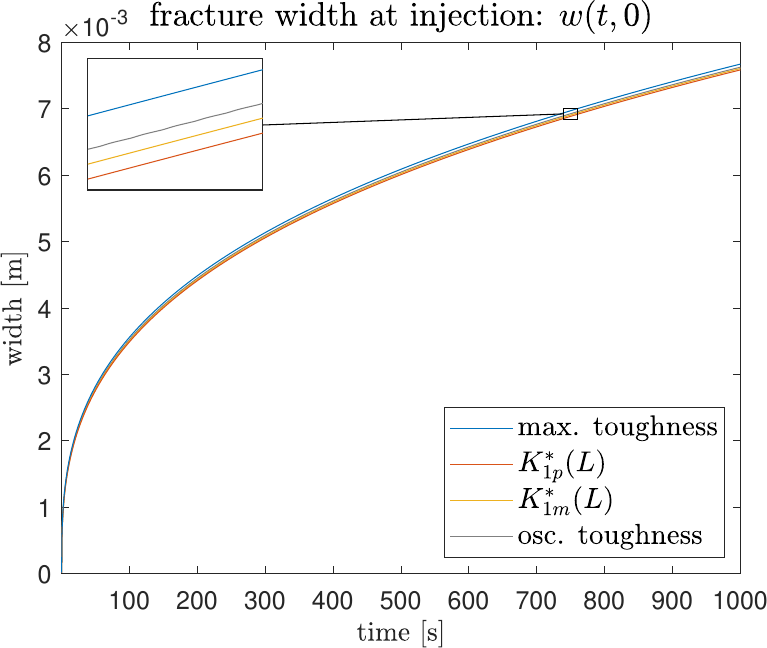}
	 \put(-205,155) {{\bf (b)}}
\\
 \includegraphics[width=0.4\textwidth]{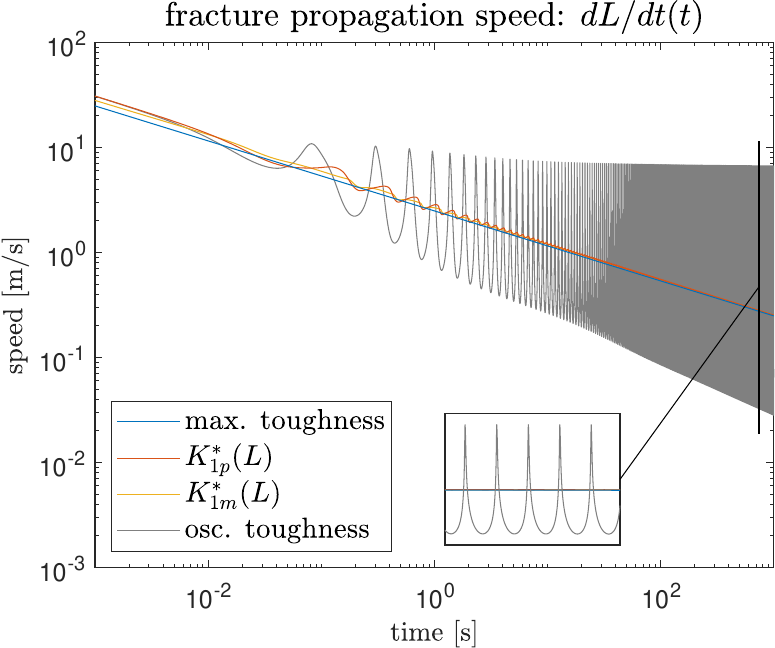}
	 \put(-205,155) {{\bf (c)}}
 \hspace{12mm}
 \includegraphics[width=0.4\textwidth]{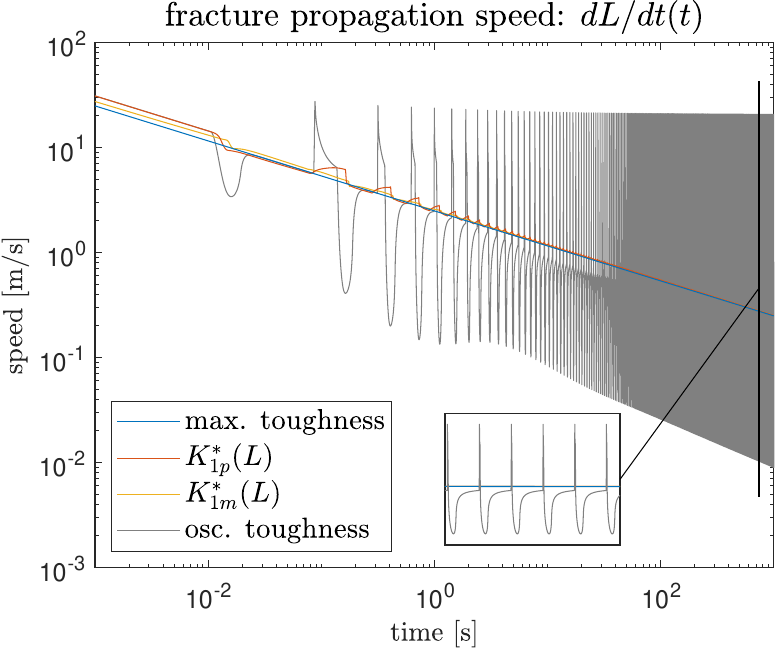}
	 \put(-205,155) {{\bf (d)}}
\caption{The {\bf (a)}, {\bf (b)} aperture at the point of injection $w(t,0)$ and {\bf (c)}, {\bf (d)} velocity of the fracture tip $v(t)$, estimated via homogenization using the maximum toughness $K_{max}$ and the temporal averages $K_{1p(m)}^*$ from \eqref{approx_K_L1v}-\eqref{approx_K_L2v}, together with the real oscillatory behaviour. Here we consider the oscillating toughness given by $\delta_{max}=10$ and $\delta_{min}=1$ with distribution: {\bf (a)}, {\bf (c)} sinusoidal, {\bf (b)}, {\bf (d)} step-wise.}
	\label{fig:12b}
\end{figure}

\begin{figure}[t!]
\centering
 \includegraphics[width=0.4\textwidth]{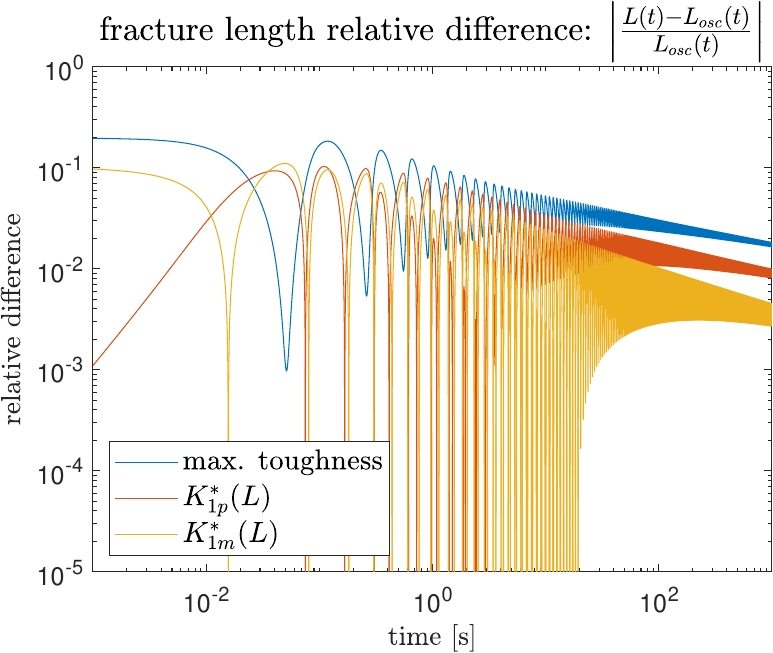}
	 \put(-205,155) {{\bf (a)}}
 \hspace{12mm}
 \includegraphics[width=0.4\textwidth]{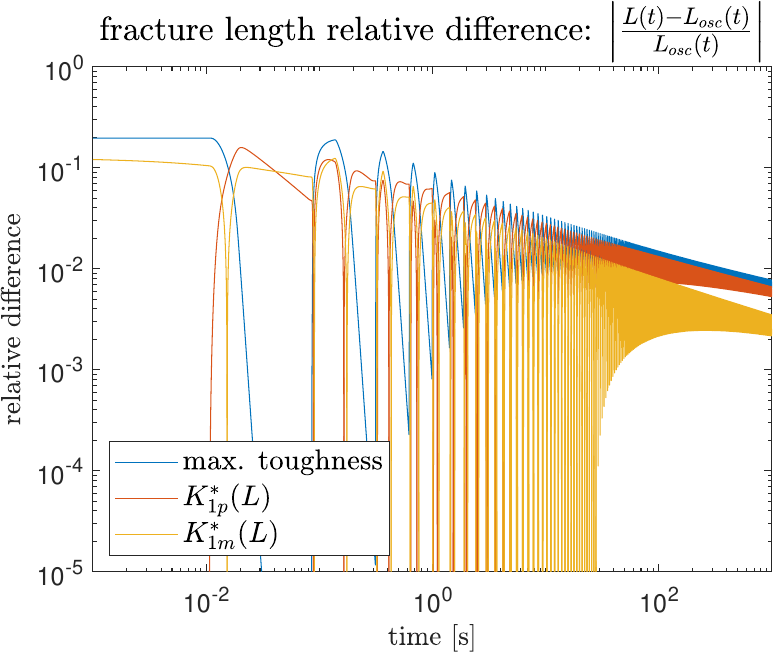}
	 \put(-205,155) {{\bf (b)}}
\\
 \includegraphics[width=0.4\textwidth]{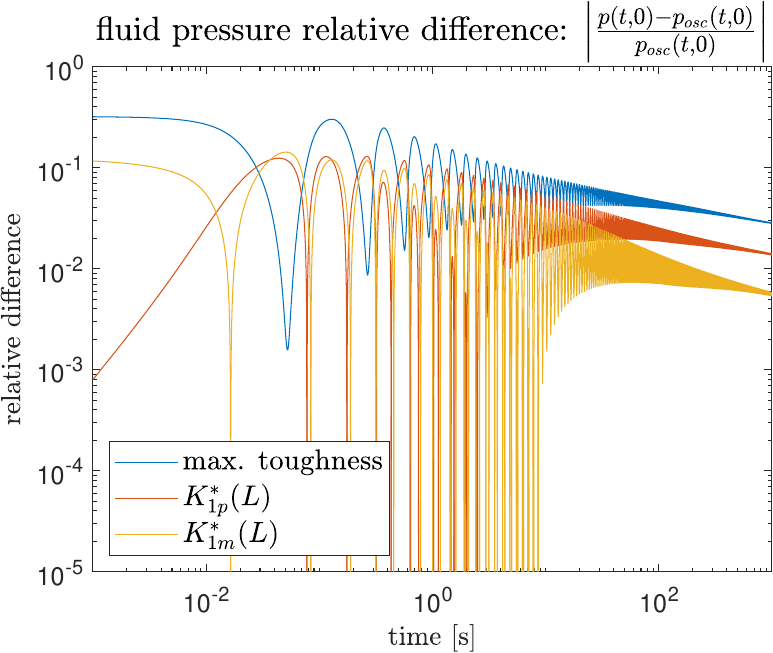}
	 \put(-205,155) {{\bf (c)}}
 \hspace{12mm}
 \includegraphics[width=0.4\textwidth]{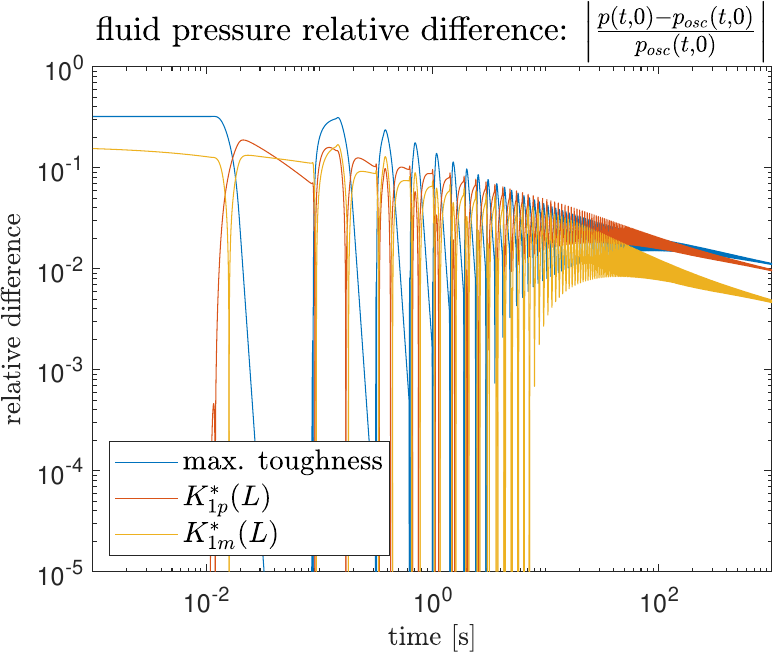}
	 \put(-205,155) {{\bf (d)}}
\\
 \includegraphics[width=0.4\textwidth]{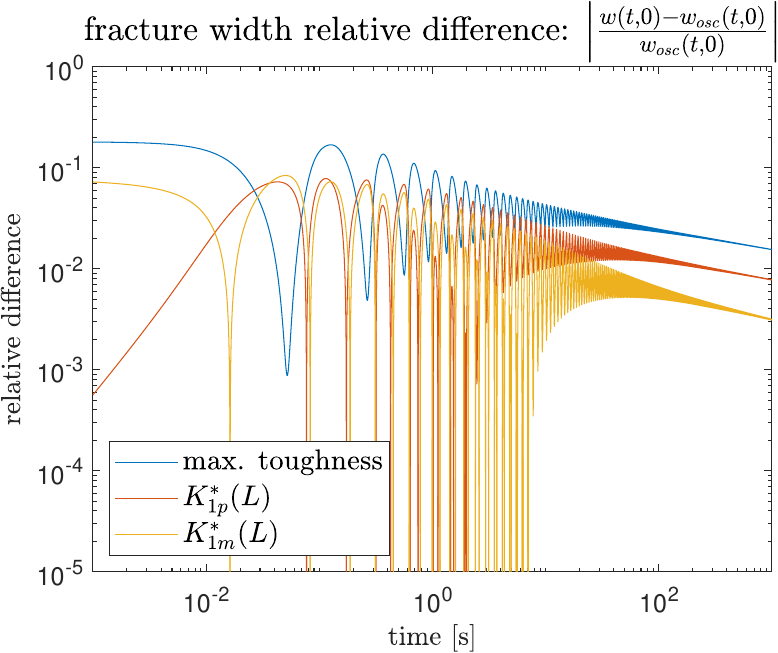}
	 \put(-205,155) {{\bf (e)}}
 \hspace{12mm}
 \includegraphics[width=0.4\textwidth]{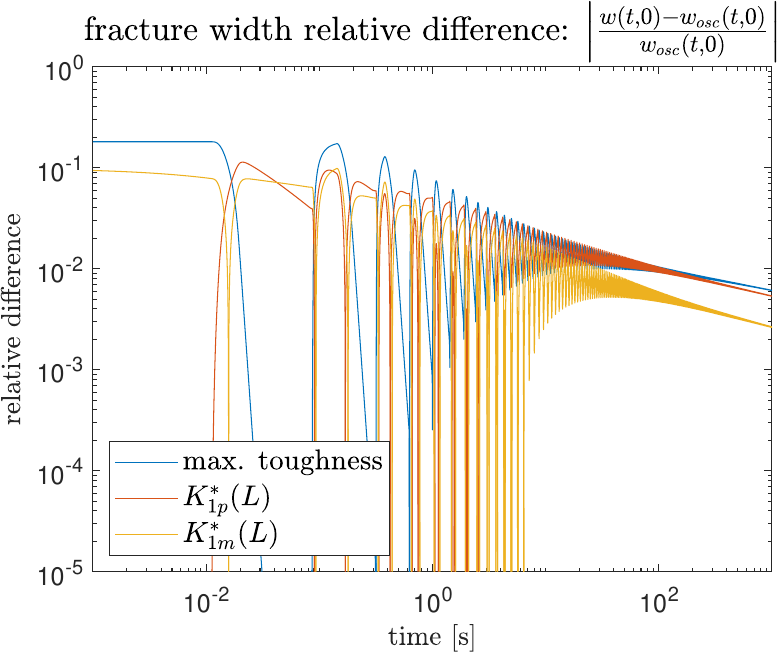}
	 \put(-205,155) {{\bf (f)}}
\caption{Relative difference between the {\bf (a)}, {\bf (b)} fracture half-length $L(t)$ and {\bf (c)}, {\bf (d)} fluid-induced pressure, {\bf (e)}, {\bf (f)} fracture aperture at the point of injection $w(t,0)$, obtained for oscillating toughness against those estimated via homogenization using the maximum toughness $K_{max}$ and the temporal averages $K_{1p(m)}^*$ from \eqref{approx_K_L1v}-\eqref{approx_K_L2v}. Here we consider the oscillating toughness given by $\delta_{max}=10$ and $\delta_{min}=1$ with distribution: {\bf (a)}, {\bf (c)}, {\bf (e)} sinusoidal, {\bf (b)}, {\bf (d)}, {\bf (f)} step-wise. }
	\label{fig:13a}
\end{figure}

To investigate this for fractures with a high toughness the case with $\delta_{max}=10$ and $\delta_{min}=1$ is considered, with the former corresponding to a distinctly toughness dominated regime but not an unrealistic one, while the latter is in the transient regime. The results of simulations for measures $K_{1p}^*$ and $K_{1m}^*$ (with $dL = X$) are provided in Figs.~\ref{fig:12a} -- \ref{fig:13a}. It is clear from the results that the temporal averaging based approach consistently obtains a lower level of error than the maximum toughness approximation for all of the process parameters considered: the fracture length, the inlet fluid pressure and the crack aperture at the point of injection. 

One crucial observation from Fig.~\ref{fig:12a} -- \ref{fig:13a} is that, while the maximum toughness approximation provides a lower bound on the fracture length, the two measures $K_{1p}^*$ and $K_{1m}^*$ provide an upper bound on this parameter. Similarly, the maximum toughness provides an upper bound on the aperture and fluid pressure at the point of injection, while the measures based on temporal averaging \eqref{approx_K_L1v}-\eqref{approx_K_L2v} provide a lower bound. As such, the two approaches (averaging and maximum) can be combined to obtain effective bounds on these key process parameters.

As for the comparative effectiveness of $K_{1p}^*$ and $K_{1m}^*$, while it is clear that both are effective it can be stated that the roaming average $K_{1m}^*$ has a small advantage over the full-length averaging. It can be seen in Fig.~\ref{fig:13a} that $K_{1m}^*$ consistently achieves a lower level of error of approximation for all process parameters, while managing to capture all of the important behaviour of the process.

Finally, as is the case for most averaging procedures, the proposed approach does not preserve local peculiarities of the process. For example, the real crack speeds and those produced with use of the averaged toughness, $K_{1p}^*$ or $K_{1m}^*$, have nothing in common apart from the fact that they produce comparable crack lengths.

However, comparing the relative errors of the different strategies, presented in Fig.~\ref{fig:13a}, we confirm that the proposed averaging does improve the results of simulations for short and moderate crack lengths  in comparison with the $K_{max}$ strategy. Moreover, they even manage to produce a slightly better approximation than the latter strategy for long cracks, although the difference is not enough to be significant in practical application.

$\quad$
\newpage

$\quad$
\newpage

$\quad$
\newpage

\subsection{A temporal energy averaging approach}

In the previous section we have analysed the toughness averaging approach based on the fact that, in the case with constant elastic parameters, even between layers with differing toughness, the Irwin and Griffith criteria locally coincide ($K_I=K_{IC}$ and ${\cal E}={\cal E}_c$, respectively, see also footnote~\ref{Foot1}, page \pageref{Foot1}). However, when these criteria are written in their global (averaging) form (that is $K_I=K^*_{1p(m)}$ and ${\cal E}={\cal E}^*_{1p(m)}$), this property is not preserved, as noted in \cite{DONTSOV2021108144}.

In order to construct an averaging-based approach that is effective for inhomogeneous elastic parameters alongside the toughness, we modify the formulation to be based on the fracture energy. To achieve this, we start by introducing:
\begin{equation}
\label{approx_K_L_energy}
{\cal E}_{1p(m)}^* (L) = \langle {\cal E}_{c} \rangle_{1p(m)} \left( t^{-1} (L) \right).
\end{equation}
It is straightforward to verify that measures in the pair $K_I,{\cal E}$, ($K_I =K^*_{2p(m)}$ and ${\cal E}={\cal E}^*_{1p(m)}$) are equivalent if we define the averaging toughness:
\begin{equation}
\label{approx_K_L2e1}
K_{2p(m)}^*(L)=\sqrt{\langle K^2_{IC} \rangle_{1p(m)} \left( t^{-1} (L) \right)}.
\end{equation}
Thus, instead of the weighted $L_1$ norm, we consider averaging in $L_2$ norm with the same weight. As such, the energy-based equivalent of measures \eqref{approx_K_L1v}-\eqref{approx_K_L2v} can be written as
\begin{equation}
\label{approx_K_L2evp}
K_{2p}^*(L)=\sqrt{\left(\int_{0}^{L}\frac{dx}{v(x)}\right)^{-1}\int_{0}^{L}K^2_{IC}(x)\,\frac{dx}{v(x)}},
\end{equation}
and
\begin{equation}
\label{approx_K_L2evm}
K_{2m}^*(L,dL)=\sqrt{\left(\int_{L}^{L+dL}
\frac{dx}{v(x)}\right)^{-1}\int_{L}^{L+dL}K^2_{IC}(x)\,\frac{dx}{v(x)}} ,
\end{equation}
where $v(L)$ is again the velocity the fracture tip takes for a given fracture length.
\\
Due to the strong similarities between these measures and the toughness averaging $K_{1p}^*$ and $K_{1m}^*$, we will not repeat all of the results and analysis provided in the previous subsections. Instead, we highlight here the relative difference between the oscillating toughness solution and that obtained when homogenising using approximations $K_{2p}^*$ and $K_{2m}^*$ for the three key process parameters: the fracture length, crack opening and pressure at the injection point (similar to those provided for the toughness averaging measures in Fig.~\ref{fig:13a}).

\begin{figure}[t!]
\centering
 \includegraphics[width=0.4\textwidth]{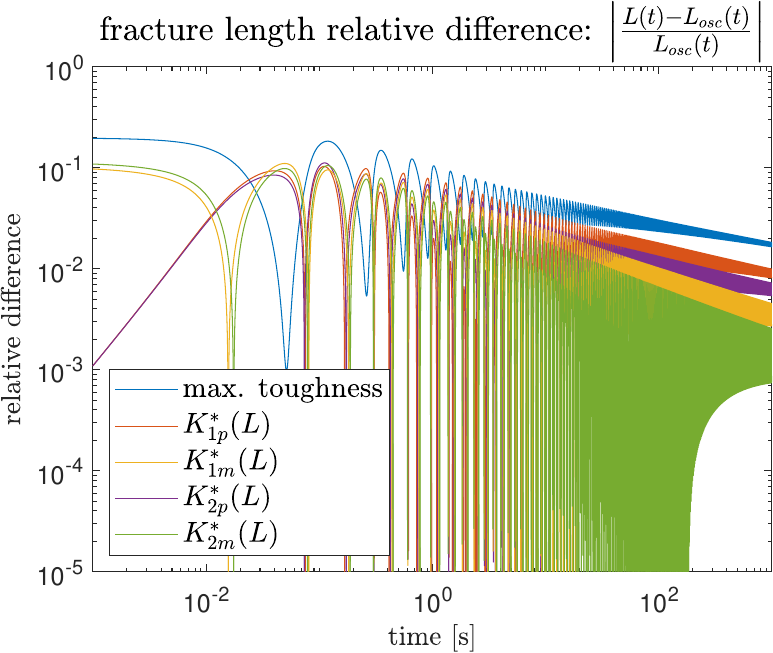}
	 \put(-205,155) {{\bf (a)}}
 \hspace{12mm}
 \includegraphics[width=0.4\textwidth]{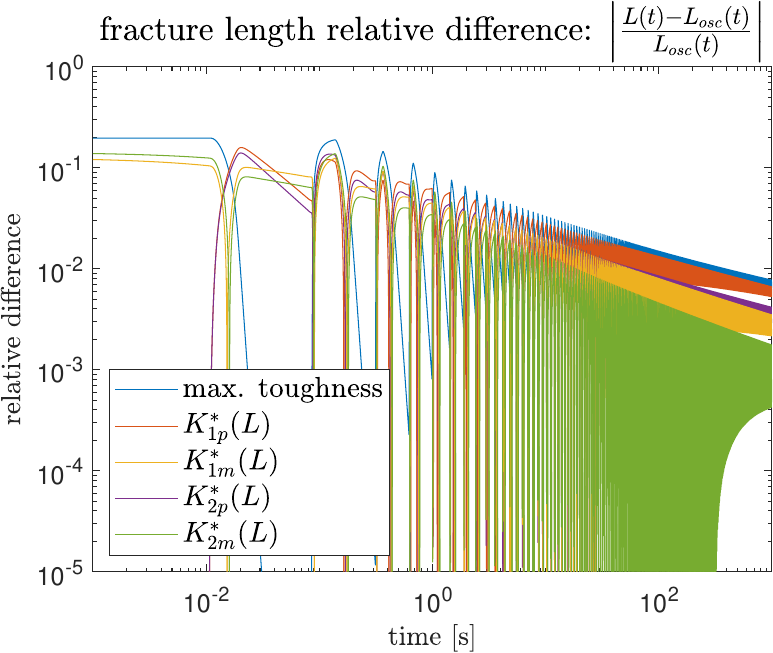}
	 \put(-205,155) {{\bf (b)}}
\\
 \includegraphics[width=0.4\textwidth]{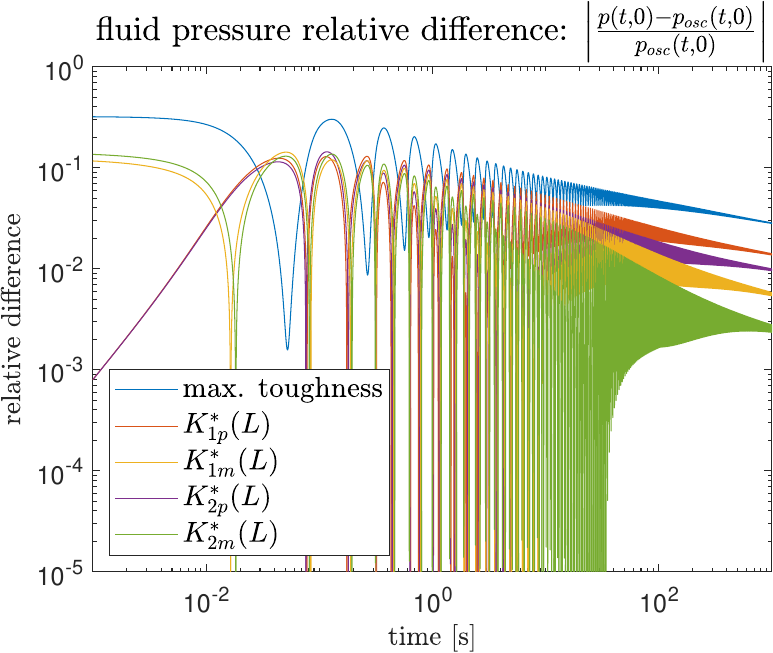}
	 \put(-205,155) {{\bf (c)}}
 \hspace{12mm}
 \includegraphics[width=0.4\textwidth]{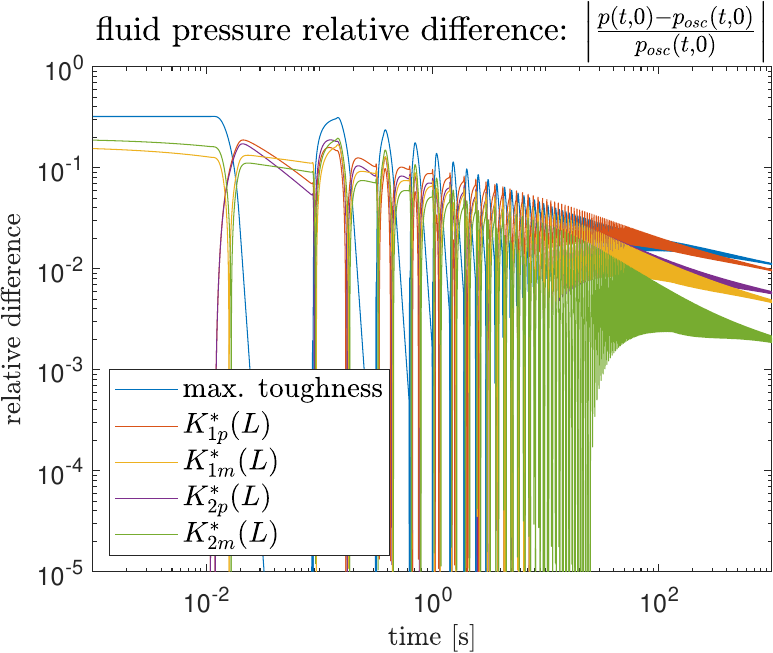}
	 \put(-205,155) {{\bf (d)}}
\\
 \includegraphics[width=0.4\textwidth]{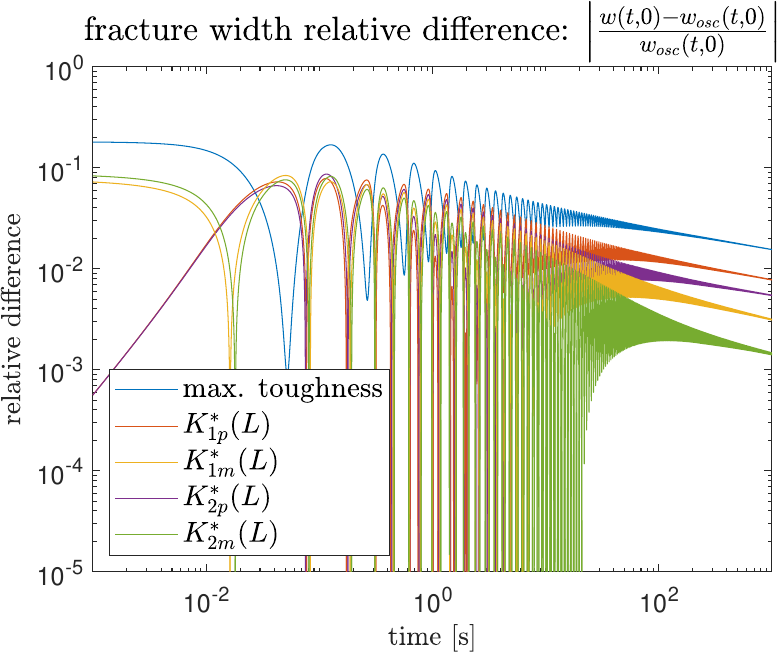}
	 \put(-205,155) {{\bf (e)}}
 \hspace{12mm}
 \includegraphics[width=0.4\textwidth]{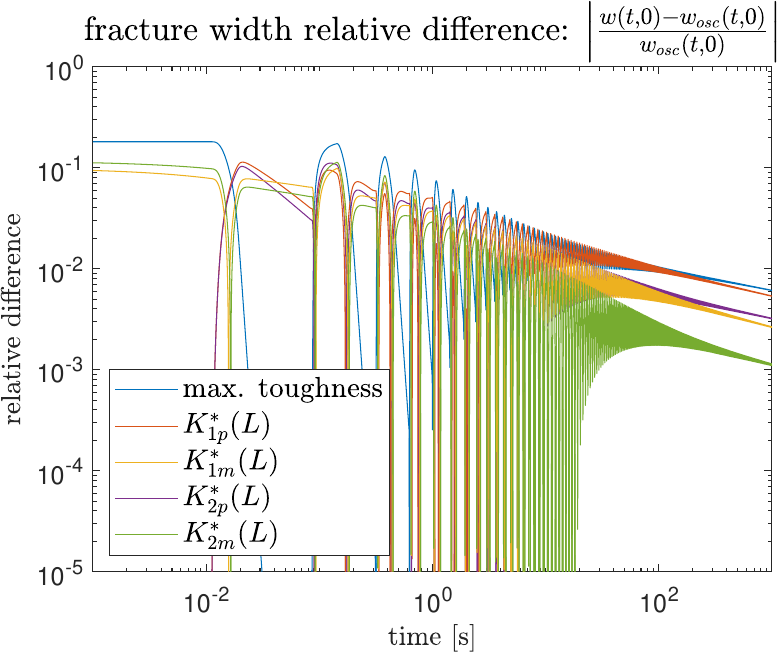}
	 \put(-205,155) {{\bf (f)}}
\caption{Relative difference between the {\bf (a)}, {\bf (b)} fracture half-length $L(t)$, {\bf (c)}, {\bf (d)} fluid-induced pressure, {\bf (e)}, {\bf (f)} fracture aperture at the point of injection $w(t,0)$, obtained for oscillating toughness against those estimated via homogenization using the maximum toughness $K_{max}$ and the temporal averages $K_{1p(m)}^*$, $K_{2p(m)}^*$ from \eqref{approx_K_L1v}-\eqref{approx_K_L2v}, \eqref{approx_K_L2evp}-\eqref{approx_K_L2evm}. Here we consider the oscillating toughness given by $\delta_{max}=10$ and $\delta_{min}=1$ with distribution: {\bf (a)}, {\bf (c)}, {\bf (e)} sinusoidal, {\bf (b)}, {\bf (d)}, {\bf (f)} step-wise.}
	\label{fig:15}
\end{figure}

Comparing the different measures in Fig.~\ref{fig:15}, we can observe that the predictions given in simulations by the new averages $K_{2p}^*$ and $K_{2m}^*$ are very similar to those obtained using the previous measures, $K_{1p}^*$ and $K_{1m}^*$, for a very short time/crack. In the case of moderate and long cracks/times, the new averages offer better predictions for the main global peculiarities of the process (length, crack opening and the pressure at the injection point). Again, local quantities (such as the instantaneous crack speed) are not representative after any averaging (when comparing the speeds between the accurate and averaged solutions).

\begin{figure}[h!]
\centering
 \includegraphics[width=0.4\textwidth]{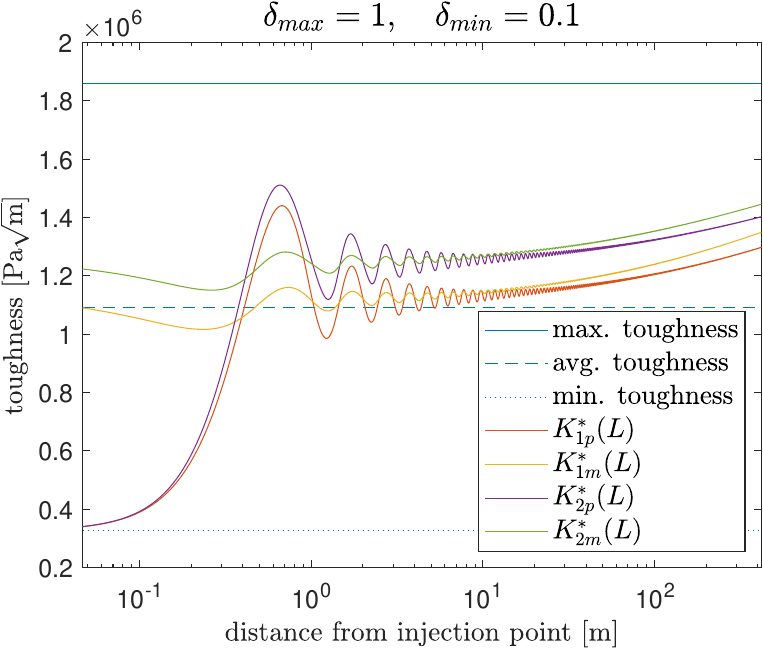}
	 \put(-205,155) {{\bf (a)}}
 \hspace{12mm}
 \includegraphics[width=0.4\textwidth]{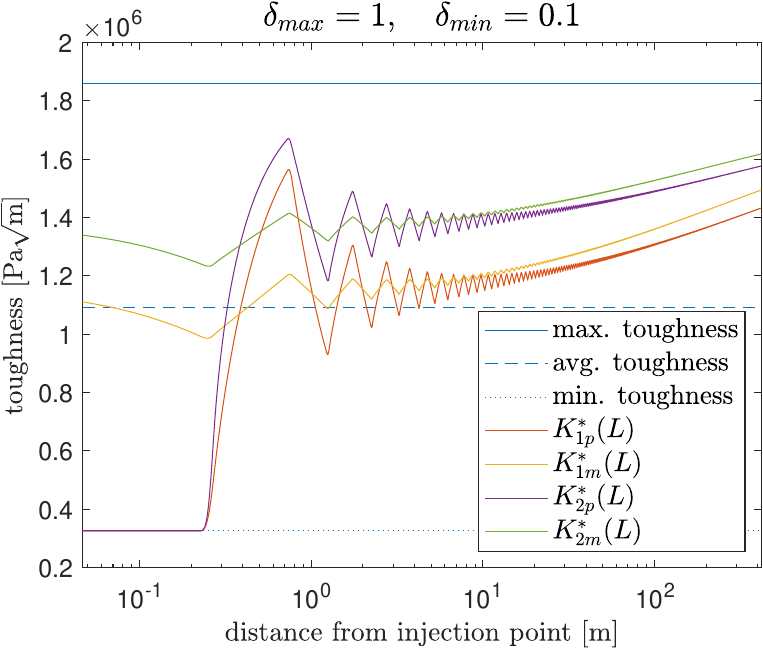}
	 \put(-205,155) {{\bf (b)}}
\\
 \includegraphics[width=0.4\textwidth]{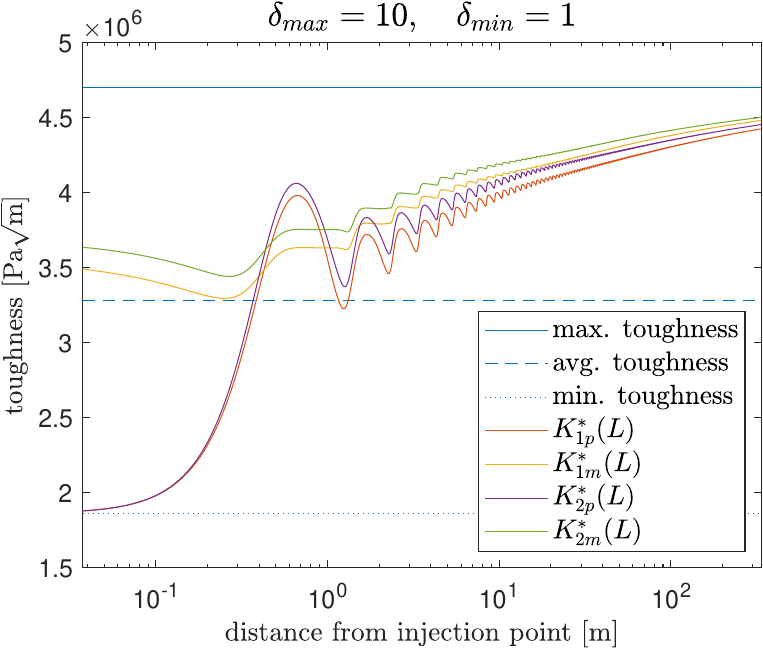}
	 \put(-205,155) {{\bf (c)}}
 \hspace{12mm}
 \includegraphics[width=0.4\textwidth]{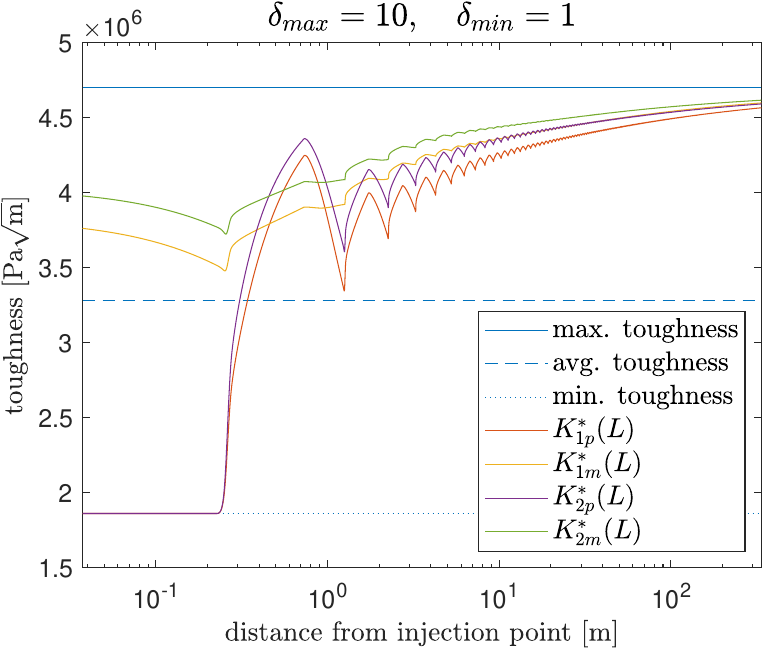}
	 \put(-205,155) {{\bf (d)}}
\\
 \includegraphics[width=0.4\textwidth]{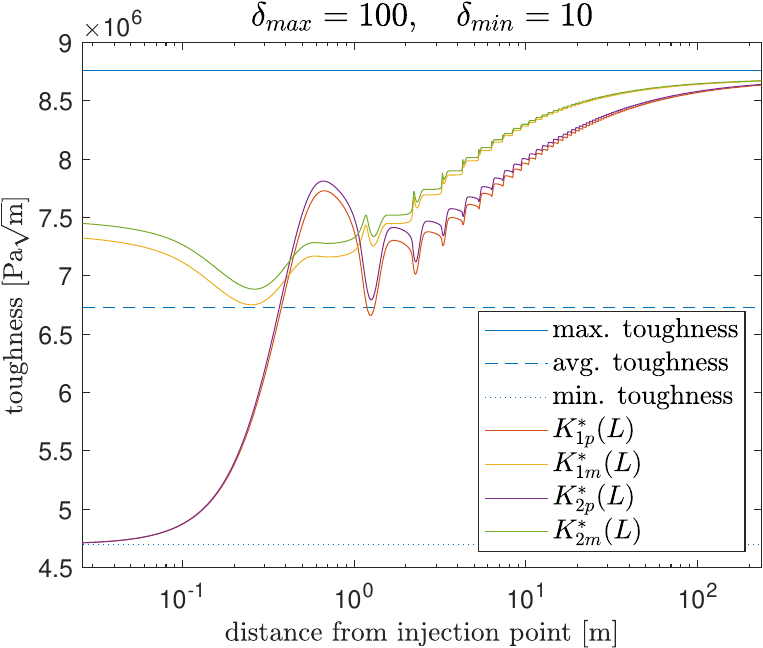}
	 \put(-205,155) {{\bf (e)}}
 \hspace{12mm}
 \includegraphics[width=0.4\textwidth]{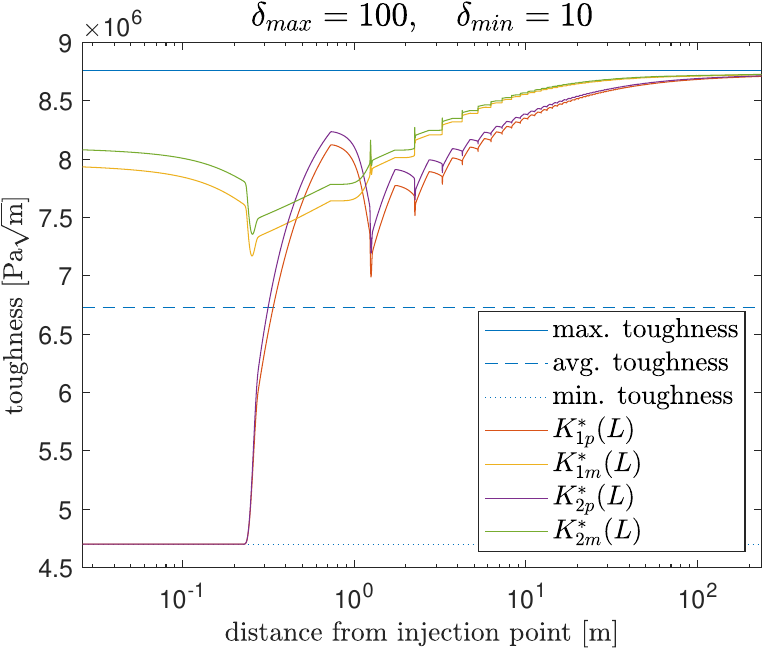}
	 \put(-205,155) {{\bf (f)}}
\caption{Various averaging strategies for oscillating toughness for different combinations of the pairs $(\delta_{min},\delta_{max})$ with distribution: {\bf (a)}, {\bf (c)} and {\bf (e)} - sinusoidal, and {\bf (b)}, {\bf (d)} and {\bf (f)} step-wise toughness distributions. Graphs on {\bf (a), (b)} correspond to the pair $\delta_{min}=0.1$, $\delta_{max}=1$.
Figures in {\bf (c), (d)} demonstrate the pair $\delta_{min}=1$, $\delta_{max}=10$, and, finally
{\bf (e), (f)} correspond to the pair $\delta_{min}=10$, $\delta_{max}=100$. }
	\label{fig:20}
\end{figure}

With the effectiveness of the temporal averaging based approach established, a comparison of the measures was undertaken for all range of the regimes under investigation in this paper. The final results in comprehensive form are presented in Fig.~\ref{fig:20}, for different combinations of the process parameters $\delta_{max}$ and $\delta_{max}$.

\subsection{Consistency of the various averaging techniques}

\subsubsection{Behaviour under repeated averaging}

The temporal averaging approach has been shown above to produce an effective approximation of the key process parameters for the cases under consideration. In order to provide further justification for the approach, and provide a comparison between the different averaging techniques proposed in this paper, let us here perform a brief examination of the consistency.

\begin{figure}[t!]
\centering
 \includegraphics[width=0.4\textwidth]{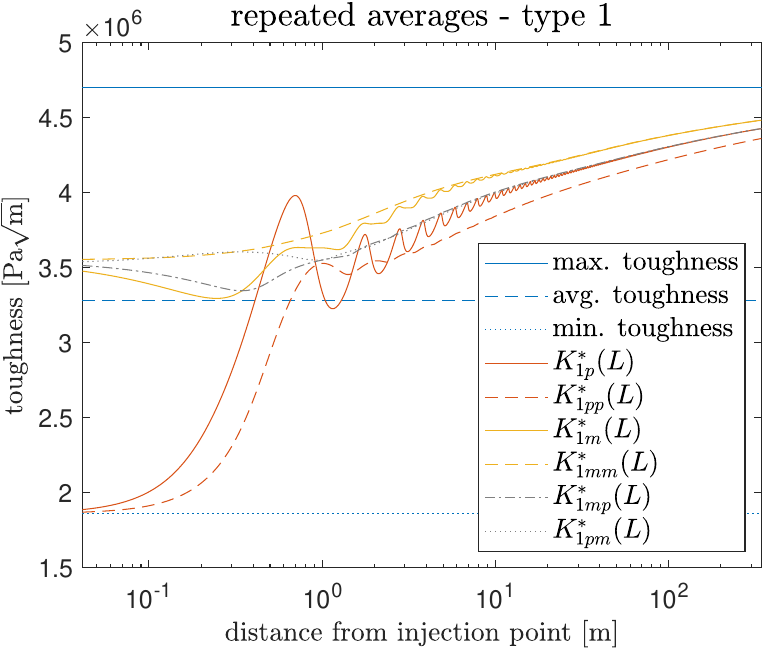}
	 \put(-205,155) {{\bf (a)}}
 \hspace{12mm}
 \includegraphics[width=0.4\textwidth]{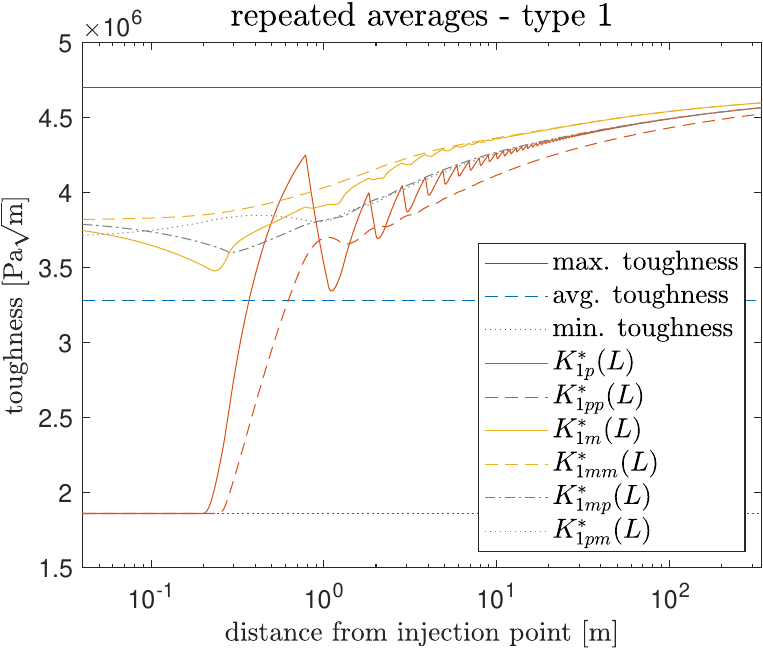}
	 \put(-205,155) {{\bf (b)}}
\\
 \includegraphics[width=0.4\textwidth]{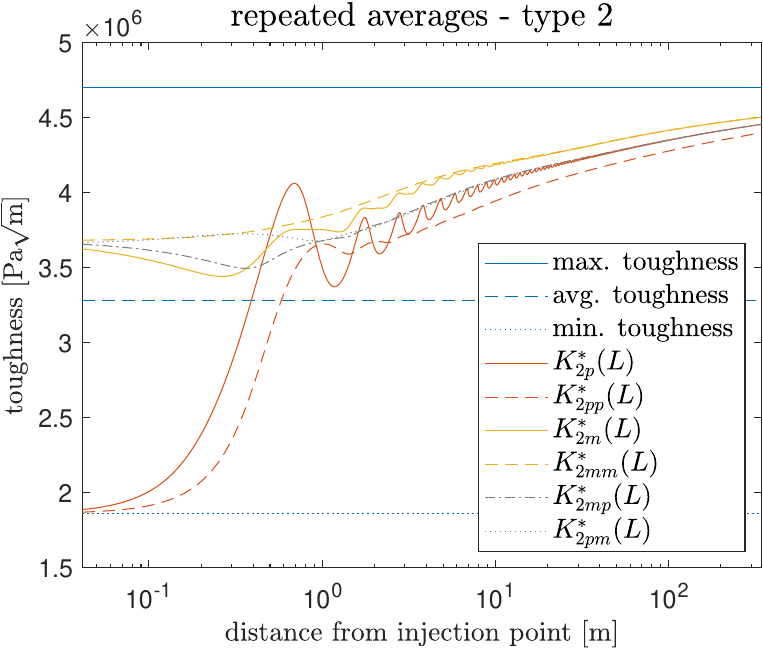}
	 \put(-205,155) {{\bf (c)}}
 \hspace{12mm}
 \includegraphics[width=0.4\textwidth]{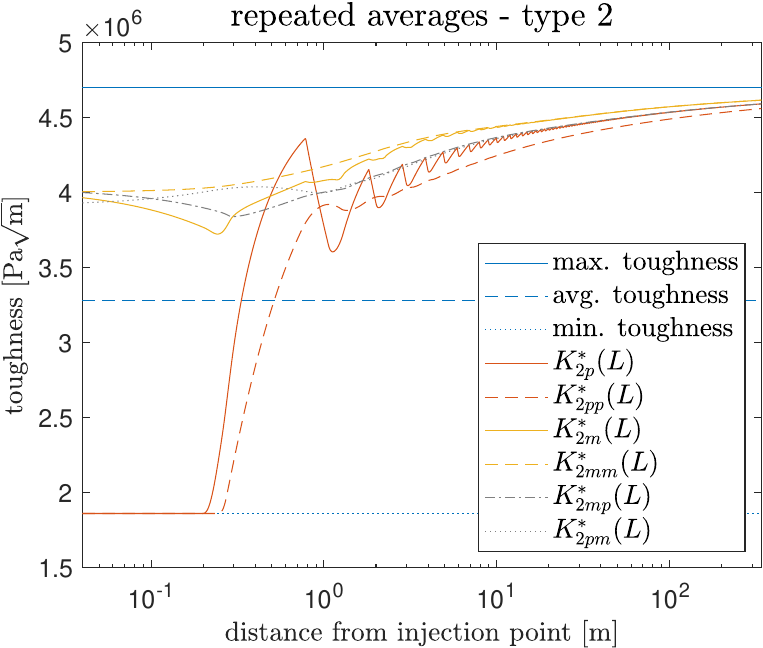}
	 \put(-205,155) {{\bf (d)}}
\caption{Repeated averaging of {\bf (a)}, {\bf (b)} $K_{1p}^*$ and $K_{1m}^*$, {\bf (c)}, {\bf (d)} $K_{2p}^*$ and $K_{2m}^*$, for $\delta_{max}=10$ and $\delta_{min}=1$, with {\bf (a), (c)} sinusoidal, {\bf (b), (d)} step-wise toughness distributions. }
	\label{fig:17old}
\end{figure}

\begin{figure}[h!]
\centering
 \includegraphics[width=0.4\textwidth]{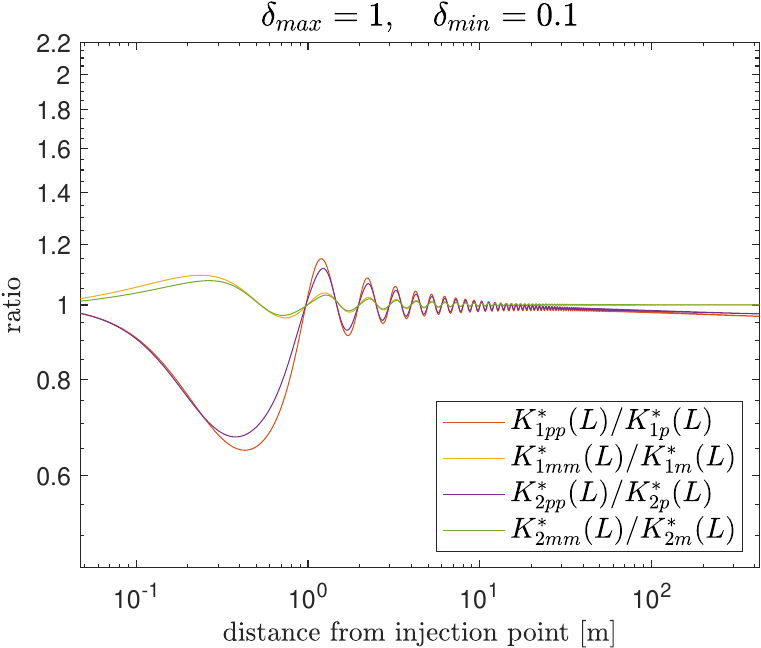}
	 \put(-205,155) {{\bf (a)}}
 \hspace{12mm}
 \includegraphics[width=0.4\textwidth]{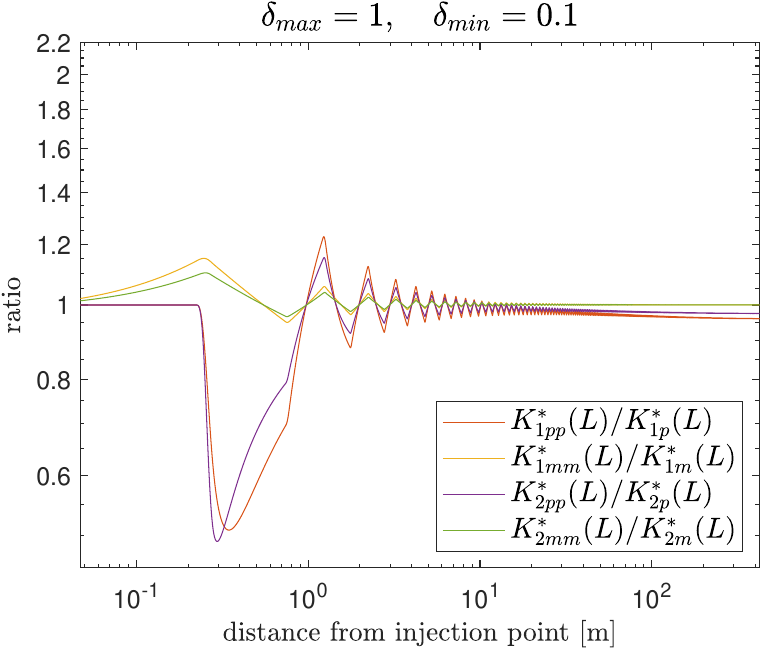}
	 \put(-205,155) {{\bf (b)}}
\\
 \includegraphics[width=0.4\textwidth]{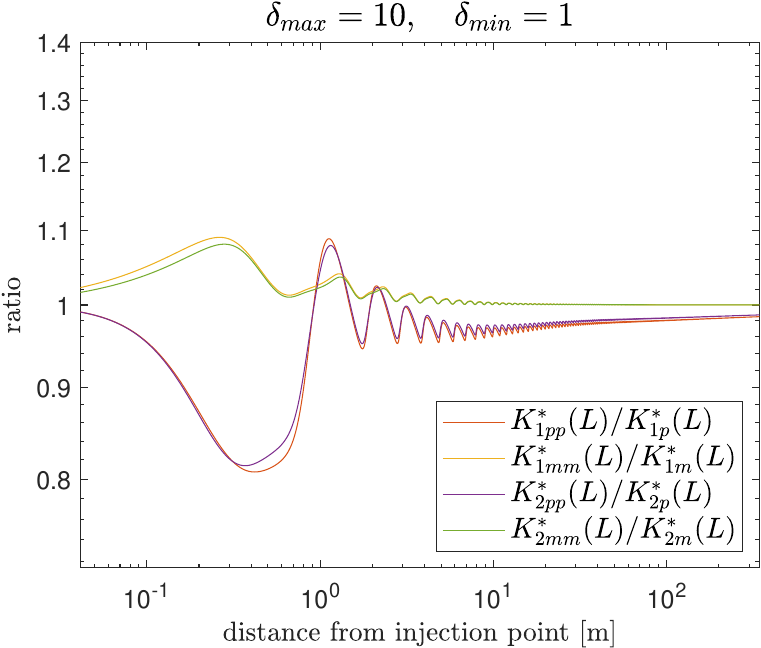}
	 \put(-205,155) {{\bf (c)}}
 \hspace{12mm}
 \includegraphics[width=0.4\textwidth]{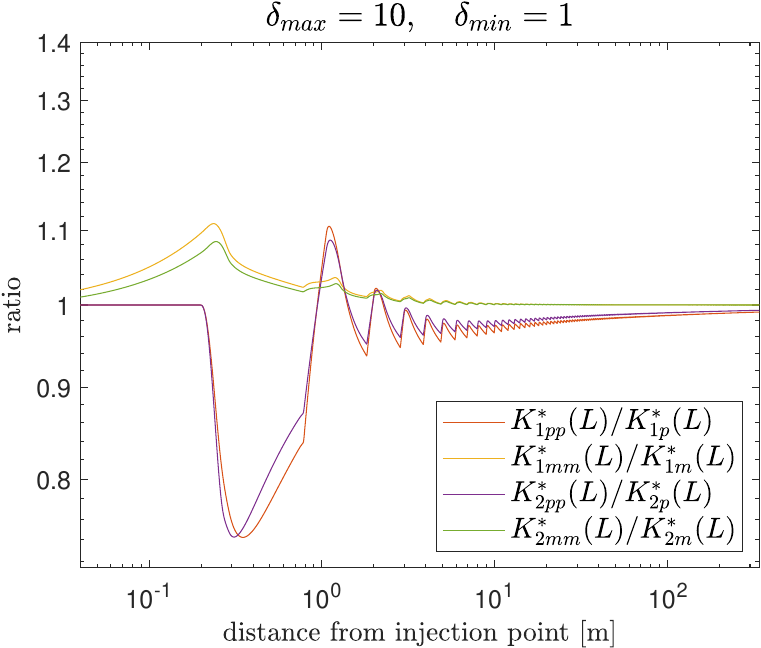}
	 \put(-205,155) {{\bf (d)}}
\\
 \includegraphics[width=0.4\textwidth]{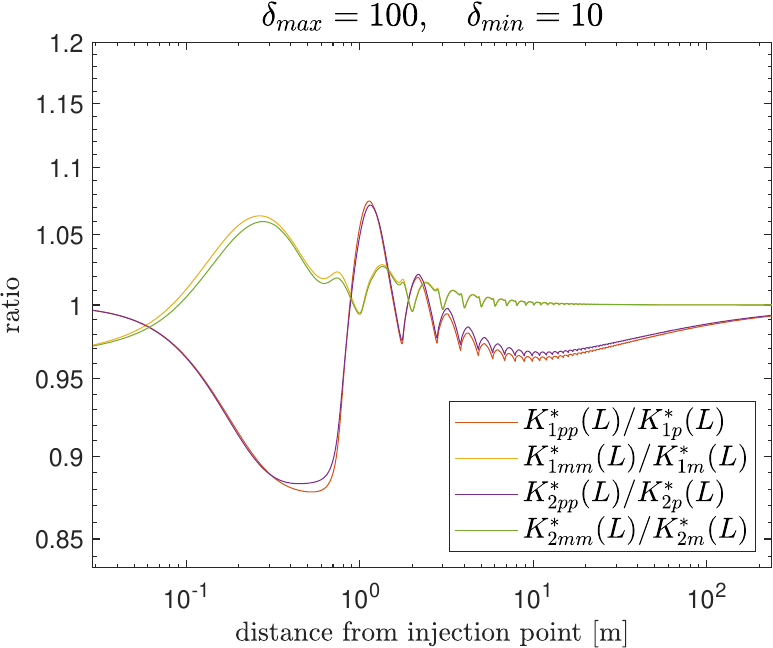}
	 \put(-205,155) {{\bf (e)}}
 \hspace{12mm}
 \includegraphics[width=0.4\textwidth]{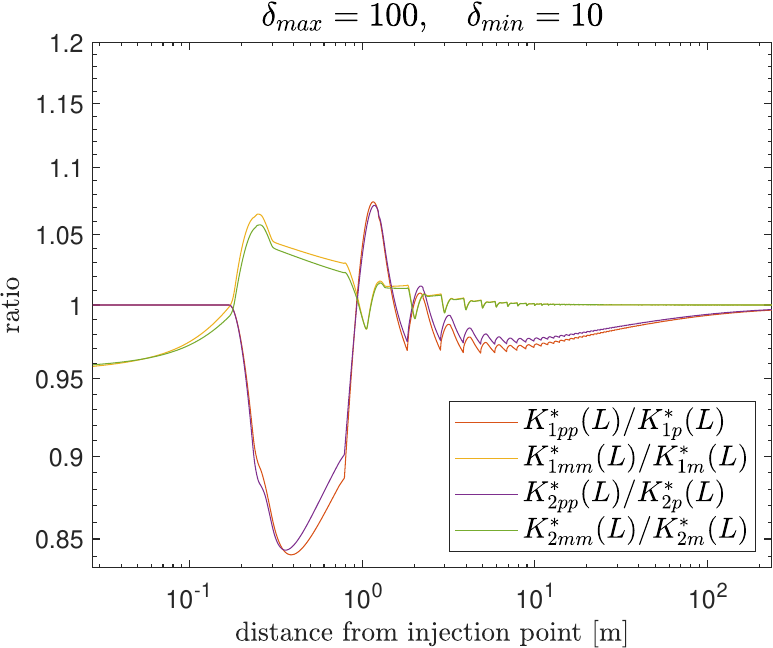}
	 \put(-205,155) {{\bf (f)}}
\caption{Ratios of the repeated averaging of $K_{jpp(mm)}^*$ and $K_{jp(m)}^*$
$j=1,2$ and various combinations of $(\delta_{min},\delta_{max})$.
{\bf (a), (c), (e)} sinusoidal, {\bf (b), (d), (f) } step-wise distributions.
Graphs on {\bf (a), (b)} correspond to the pair $\delta_{min}=0.1$, $\delta_{max}=1$,
Figures in {\bf (c), (d)} demonstrate the pair $\delta_{min}=1$, $\delta_{max}=10$, and, finally
{\bf (e), (f)} correspond to the pair $\delta_{min}=10$, $\delta_{max}=100$.
}
\label{fig:17}
\end{figure}

Some initial observations regarding the consistency of the proposed measures are easy to verify. For example, in the case with of a homogeneous material with constant toughness $K_{IC}$, we conclude that $K^*_{jp(m)} \equiv K_{IC}$, $j=1,2$, regardless of the crack propagation speed. Secondly, if a fracture is propagating with a constant speed in a material with a heterogeneous toughness (while it is highly difficult to imagine), then we have $K^*_{jp(m)}\equiv \langle K_{IC} \rangle_{jp}$, $j=1,2$.

Towards a more thorough investigation, we pose the following question: if the approach has been applied twice, does it produce a comparable outcome\footnote{Note that when using homogeneous techniques, in most cases the homoginised material will "reproduce itself" after repeated averaging}? In other words, we proceed as indicated in the following scheme:
\[
K_{IC}(x) \quad \overrightarrow{\mbox{simulation}} \quad w(x,t), p(x,t), v(t), L(t) \quad \Rightarrow  \quad v(L) \quad \rightsquigarrow \mbox{averaging} \rightsquigarrow \quad K^*_{jp(m)}(x)
\]
\[
K^*_{jp(m)}(x) \quad \overrightarrow{\mbox{sim.}} \quad w^*(x,t), p^*(x,t), v^*(t), L^*(t) \quad  \Rightarrow \quad  v^*(L) \quad \rightsquigarrow \mbox{aver.} \rightsquigarrow \quad K^*_{jpp(mm)}(x),
\]
and try to estimate the outcome of the process.
The results of applying the averaging twice are provided in Fig.~\ref{fig:17old}, while the ratio between the single/double averaging are given in Fig.~\ref{fig:17}. Here, we take the notation:
$K^*_{jp(m)}$, ($j=1,2$) as shown in the scheme.

It is clear from Fig.~\ref{fig:17old} that the averaging procedures based on the moving frame approach ($K^*_{1m}$, $K^*_{2m}$ and then, consequently, $K^*_{1mm}$, $K^*_{2mm}$) appear stable, with the second application of the averaging procedure producing a close but `smoother' approximation of the first. However, the same can not be said of the progressive averaging process ($K^*_{1p}$, $K^*_{2p}$ and then, consequently, $K^*_{1pp}$, $K^*_{2pp}$), with the result $K_{jpp}^*$ consistently taking a significantly lower value than that for $K_{jp}^*$. This difference in behaviour can be clearly seen in Fig.~\ref{fig:17}, where the ratio between results when applying repeated averaging against only applying the process once are presented. For the moving frame approach it is clear that the ratio tends to one as the distance increases, as should be expected, however it is not immediately apparent if this is true for the progressive procedure (as seen most clearly in Fig.~\ref{fig:17}a,b).

For the sake of curiosity, we also test the mixed iterations ($K^*_{jpm}$ and $K^*_{jmp}$). These exhibit similar behaviour, as can be seen in Fig.~\ref{fig:17old}, with each $p$-iteration decreasing the value of the `average', while subsequent $m$-iterations preserve the behavior (while each iteration makes the average toughness also smoother). Moreover, they are very close to one another
$K^*_{jpm}\approx K^*_{jmp}$ and situated between the respective averages $K^*_{jpp}$ and $K^*_{jmm}$ ($j=1,2$).

Overall behaviour of the measures, combined with the lower error of approximation obtained using $K_{jm}^*$ reported in the previous subsections, indicates that the `roaming' averages \eqref{approx_K_L2v} and \eqref{approx_K_L2evm} seem to be the more reliable of the two strategies. Meanwhile, over the total process time, the averaging measure based on the energy arguments, $K^*_{2m}$, demonstrates/delivers the best performance in predicting the real process parameters in comparison with all other measures.

\subsubsection{Changing the order of the layers}

As a final consistency check, we change the order of the layers in the periodic toughness distribution. Namely, so far we have had the crack starting from the interior of the weakest layer, while in this subsection we consider the case when the injection point is situated inside the tougher layer\footnote{In practice, this is equivalent to a translation of the toughness distribution by $dL / 2$. Note that the symmetry of the distribution about $x=0$ is preserved.}. Throughout this analysis, the periodic nature and distribution of the toughness remains the same as considered previously (as outlined in Sect.~\ref{ToughnessDistExp}).

\begin{figure}[h!]
\centering
 \includegraphics[width=0.4\textwidth]{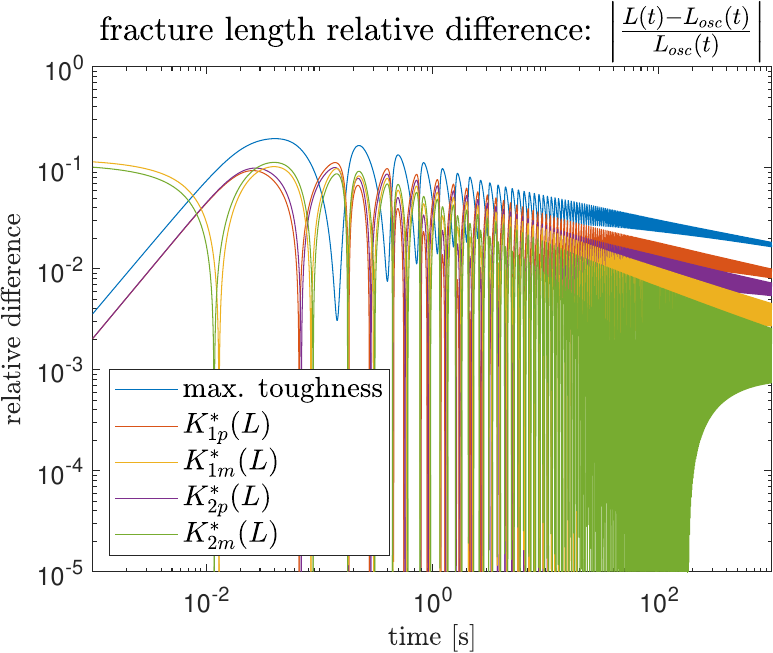}
	 \put(-205,155) {{\bf (a)}}
 \hspace{12mm}
 \includegraphics[width=0.4\textwidth]{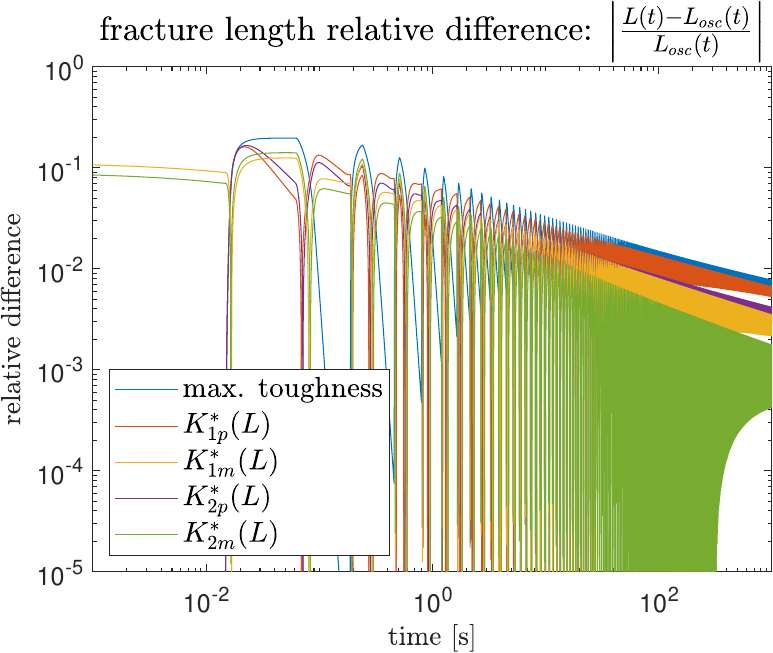}
	 \put(-205,155) {{\bf (b)}}
\\
 \includegraphics[width=0.4\textwidth]{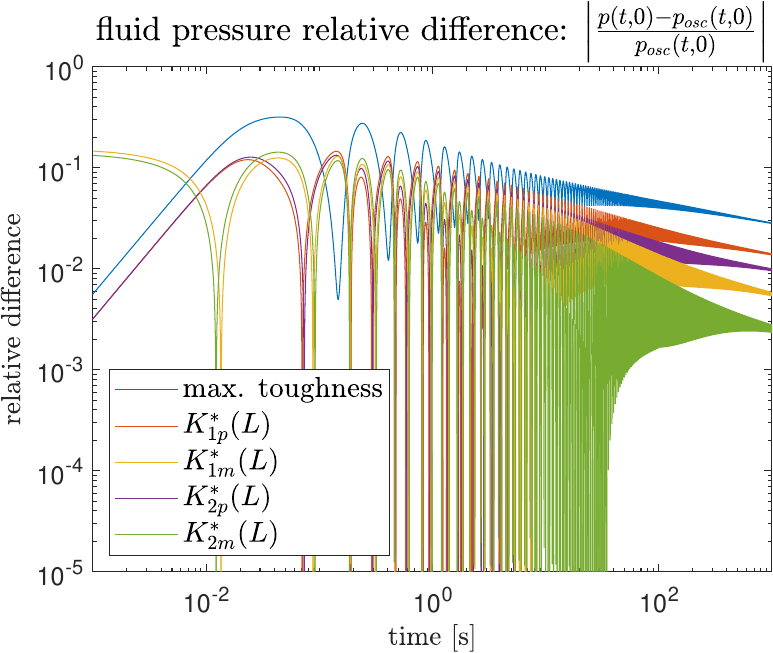}
	 \put(-205,155) {{\bf (c)}}
 \hspace{12mm}
 \includegraphics[width=0.4\textwidth]{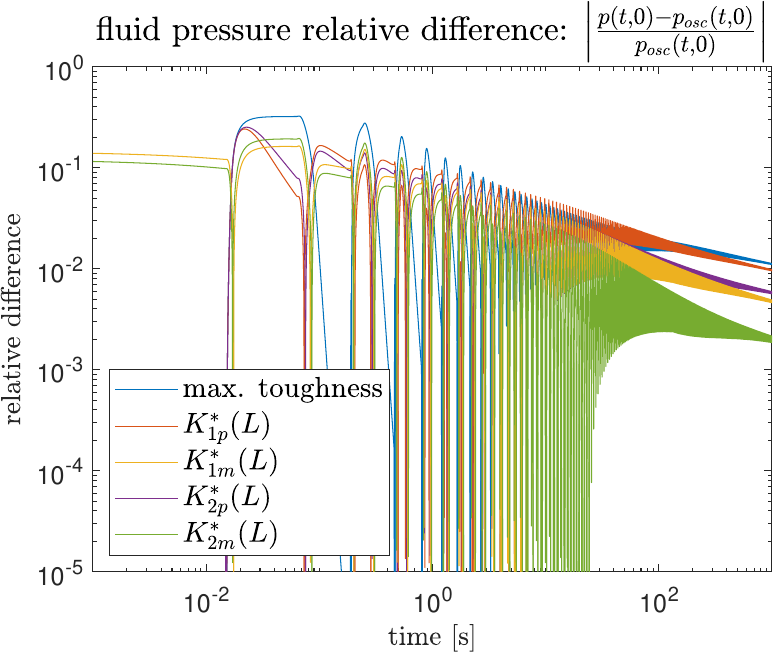}
	 \put(-205,155) {{\bf (d)}}
\\
 \includegraphics[width=0.4\textwidth]{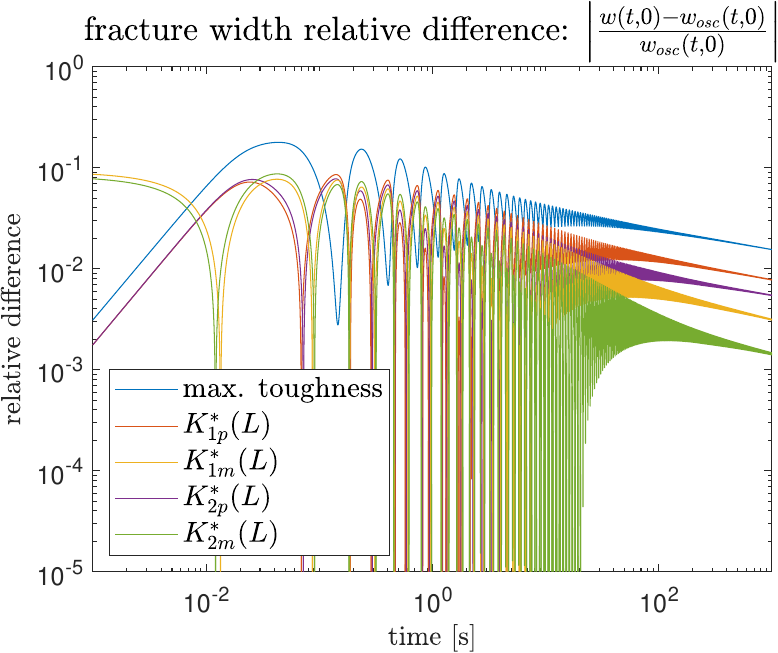}
	 \put(-205,155) {{\bf (e)}}
 \hspace{12mm}
 \includegraphics[width=0.4\textwidth]{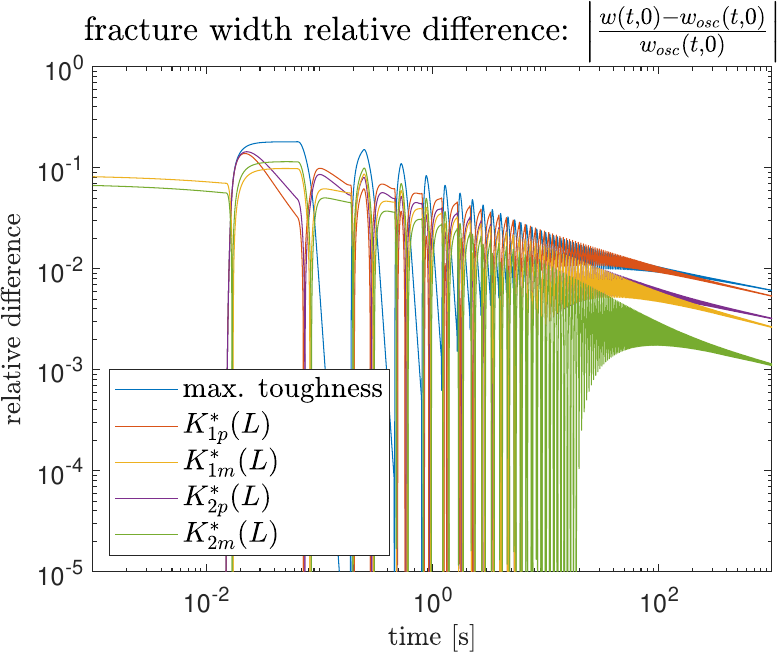}
	 \put(-205,155) {{\bf (f)}}
\caption{The case when crack initiation is in the middle of the tougher layer. Relative difference between the {\bf (a)}, {\bf (b)} fracture half-length $L(t)$, {\bf (c)}, {\bf (d)} fluid-induced pressure, {\bf (e)}, {\bf (f)} fracture aperture at the point of injection $w(t,0)$, obtained for oscillating toughness against those estimated via homogenization using the maximum toughness $K_{max}$ and the temporal averages $K_{1p(m)}^*$, $K_{2p(m)}^*$ from \eqref{approx_K_L1v}-\eqref{approx_K_L2v}, \eqref{approx_K_L2evp}-\eqref{approx_K_L2evm}. Here we consider the oscillating toughness given by $\delta_{max}=10$ and $\delta_{min}=1$ with distribution: {\bf (a)}, {\bf (c)}, {\bf (e)} sinusoidal, {\bf (b)}, {\bf (d)}, {\bf (f)} step-wise.}
	\label{fig:21}
\end{figure}

In Fig.~\ref{fig:21} the relative difference between the oscillating toughness solution and that obtained when homogenising using the four measures are given for three key process parameters: the fracture length, crack opening and pressure at the injection point. It is evident that the relative differences in these figures are practically identical to those presented previously in Fig.~\ref{fig:13a} and Fig.~\ref{fig:15} (for the fracture starting in the weaker layer).

\begin{figure}[h!]
\centering
 \includegraphics[width=0.4\textwidth]{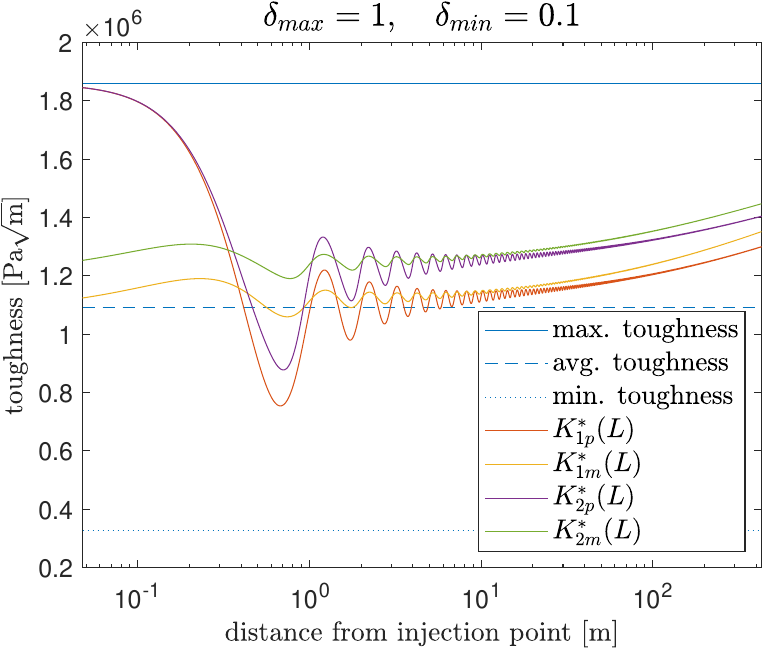}
	 \put(-205,155) {{\bf (a)}}
 \hspace{12mm}
 \includegraphics[width=0.4\textwidth]{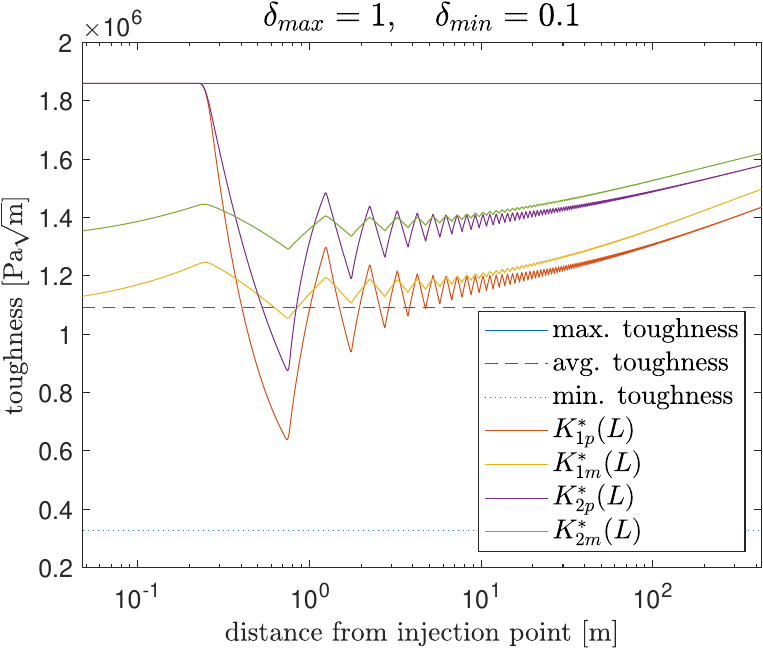}
	 \put(-205,155) {{\bf (b)}}
\\
 \includegraphics[width=0.4\textwidth]{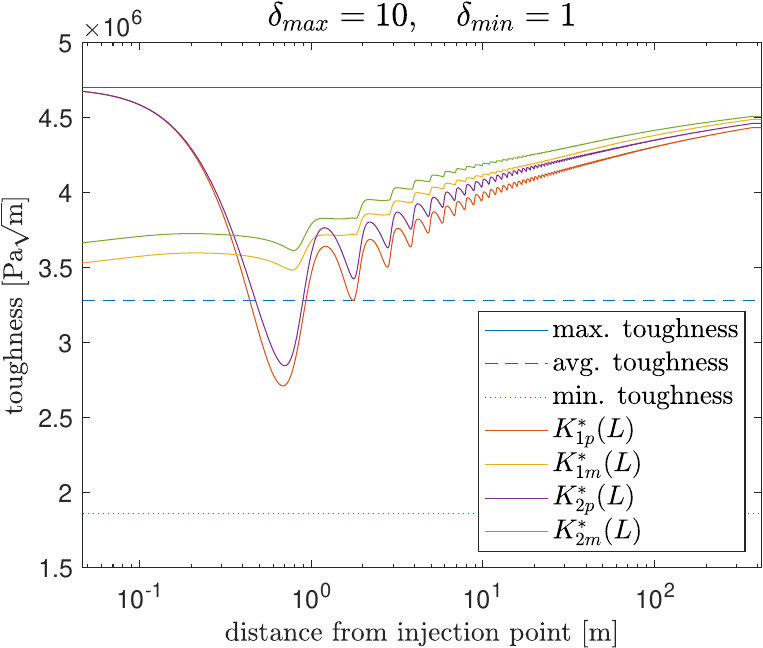}
	 \put(-205,155) {{\bf (c)}}
 \hspace{12mm}
 \includegraphics[width=0.4\textwidth]{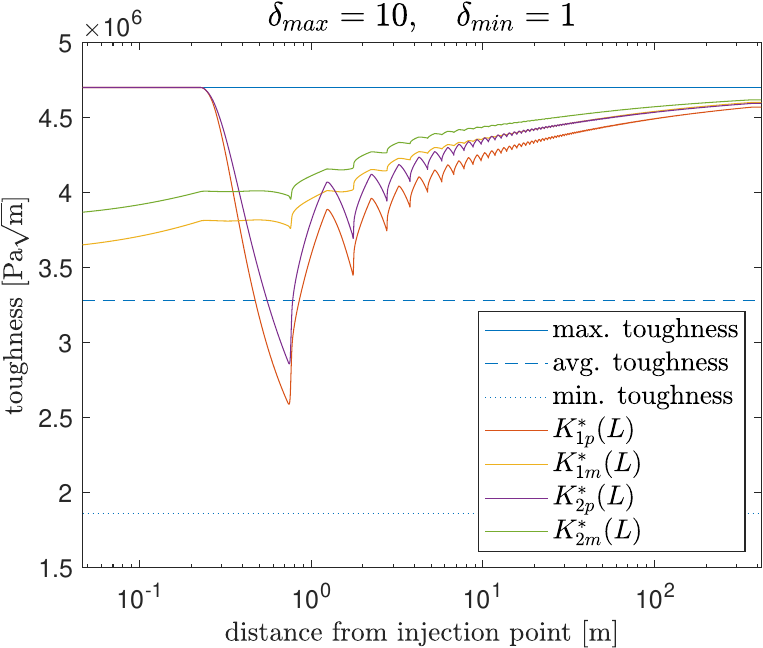}
	 \put(-205,155) {{\bf (d)}}
\\
 \includegraphics[width=0.4\textwidth]{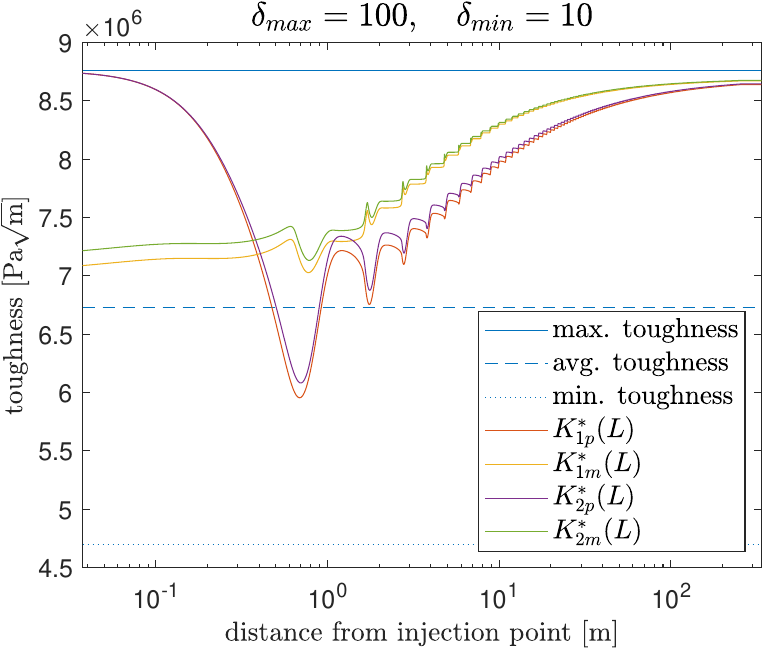}
	 \put(-205,155) {{\bf (e)}}
 \hspace{12mm}
 \includegraphics[width=0.4\textwidth]{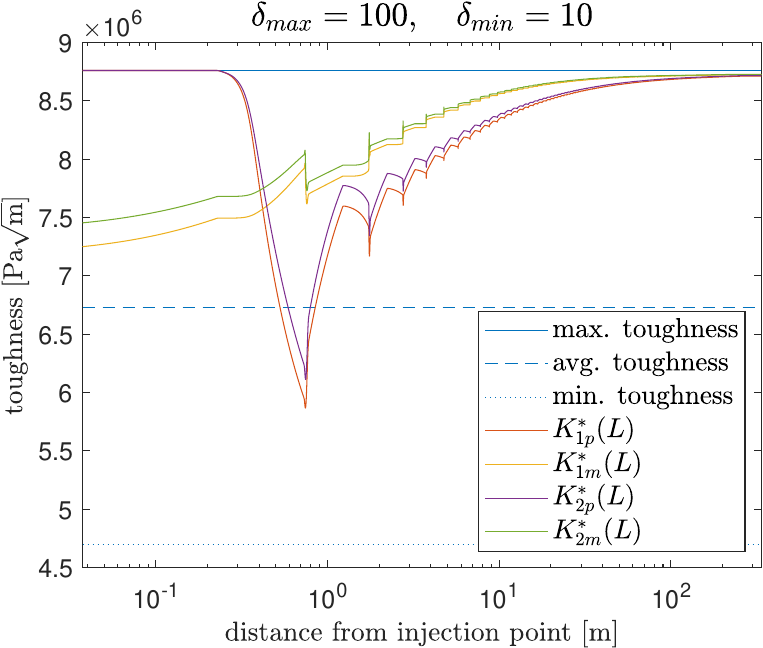}
	 \put(-205,155) {{\bf (f)}}
\caption{The case when crack initiation is in the middle of the tougher layer. Various averaging strategies for oscillating toughness for different combinations of the pairs $(\delta_{min},\delta_{max})$ with distribution: {\bf (a)}, {\bf (c)}, {\bf (e)} sinusoidal, and {\bf (b)}, {\bf (d)}, {\bf (f)} step-wise toughness distributions. Graphs on {\bf (a), (b)} correspond to the pair $\delta_{min}=0.1$, $\delta_{max}=1$.
Figures in {\bf (c), (d)} give the pair $\delta_{min}=1$, $\delta_{max}=10$, and, finally
{\bf (e), (f)} correspond to the pair $\delta_{min}=10$, $\delta_{max}=100$. }
	\label{fig:16}
\end{figure}

Meanwhile, in Fig.~\ref{fig:16}, we provide the values for all of the proposed measures for different combinations of the process parameters $\delta_{max}$ and $\delta_{min}$ (compare with Fig.~\ref{fig:20}). As should be expected, the progressive measures now overestimate the toughness in the initial moments of the fracture, while the moving average is not so strongly effected. However, it is apparent even from first glance that over larger distances (after passing through roughly 10 rock layers) the difference between the respective averages is not visible for any of the measures. 

\begin{figure}[h!]
\centering
 \includegraphics[width=0.4\textwidth]{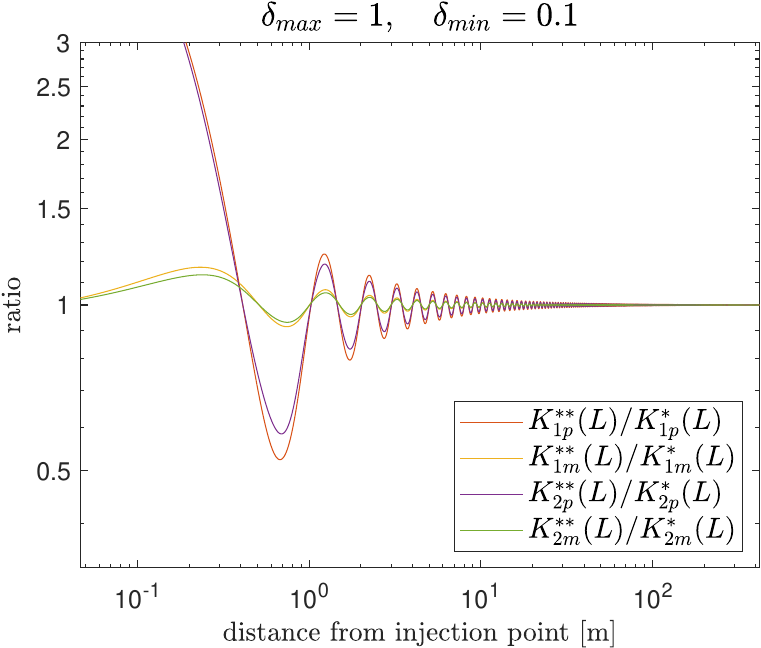}
	 \put(-205,155) {{\bf (a)}}
 \hspace{12mm}
 \includegraphics[width=0.4\textwidth]{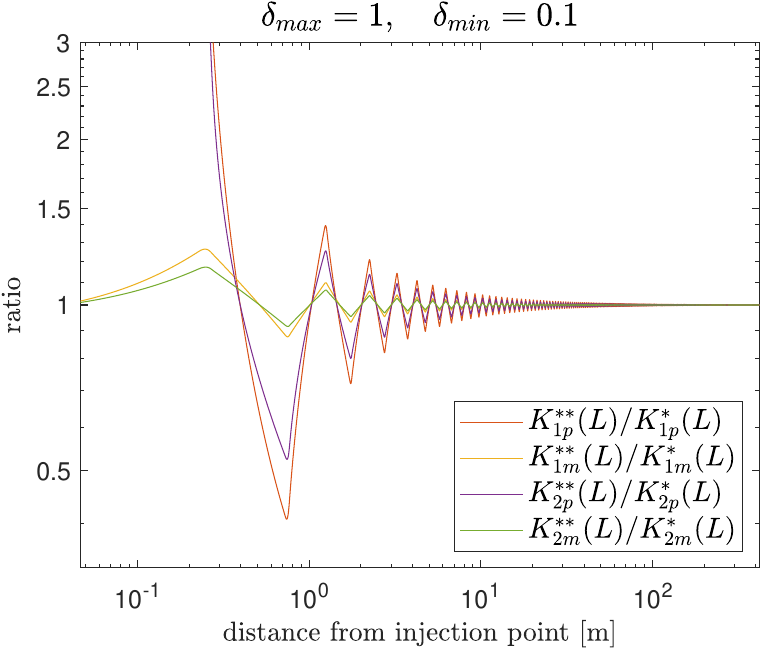}
	 \put(-205,155) {{\bf (b)}}
\\
 \includegraphics[width=0.4\textwidth]{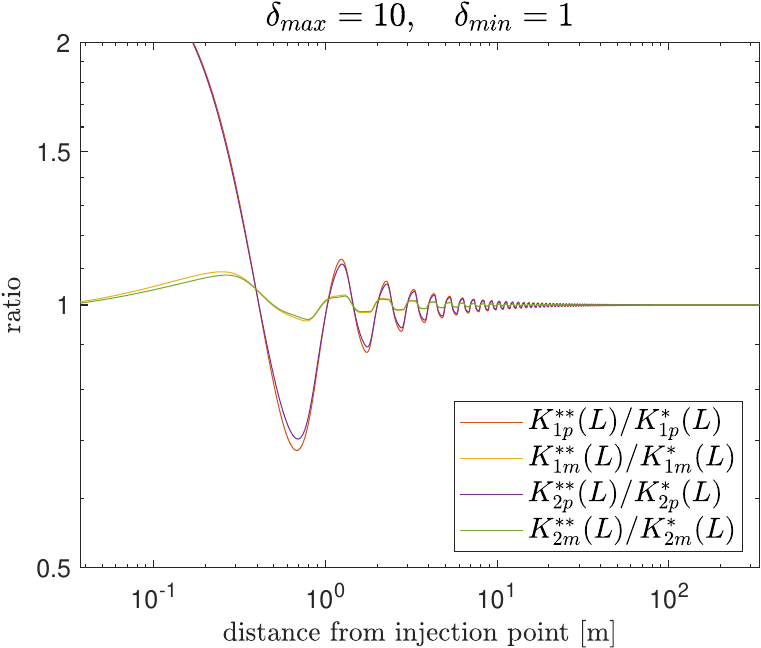}
	 \put(-205,155) {{\bf (c)}}
 \hspace{12mm}
 \includegraphics[width=0.4\textwidth]{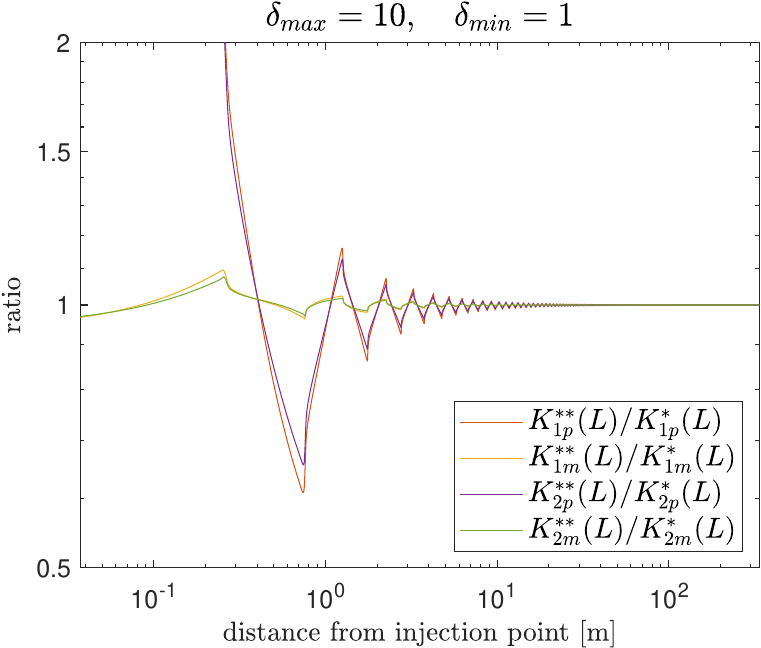}
	 \put(-205,155) {{\bf (d)}}
\\
 \includegraphics[width=0.4\textwidth]{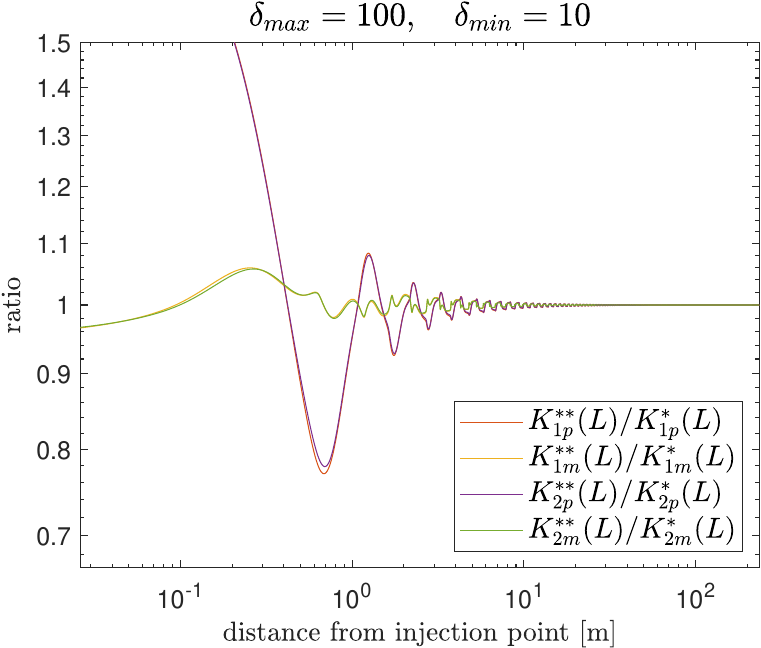}
	 \put(-205,155) {{\bf (e)}}
 \hspace{12mm}
 \includegraphics[width=0.4\textwidth]{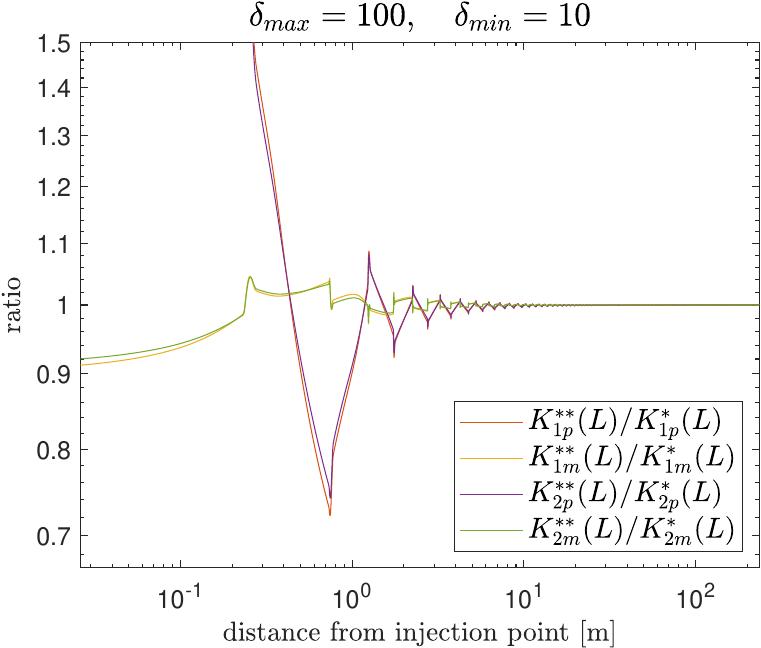}
	 \put(-205,155) {{\bf (f)}}
\caption{Ratio between the respective averages for the crack initiated in the middle of the tougher layer, $K^{**}_{jp(m)}$, to that for the crack initiated inside the weaker layer, $K^{*}_{jp(m)}$. All averaging strategies for oscillating toughness for different combinations of the pairs $(\delta_{min},\delta_{max})$ with distribution: {\bf (a)}, {\bf (c)}, {\bf (e)} sinusoidal, and {\bf (b)}, {\bf (d)}, {\bf (f)} step-wise toughness distributions. Graphs on {\bf (a), (b)} correspond to the pair $\delta_{min}=0.1$, $\delta_{max}=1$.
figures in {\bf (c), (d)} give the pair $\delta_{min}=1$, $\delta_{max}=10$, and, finally
{\bf (e), (f)} correspond to the pair $\delta_{min}=10$, $\delta_{max}=100$. }
	\label{fig:19}
\end{figure}

To highlight this more clearly, in Fig.~\ref{fig:19} we show the ratio between the averages obtained by each of the measures in each of the two cases, when it is initiated within the high toughness layer and when it starts within a weakened layer. It is clear that within the first period the results of the two are different, particularly for the progressive measures. This primarily concerns the location and nature of the peaks for each line, which correspond to the differing interfaces (high gradients) between the various rock layers. However, this difference gradually decreases in all cases, until becoming almost unrecognizable after a distance of roughly $10$ meters, or $10$ periods of the toughness. Furthermore, it is immediately apparent that the ratio tends to unity as the distance increases. 

While this helps to demonstrate the consistency of the approach it also demonstrates that, in the case where an almost exact solution is needed from the very start of computations, then the precise layering of the rock stata will need to be incorporated into the initial computations and the averaging-based homogenisation only incorporated during a later point in computations. Here, the results from Fig.~\ref{fig:19} give a good indication of the length-scale over which this effect would need to be accounted for. 

Combining these observations with the results of the previous subsection, it is clear from the consistency analysis that the roaming averages \eqref{approx_K_L2v}, \eqref{approx_K_L2evm} provide a more representative approach, with them both reproducing themselves under repeated averaging and showing less dependency on the ordering of the rock layers. In addition, the energy-based average $K_{2m}$ appears the slightly more effective of the two moving measures, whilst having the advantage that it is also viable in the case with heterogeneous elastic constants.  

\newpage

$\quad$

\newpage

\section{Discussions and conclusions}

A few strategies for handling the toughness heterogeneity of rock in hydraulic fracture (HF) based on temporal averaging have been considered. Four measures were introduced in \eqref{approx_K_L1v}-\eqref{approx_K_L2v} and \eqref{approx_K_L2evp}-\eqref{approx_K_L2evm}, with the different forms being based on the toughness or fracture energy, and for the global and roaming averages. These measures were formulated to directly incorporate aspects of the process behaviour, or more specifically, the instantaneous crack speed.
All those measures coincide for a constant toughness, while the fracture energy-based approaches have the advantage of being able to incorporate heterogeneous elastic constants.

An in-house developed time-space adaptive solver allowed us to produce extremely accurate computations with a guaranteed tolerance. These four measures were compared with the maximum toughness strategy 
\cite{DONTSOV2021108144} in a variety of different fracture regimes for an impermeable rock with periodic toughness.

It was demonstrated that the roaming measure $K_{jm}^*$, $j=1,2$ (where $j=1$ is the toughness-derived average, while $j=2$ represents the fracture energy-based measure), was not only more accurate than the global average $K_{jp}^*$, $j=1,2$, but also more consistent under repeated application and changes to the positioning of rock layers with respect to the injection point.
The following conclusions were also apparent:
\begin{itemize}
 \item For long fractures, or those where both rock layers permit the crack to remain in the viscosity dominated regime, the maximum toughness strategy is highly effective, with a typical error of the order of a few percent at the end of the process.
  \item For short and moderate fractures, or during the initial stages of propagation, the maximum toughness strategy is rather inaccurate. The level of error is process dependent, alongside depending on the maximum toughness of the rock, however may be of the order $20$\% or above (as observed even after $10$ seconds in Fig.~\ref{fig:1}).
  \item All of the averaging based approaches proposed in this paper consistently produced more accurate approximations than the maximum toughness strategy (see Sect.~\ref{Sect:Effective}). The highest gain is seen for short to moderate length fractures, where the maximum toughness strategy is at it's least effective. Results obtained by the authors so far indicate that, after five toughness periods, the averaging-based homogenisation achieves an up to $50$\% reduction in the relative error of the key process parameters (fracture length, width and inlet pressure) compared to the maximum toughness strategy for short, high-toughness fractures ($\delta_{max}=100$, $\delta_{min}=10$). Similarly, for such cracks the relative error obtained by the averaging approaches for these parameters never exceeds $10$\% after 5 cycles. The improvement over the maximum toughness approach decreases when considering the toughness-transient case ($\delta_{max}=10$, $\delta_{min}=1$, see Fig.~\ref{fig:13a}), however the relative error of these key process parameters again remains below $10$\% after three or four periods for all averaging-based approaches. As such, the approach of temporal averaging fulfills the aim of providing an improved homogenisation technique for short cracks, such as those considered in mini-frac tests.
  \item The maximum toughness strategy and temporal averaging approach were found to produce opposite bounds for the three key process parameters $L(t)$, $p(0,t)$ and $w(0,t)$ in the cases considered over most fracture lengths (i.e.\ if the $K_{max}$ solution provides an upper bound then $K_{jp}^*$ and $K_{jm}^*$ ($j=1,2$) provide a lower bound, and visa versa). As a result, combining the two approaches provides effective bounds on the behaviour of key process parameters.
\item Since $K_{jp}^*(0)=K_{IC}(0)$, the strategies based on $K_{1p}^*$ and $K_{2p}^*$ are more relevant for the initial part of the process (or short cracks), while $K_{1m}^*$ and $K_{2m}^*$ (especially the latter) provide the most accurate results afterwards. Moreover, as it clear from the Fig.~\ref{fig:20}, the averages intersect in the initial time (domain) of the crack propagation. Thus a mixed strategy can be useful where $K_{2p}^*$ average is utilised until the last intersection with $K_{2m}^*$ and then the latter is applied. However, this strategy would be advisable only when considering processes in a short-intermediate crack.
\item Consistency analysis performed on the measures demonstrated that the moving averages $K_{1m}$, $K_{2m}$ were more reliable, reproducing themselves under repeated averaging and displaying less dependence on the ordering of the rock layers. However, there was almost no dependence of the average on the positioning of the (periodic) rock layers with respect to the injection point for any measure after $10$ cycles of the toughness. In the event of using a mixed strategy, as discussed above, this also gives an indication of where the switch between strategies should take place.
\item When implementing these averaging procedures into commercial solvers, the roaming averages $K_{1m}$ and $K_{2m}$ should have the integration interval $dL$ linked to the minimal size of the finite element, to allow for effective computations (see footnote~\ref{Footy5} on page~\pageref{Footy5}).
\item All of the results demonstrate that an averaging based approach (using any of the outlined measures) has great potential to form part of an effective homogenisation procedure constructed around mimicking the real instantaneous (local) crack speed.

\end{itemize}

In light of the above, we can conclude that an averaging based approach can be successfully used to approximate the material toughness in the HF process. The new averaging-based homogenisation achieves the aim of this work, significantly reducing the error for crucial process parameters when working with short and moderate cracks, and even improves upon existing prediction strategies for long fracture lengths. The crucial change is that the averaging must be specifically weighted in order to incorporate the influence of both the rock heterogeneity and the process peculiarities. While the focus in this case was on the HF process, this approach to homogenising utilizing averaging could potentially be employed in other areas of application.

There does, however, remain one crucial obstacle to successful application of this technique, namely that the measures require prior knowledge of the (unknown) instantaneous fracture velocity. This can, in principle, be approximated sufficiently accurate utilizing known properties of the process behaviour, which will be the subject to a future paper by the authors. 

Implementation of this averaging-based homogenisation strategy on a wider scale also necessitates investigating the effect of other heterogeneous effects. This includes many of the measures outlined in the introduction, including crack redirection, either to a tortuous path or permanent deflection, or other effects of the elastic heterogeneity like stress shielding or other hardening effects, amongst others. While the direct homogenisation of elastic heterogeneity can already be incorporated for the temporal energy averaging approach, the ability to incorporate these wider effects needs to be demonstrated.

For a wider averaging-based homogenisation procedure to be effective in HF, there also needs to be further investigation into the effect of other process inhomogeneities on the proposed measures. In addition, there has recently been a greatly increased interest in the driving of HF processes utilizing ``pulse loading'' techniques, with the aim of introducing a greater damaged area than via traditional static loading (a good overview of the topic can be found in the introduction of \cite{Xi2021}). Another area where non-uniform crack propagation has been observed is hydraulic fracture in poro-elastic media with saturation \cite{CAO201724,Schrefler2019a}.

Note that in this paper we have discussed only a few 'balanced' toughness periodic distributions where the toughness heterogeneity is normal to the direction of fracture propagation. As was made clear by the preliminary research \cite{ARMA2021}, the nature of this distribution will play an important role and will manifest itself during the averaging. The extent to which the approach will be effective when considering differing orientations of the toughness inhomogeneity, such as the case when rock layers are instead aligned at some angle to the direction of fracture propagation needs to be demonstrated. Another question that is now under analysis is the case of random toughness distribution, where the notion of the maximal toughness is not well defined at all, and which has been shown to significantly effect the resulting fracture process and geometry \cite{HUANG2019207}.

Finally, fracture propagation, in the context of the observed step-wise nature of the crack advancement (see e.g. \cite{CAO201724,Schrefler2019a,Schrefler2021}), is crucial for understanding the nature of this important technological and natural process (material destruction). Various scenarios are possible here, not only a monotonic (but non uniform in time) crack propagation but also branching, clustering and forerunning like behaviours have been already observed and analysed \cite{MiMoSle2008,MiMoSle2009,SleAyMi2015,NiMiSle2016,NI2021104024,Schrefler2021}. While  having their own peculiarities, their impact on the effectiveness of any averaging approach (if feasible) needs to be considered.

\section*{Funding}
The authors have been funded by Welsh Government via S$\hat{\mbox {e}}$r Cymru Future Generations Industrial Fellowship grant AU224 and the European Union's Horizon 2020 research and innovation programme under the Marie Sklodowska-Curie grant agreement EffectFact No 101008140.

\section*{Acknowledgements}
GM is thankful to the Royal Society for the Wolfson Research Merit Award, while MD acknowledges the Royal Academy of Engineering for the Industrial Fellowship. The authors are very grateful to the (anonymous) reviewers for their helpful comments.


\bibliography{Penny_Bib}
\bibliographystyle{abbrv}

\end{document}